\documentclass[11pt]{article}
\usepackage{epsf,amsmath,amssymb,axodraw,graphicx,scalefnt}
\usepackage{subfigure}
\usepackage{color}
\usepackage{rotating}
\newcommand{\llx}{{\abbrev LL}$x$}
\newcommand{\eft}{{\abbrev EFT}}
\newcommand{\qcd}{{\abbrev QCD}}
\newcommand{\lhc}{{\abbrev LHC}}

\newcommand{\lo}{{\abbrev LO}}
\newcommand{\nlo}{{\abbrev NLO}}
\newcommand{\nnlo}{{\abbrev NNLO}}
\newcommand{\msbar}{\overline{\mbox{\abbrev MS}}}

\newcommand{\abbrev}{\scalefont{.9}\rm}
\newcommand{\lt}{l_{t}}

\newcommand{\lFR}{l_{\rm FR}}
\newcommand{\lhF}{l_{\rm F}}
\newcommand{\lhR}{l_{\rm R}}
\newcommand{\muF}{\mu_{\rm F}}
\newcommand{\muR}{\mu_{\rm R}}
\newcommand{\mhiggs}{M_{\rm H}}
\newcommand{\mtop}{M_{t}}

\newcommand{\api}{\frac{\alpha_s}{\pi}}
\newcommand{\eqn}[1]{Eq.\,(\ref{#1})}
\newcommand{\fig}[1]{Fig.\,\ref{#1}}

\newcommand{\dd}{{\rm d}}

\newcommand{\order}[1]{{\cal O}(#1)}
\newcommand{\bld}[1]{\boldmath{$#1$}}

\date{}
\title{\vspace*{-6em}
  \begin{flushright}
    {\sf\small December 2009 --- WUB/09-18
      \\MAN/HEP/2009/44}
  \end{flushright}
  \vspace*{2em} Higgs production in gluon fusion at
  next-to-next-to-leading order QCD for finite top mass} 
\author{%
  Robert V. Harlander$^{(1)}$, 
  Hendrik Mantler$^{(1)}$, 
  Simone Marzani$^{(2)}$, Kemal J.
  Ozeren$^{(1)}$\\[2em] 
  {\it $^{(1)}$Fachbereich C, Bergische Universit\"at
    Wuppertal}\\{\it 42097 Wuppertal, Germany}\\
  {\it $^{(2)}$School of Physics \& Astronomy,
    University of Manchester,}\\
  {\it Manchester M13 9PL, United Kingdom}
  }

\begin{document}
\maketitle

\begin{abstract}
The inclusive Higgs production cross section from gluon fusion is
calculated through \nnlo{} \qcd{}, including its top quark mass
dependence. This is achieved through a matching of the $1/\mtop$
expansion of the partonic cross sections to the exact large-$\hat s$
limits which are derived from $k_T$-factorization. The accuracy of this
procedure is estimated to be better than $1\%$ for the hadronic cross
section. The final result is shown to be within $1\%$ of the commonly
used effective theory approach, thus confirming earlier findings.
\end{abstract}

\section{Introduction}
It is well-known that a reliable quantitative prediction of the gluon
fusion production cross section for Higgs bosons requires a
next-to-next-to-leading order (\nnlo{}) calculation~(for a review on
Higgs physics, see
Refs.\,\cite{Djouadi:2005gi,Djouadi:2005gj}). However, since it is a
loop-induced process, its NNLO correction requires a three-loop
calculation of a $2\to 1$ process. Fortunately, it was found at
next-to-leading order
(\nlo{})~\cite{Dawson:1990zj,Graudenz:1992pv,Spira:1995rr} that the
perturbative K-factor is very well reproduced in the so-called {\it
  effective field theory {\rm (\eft{})} approach}, where the gluon-Higgs
coupling is taken into account by an effective Lagrangian
\begin{equation}
\begin{split}
{\cal L}_{\rm eff} &= -\frac{H}{4v}C_1G_{\mu\nu}G^{\mu\nu}\,,
\end{split}
\end{equation}
with $H$ the Higgs field, $G_{\mu\nu}$ the gluonic field strength
tensor, $v=246$\,GeV the vacuum expectation value of the Higgs field,
and $C_1$ a perturbatively evaluated Wilson coefficient (see, e.g.,
Ref.\,\cite{Schroder:2005hy,Chetyrkin:2005ia}). The \nlo{} cross section
in the \eft{} approach is then obtained by scaling the \lo{} cross
section (obtained in the full theory) with the effective \nlo{}
K-factor.

Although, to our knowledge, a quantitative understanding of the accuracy
of this approach is still missing (in the sense that there is no
  error estimate), the observed difference of less than 1\% to the full
\nlo{} cross section (which is known in numerical
form~\cite{Spira:1995mt}) for $\mhiggs<2\mtop$ was considered to be
sufficiently convincing in order to trust the \eft{} approach also at
\nnlo{}.

Apart from the inclusive \nnlo{}
calculation~\cite{Harlander:2002wh,Anastasiou:2002yz,Ravindran:2003um},
the heavy-top limit has also been used for distributions, resummations,
and even fully differential quantities at \nnlo{} (for a review,
see~Ref.\,\cite{Harlander:2007zz}). It is therefore of the utmost
importance to justify the validity of the \eft{} approach.  This has
been first achieved at \nnlo{} in
Ref.\,\cite{Harlander:2009mq,Harlander:2009bw,Pak:2009dg,Pak:2009bx} by
an expansion of the relevant Feynman diagrams in the limit
$\mhiggs^2,\hat s\ll \mtop^2$, where $\sqrt{\hat s}$ is the partonic
center-of-mass energy. The apparent failure of this expansion for large
$\hat s$ (which is only restricted by the hadronic center-of-mass energy
squared, $s$) is only partly cured by the strong suppression of the
parton luminosity. The prediction of the $gg$ channel contribution,
which accounts for more than 95\% of the \nlo{} hadronic cross section,
was additionally treated by matching to the known large-$\hat s$
behaviour~\cite{Marzani:2008az}. The other channels, for which the
large-$\hat s$ behaviour is not known, were treated by including only
those terms in the $1/\mtop$ expansion up to which the series was
observed to converge. It was found that the resulting cross section
agrees with the \eft{} result to better that $1\%$ over the relevant
mass range between $100$ and $300$\,GeV.

In this paper, we extend the analysis of
Refs.\,\cite{Harlander:2009mq,Pak:2009dg,Marzani:2008az} by deriving the
high-energy limits of the other channels as well. This leads to a
significant stabilization of the $qg$ channel which contributes about
2-5\% to the total \nlo{} cross section. The peculiar threshold
behaviour of the quark--anti-quark channel prohibits a reasonable
approximation from the high- and the low-energy information alone, but
its contribution is in any case only at the per-mille level. We can
therefore safely claim to present a stable prediction for the total
cross section including top quark mass effects for Higgs masses between
100 and 300~GeV.

\section{Preliminaries}

\subsection{Notation}

For the convenience of the reader, let us outline the notation at the
very beginning. The Higgs mass is denoted by $\mhiggs$, the on-shell top
quark mass by $\mtop$, and the hadronic and the partonic center of mass
energies are $s$ and $\hat s$, respectively. Unless indicated otherwise,
$\alpha_s\equiv \alpha_s^{(5)}(\muR^2)$ denotes the strong coupling in
the $\msbar$ scheme for
five active flavours at the renormalization scale $\muR$. The following
variables will turn out to be useful throughout the text:
\begin{equation}
\begin{split}
&z = \frac{\mhiggs^2}{s}\,,\qquad
x = \frac{\mhiggs^2}{\hat s}\,,\qquad
\tau = \frac{4\mtop^2}{\mhiggs^2}\,,\qquad \omega=\frac{\hat s}{s}\,.
\\
&\lhF = \ln\frac{\muF^2}{\mhiggs^2}\,\qquad
\lhR = \ln\frac{\muR^2}{\mhiggs^2}\,\qquad
\lt = \ln\frac{\mtop^2}{\mhiggs^2}\,,
\end{split}
\end{equation}
with the factorization scale $\muF$.

The inclusive hadronic cross section $\sigma_{pp'}$ for Standard Modell
Higgs production in proton--(anti-)proton collisions is obtained by
convoluting the partonic cross section $\hat\sigma_{\alpha\beta}$ for
the scattering of parton $\alpha$ with parton $\beta$ by the
corresponding parton density functions $\phi_{\alpha/p}(x)$ ({\abbrev
  PDF}s):
\begin{equation}
\begin{split}
\sigma_{\alpha\beta}(z,\tau,\lhF) &= \int_{z}^1\dd \omega \, {\cal
  E}_{\alpha\beta}(\omega,\muF)\,\hat\sigma_{\alpha\beta}(z/\omega,\tau,\lhF)\,,\\ \sigma_{pp'}(z,\tau)
&= \sum_{\alpha,\beta\in\{q,\bar
  q,g\}}\sigma_{\alpha\beta}(z,\tau,\lhF)\,,\qquad p'\in \{p,\bar
p\}\,,\\ {\cal E}_{\alpha\beta}(\omega,\muF) &\equiv
\int_{\omega}^1\frac{\dd y}{y}\left[
  \phi_{\alpha/p}(y,\muF)\phi_{\beta/p'}(\omega/y,\muF)\right]\,.
\label{eq::sigmapp}
\end{split}
\end{equation}
Note that the $\sigma_{\alpha\beta}$ depend on the factorization scheme
(we use $\msbar$ throughout this paper); only their sum $\sigma_{pp'}$
is physical and thus formally independent of the factorization
scale. Nevertheless, it will be useful to study the individual
contributions to the total cross section separately as they have very
different characteristics. The ${\cal E}_{\alpha\beta}$ are parton
luminosities and will be discussed in more detail in
Section~\ref{sec::soutline}.

We write the top quark induced partonic cross section as
\begin{equation}
\begin{split}
\hat\sigma_{\alpha\beta}(x,\tau,\lhF) &=
\sigma_0(\tau)\,\Delta_{\alpha\beta}(x,\tau,\lhF)\,,
\label{eq::sigmahat}
\end{split}
\end{equation}
with
\begin{equation}
\begin{split}
\sigma_0(\tau) &= \frac{\pi\sqrt{2}G_{\rm F}}{256}\left(\api\right)^2
\tau^2\left|1+(1-\tau)\arcsin^2\frac{1}{\sqrt{\tau}}\right|^2\,,
\label{eq::sigma0}
\end{split}
\end{equation}
where $G_{\rm F}\approx 1.16637\cdot 10^{-5}$\,GeV$^{-2}$ is Fermi's
constant. The kinetic terms assume the form
\begin{equation}
\begin{split}
\Delta_{\alpha\beta}(x,\tau,\lhF) = \delta_{\alpha g}\delta_{\beta
  g}\,\delta(1-x) + \sum_{n\geq
  1}\left(\api\right)^n
\Delta^{(n)}_{\alpha\beta}(x,\tau,\lhF,\lhR)\,.
\label{eq::deltapert}
\end{split}
\end{equation}
At \nlo{}, the full $\mtop$ dependence is known in numerical
form~\cite{Graudenz:1992pv} (the virtual terms are known
analytically~\cite{Harlander:2005rq,Anastasiou:2006hc,Aglietti:2006tp}).

A fully general result for the partonic cross section at \nnlo{} is as
of yet unknown. In
Refs.\,\cite{Harlander:2009mq,Harlander:2009bw,Pak:2009dg,Pak:2009bx}, it
was evaluated in terms of an expansion of the form
\begin{equation}
\begin{split}
\Delta_{\alpha\beta}(x,\tau,\muF) &= \sum_{i\geq
  0}\left(\frac{\mhiggs^2}{\mtop^2}\right)^i
\Omega_{\alpha\beta,i}(x,\lt,\muF)\,,
\label{eq::mtexp}
\end{split}
\end{equation}
with the analogous perturbative expansion as in \eqn{eq::deltapert}.  At
\nnlo{}, the first four terms ($i\leq 3$) have been
evaluated~\cite{Harlander:2009mq,Harlander:2009bw}. The so-called \eft{}
approach which has been used in all higher order analyses up to now, can
be derived from the leading term of this expansion:
\begin{equation}
\begin{split}
\sigma_{pp',\infty}(z,\lt) &\equiv \sum_{\alpha,\beta\in\{q,\bar
  q,g\}}\int_{z}^1\dd \omega \, {\cal
  E}_{\alpha\beta}(\omega,\muF)\,
\hat\sigma_{\alpha\beta,\infty}(z/\omega,\lt,\lhF)\,,\\ 
&\hat\sigma_{\alpha\beta,\infty}(x,\lt,\lhF)
\equiv \sigma_0(\tau)\,\Omega_{\alpha\beta,0}(x,\lt,\lhF)\,,
\label{eq::heavytop}
\end{split}
\end{equation}
where $\sigma_0$ is given in \eqn{eq::sigma0}.

\begin{figure}
  \begin{center}
      \subfigure[]{\includegraphics[bb=110 265 465 560,width=.45\textwidth]{%
        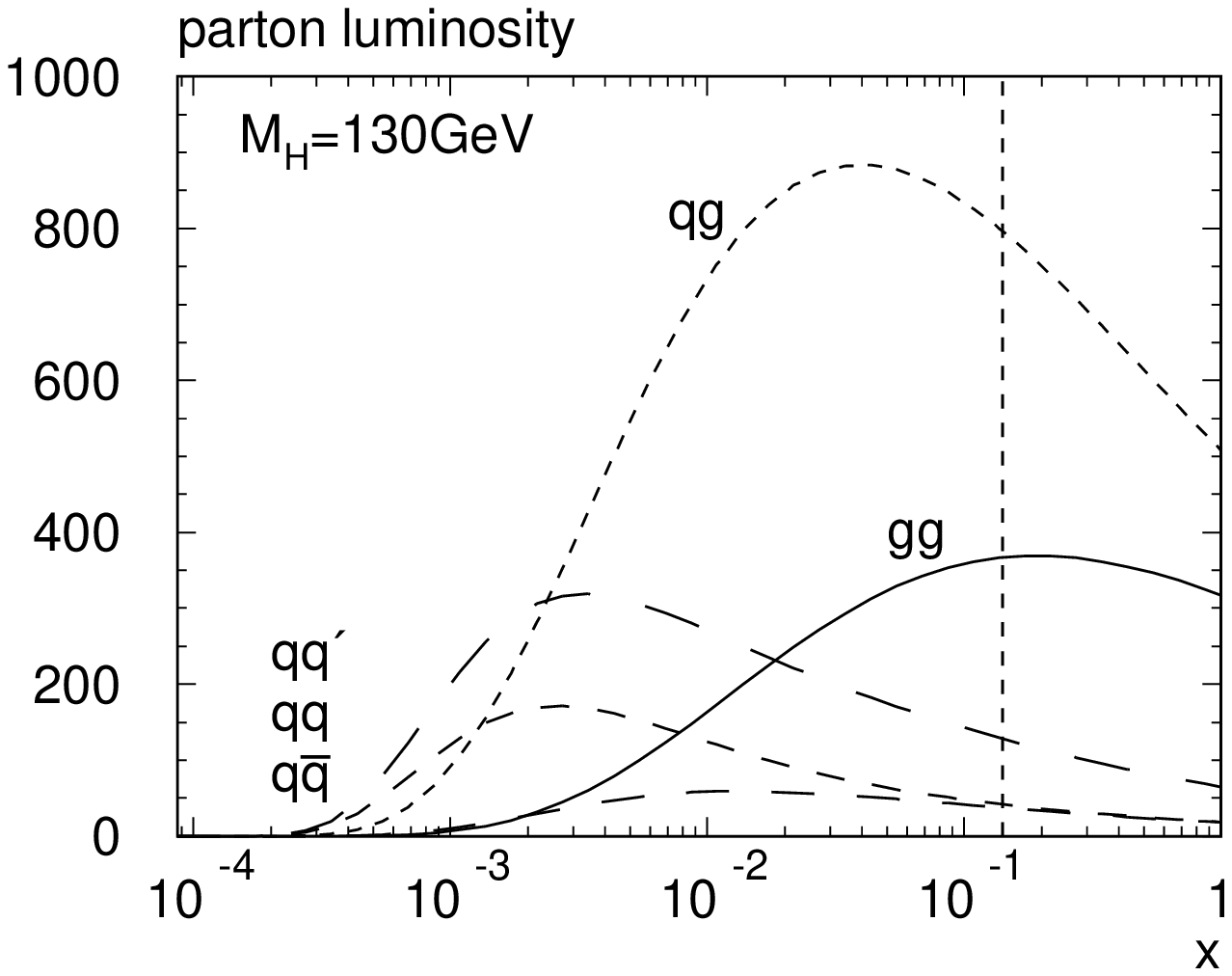}}\quad
      \subfigure[]{\includegraphics[bb=110 265 465 560,width=.45\textwidth]{%
        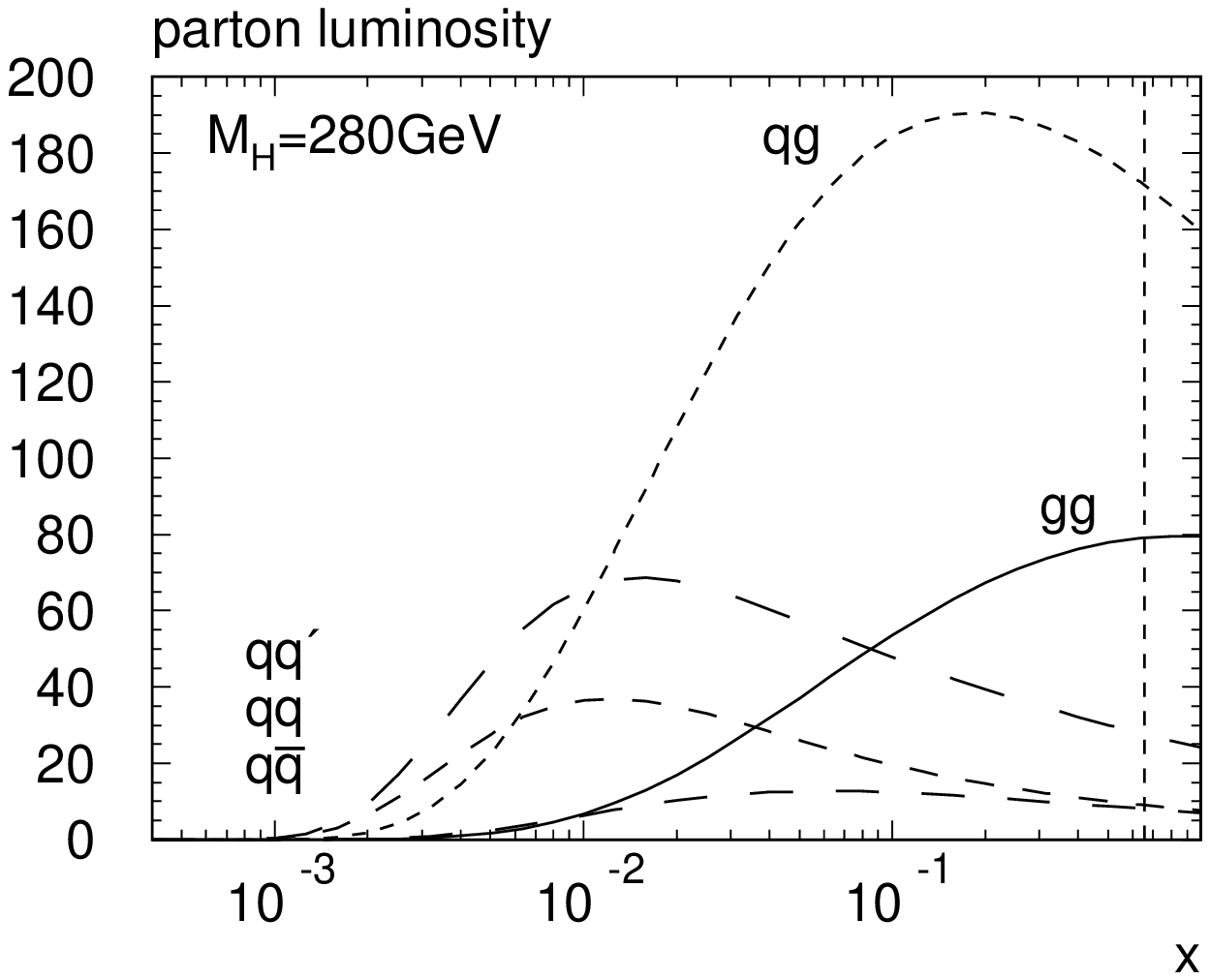}}
    \parbox{.9\textwidth}{
      \caption[]{\label{fig::lumis}\sloppy Parton luminosities ${\cal
          E}(\omega=z/x)$ at the \lhc{} for $\mtop=170.9$\,GeV at
        (a)~$\mhiggs=130$\,GeV and (b)~$\mhiggs=280$\,GeV, plotted as
        functions of $x=\mhiggs^2/\hat s$. The vertical line denotes the
        threshold $\hat s=4\mtop^2$. }}
  \end{center}
\end{figure}

\begin{figure}
  \begin{center}
      \subfigure{%
        \includegraphics[bb=100 510 470 760,width=.49\textwidth]{%
          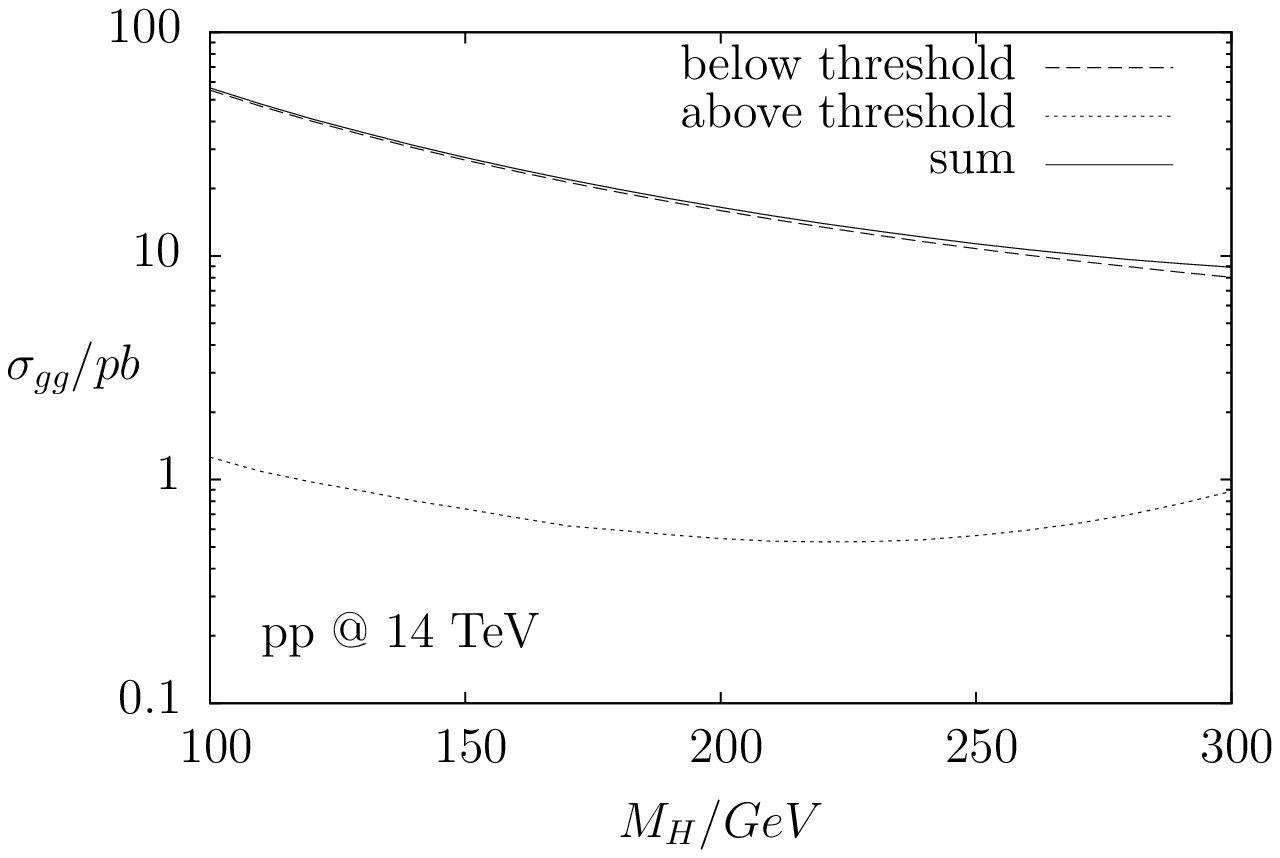}}
      \subfigure{%
        \includegraphics[bb=100 510 470 760,width=.49\textwidth]{%
          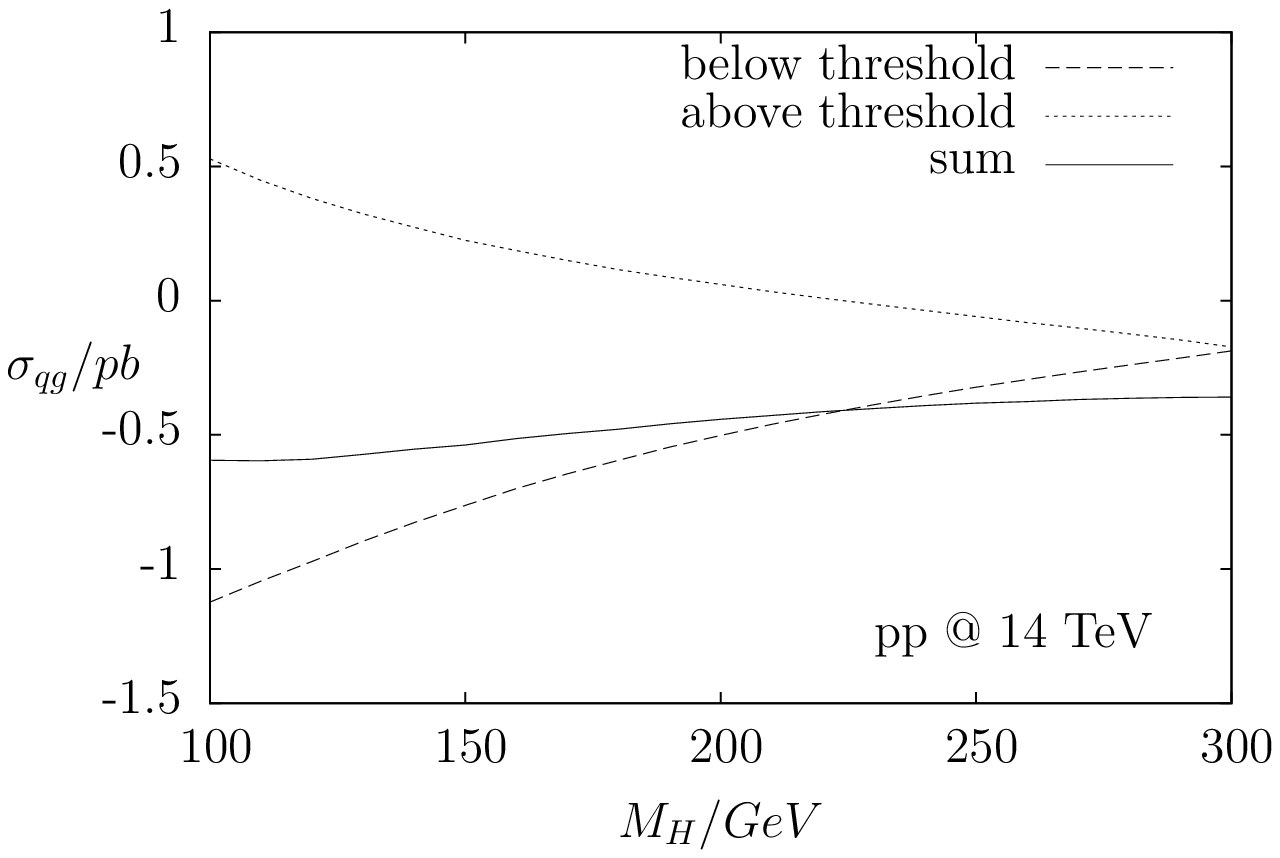}}
      \subfigure{%
        \includegraphics[bb=100 510 470 760,width=.49\textwidth]{%
          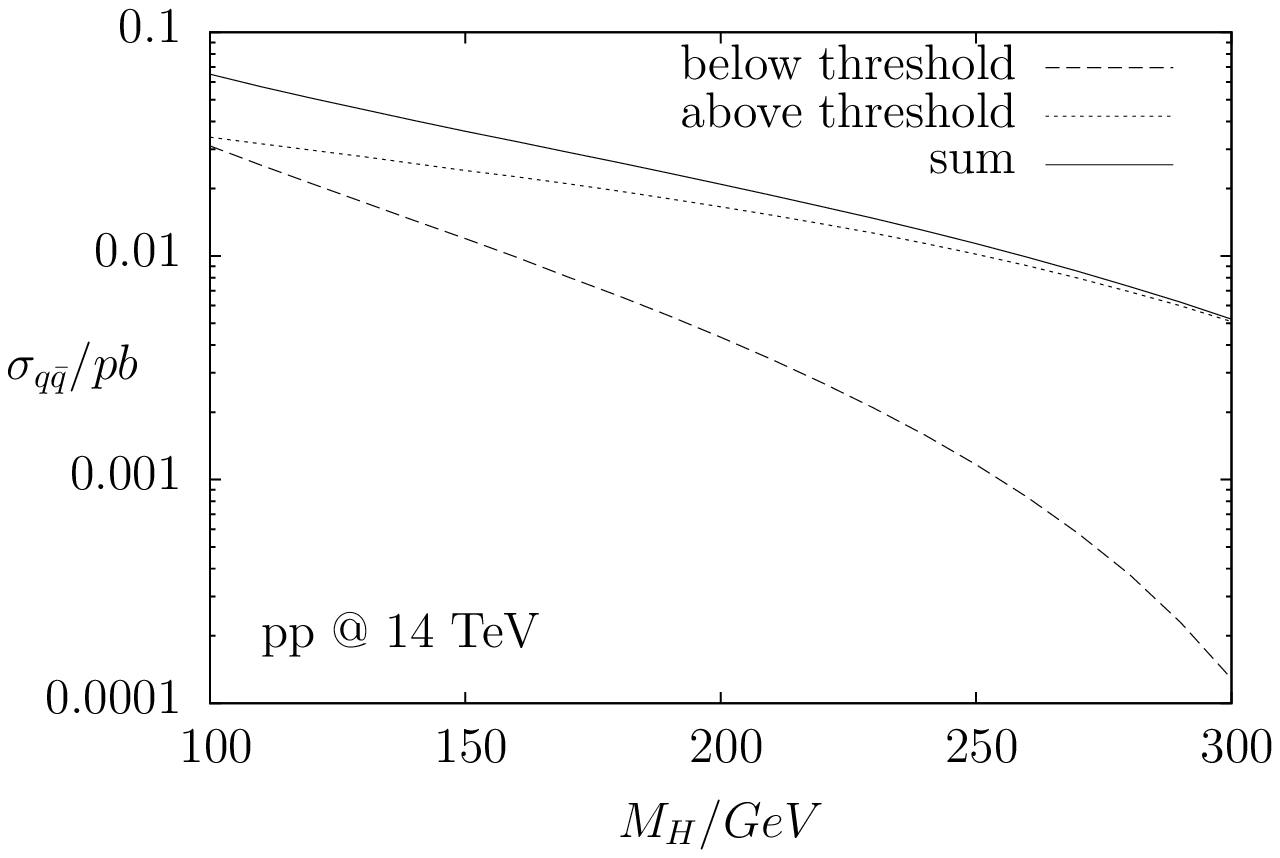}}
    \parbox{.9\textwidth}{
      \caption[]{\label{fig::thrrat}\sloppy Contributions of the
        partonic to the hadronic cross section from below ($\hat
        s<4\mtop^2$; dashed) and above ($\hat s>4\mtop^2$; dotted)
        threshold, for the $gg$, $qg$, and the $q\bar q$ channel at
        \nlo{} ($gg$ includes the \lo{} contribution).  Note that $qg$
        uses a linear scale, while for the $gg$ and the $q\bar q$ it is
        logarithmic.}  }
  \end{center}
\end{figure}

\section{Large-$\hat s$ limit}

\subsection{Outline of the problem}\label{sec::soutline}

The expansion of \eqn{eq::mtexp} is expected to converge within $\hat
s,\mhiggs^2\lesssim 4\mtop^2$. While the Higgs mass range implied by
electro-weak precision measurements lies comfortably in this range, the
partonic center-of-mass energy $\sqrt{\hat s}$ reaches values far beyond
it, both at the \lhc{} and the Tevatron. The corresponding breakdown of
convergence manifests itself in inverse powers of $x=\mhiggs^2/\hat
s$, arising from
\begin{equation}
\begin{split}
\frac{\hat s}{\mtop^2} = \frac{\mhiggs^2}{\mtop^2}\cdot\frac{1}{x}\,.
\end{split}
\end{equation}
Thus, in general,
\begin{equation}
\begin{split}
\Omega_{\alpha\beta,i} \sim \frac{1}{x^i}\qquad\mbox{as}\quad x\to 0\,.
\end{split}
\end{equation}
Note, however, that at small $x=\mhiggs^2/\hat s$ there is a strong
suppression by the parton luminosity ${\cal E}_{\alpha\beta}$ of
\eqn{eq::sigmapp} which we display for the various sub-channels
in~\fig{fig::lumis}. This, together with the fact that
$\Omega_{\alpha\beta,0}$ has no power singularities as $x \to 0$, are
the main reasons that the heavy-top limit defined in \eqn{eq::heavytop}
works so well. 

A further illustration of this observation is shown in \fig{fig::thrrat}
which compares the contributions to the hadronic cross section arising
from below ($\hat s\le 4\mtop^2$) and above threshold for the various
subchannels at \nlo{}. For the dominant $gg$ channel, the region above
threshold contributes only of the order of 2\%.

However, the spurious $1/x$ singularities described before imply that
in order to improve on the heavy-top limit by including higher terms in
$1/\mtop$, one needs to incorporate information on the large-$\hat s$
region. Fortunately, the leading terms can be obtained from general
considerations. In the case of the dominant $gg$-channel, this was done
in Ref.~\cite{Marzani:2008az}. This result was then combined with the
$1/\mtop$ expansion in Refs.\,\cite{Harlander:2009mq,Pak:2009dg}.

Considering \fig{fig::lumis}, it appears that the center of the $qg$
luminosity is at significantly lower values of $x=\mhiggs^2/\hat s$ than
for $gg$. Correspondingly, the influence of the region above threshold
is larger, as can also be seen in \fig{fig::thrrat}.  The proper
treatment of this region is thus much more relevant in the $qg$ case. In
addition, it is clear {\it a priori} that the \eft{} approach, which
assumes that the top mass dependence at higher orders is determined by
the \lo{} one, cannot work as well in the $qg$ channel which occurs only
at \nlo{}. In fact, the contribution of the $qg$ channel to the total
cross section in the \eft{} differs from the exact result by roughly a
factor of two in the mass range between $\mhiggs = 100$ and
$300$\,GeV~\cite{Spira:1995mt}.

In the next section, we extend the analysis of
Ref.\,\cite{Marzani:2008az} to the $qg$ and pure quark channels ($q\bar
q$, $qq$, $qq'$) at \nlo{} and \nnlo{}. The combination with the results
of the $1/\mtop$ expansion is done in Section~\ref{sec::merging}.

\subsection{Derivation of the leading high-energy
  behaviour}\label{sec::smallxderiv} The procedure to compute the
leading logarithmic behaviour (\llx{}) of the partonic coefficient
function to all orders in the strong coupling $\alpha_s$ is based on
$k_T$-factorization~\cite{CCH}. This technique has been used to resum
coefficient functions for a few processes, e.g. heavy quark
production~\cite{ellis-hq, camici-hq}, deep inelastic
scattering~\cite{CH}, Drell-Yan processes~\cite{marzaniballDY} and
direct photon production~\cite{Diana:2009xv}. The small $x$ behaviour of Higgs
production in gluon fusion was first computed
in Ref.\,\cite{Hautmann:2002tu}, in the heavy top approximation. The case of
finite top mass was considered in Ref.\,\cite{Marzani:2008az}, where it was
shown that, as expected, the coefficient function has only single high
energy logarithms, while double logarithms appear in the effective
theory. In Ref.~\cite{Marzani:2008az}, and in the phenomenological
analysis of Ref.\,\cite{Marzani:2008ih}, only the gluon-gluon channel was
considered. In the following the small $x$ behaviour of all the other
channels is computed, using high energy colour charge relations.
For the sake of clarity, we set $\muF=\muR$ throughout this
derivation.

The partonic cross section which enters the $k_T$-factorization formula
is the leading order cross section for the process $gg \to H$, computed
with two incoming off-shell gluons of momenta $k_{1,2}$, with 
$k_{1,2}^2 = - | {\rm {\bf k}}_{1,2}|^2$, contracted with eikonal
polarizations. The impact factor is defined as the triple Mellin
transform of the off-shell cross section
\begin{eqnarray}
h (N,\tau, M_1,M_2) &=& M_1 M_2 \int_0^1 d \zeta \zeta^{N-1}
\int_0^\infty d \xi_1 \xi_1^{M_1-1} \int_0^\infty d \xi_2 \xi_2^{M_2-1}
\int_0^{2 \pi} \frac{d \varphi}{2 \pi} \\ \nonumber && M_H^2 \sigma^{\rm
  off} (\zeta, \tau, \xi_1,\xi_2, \varphi),
\end{eqnarray}
where 
\begin{equation}
\xi_{i} = \frac{| {\rm {\bf k}}_{i}|^2}{M_H^2}, \quad \zeta =
\frac{M_H^2}{2(k_1\cdot k_2 - {\rm \bf k}_1\cdot {\rm \bf k}_2 )}
\end{equation}
and $\varphi$ is the angle between the transverse polarization vectors $
{\rm \bf k}_1$ and ${\rm \bf k}_2$.  In Mellin space the high energy
limit corresponds to $N\to 0$; moreover it is easy to see that $M_i \to
0 $ is the collinear limit.  The leading high energy behaviour of the
coefficient function in the gluon-gluon channel is then found by
identifying
\begin{equation}
M_1 = M_2 = \gamma_s \left(\frac{\alpha_s}{N} \right),
\end{equation} 
where $\gamma_s$ is the anomalous dimension which is dual to the \lo{}
{\abbrev BFKL} kernel $\chi_0$, i.e.
\begin{equation}
\chi_0(\gamma_s(\alpha_s/N)) = \frac{N}{\alpha_s},  
\end{equation}
\begin{equation}
\gamma_s\left(\frac{\alpha_s}{N} \right) = \sum_{k= 1}^{\infty}
c_k\left(\frac{C_A\alpha_s}{\pi N} \right)^k, \quad c_k = 1,0,0,2
\zeta(3), \dots
\end{equation}

To all orders in perturbation theory, the leading logarithmic
contribution to the $ \msbar$ coefficient function is
\begin{equation}
\Delta_{gg}(N, \tau, \muF) = h(0,\tau,\gamma_s, \gamma_s) R^2(\gamma_s)
\left(\frac{M_H^2}{\muF^2}\right)^{2 \gamma_s}.
\end{equation}
The factor $R$ is a scheme dependent function, first computed for
$\msbar$ in~\cite{CH}. A recent calculation~\cite{Kirschner:2009qu} has
questioned that result. Although this issue must be solved for the
resummation of the small $x$ logarithms, it is not relevant for our
present discussion. Our target is to compute the \llx{} behaviour of the
coefficient function through \nnlo{}, but the scheme dependence starts
only one order higher:
\begin{equation}
R = 1 + \mathcal{O} \left(\left(\frac{\alpha_s}{N}\right)^3\right).
\end{equation}
The high energy behaviour of the other partonic channels can be derived
from the gluon-gluon one by noticing that at \llx{} we have
\begin{equation}
\begin{split}
\gamma_{gg} &\sim \gamma_s\,,\qquad
\gamma_{gq} \sim\frac{C_F}{C_A} \gamma_s\,, \qquad
\gamma_{qq} \sim \gamma_{qg} \sim 0\,.
\end{split}
\end{equation}
This means that, at \llx{}, a quark may turn into a gluon, but, because
$\gamma_{qg}$ is next-to-\llx{}, a gluon cannot turn into a quark. This leads to
the following relations between the partonic coefficient functions and
the gluonic impact factor~\cite{CH}:
\begin{eqnarray} \label{coeffuncsN}
\Delta_{qg}(N, \tau, \muF) &=& \frac{C_F}{C_A} \Big[ h(0,\tau,\gamma_s,
  \gamma_s) R^2(\gamma_s) \left(\frac{M_H^2}{\muF^2}\right)^{2
    \gamma_s} \\ \nonumber &&-h(0,\tau,\gamma_s,0) R(\gamma_s)
  \left(\frac{M_H^2}{\muF^2}\right)^{ \gamma_s} \Big]\,, \\  \label{coeffuncsN1} \Delta_{qq}(N,
\tau, \muF) &=& \left(\frac{C_F}{C_A} \right)^2 \Big[
  h(0,\tau,\gamma_s, \gamma_s) R^2(\gamma_s)
  \left(\frac{M_H^2}{\muF^2}\right)^{2 \gamma_s} \\ \nonumber &&-2
  h(0,\tau,\gamma_s,0) R(\gamma_s) \left(\frac{M_H^2}{\muF^2}\right)^{
    \gamma_s} + h(0,\tau,0,0)\Big]\,.
\end{eqnarray}
Notice that in the high energy limit $\Delta_{qq}=\Delta_{qq'} =
\Delta_{q\bar{q}}= \mathcal{O}\left(\frac{\alpha_s^2}{N^2} \right)$,
where $qq$ refers to the identical and $qq'$ to the distinct flavour
case.

The impact factor can be expanded in powers of $M_i$, which corresponds
to an expansion in powers of $\alpha_s$:
\begin{eqnarray} \label{hexp}
h(0,\tau,M_1,M_2) &=& M_H^2 \sigma_0(\tau) \Big[1+
  h^{(1)}(\tau)(M_1+M_2) \\ \nonumber &+&h^{(2)}(\tau)(M_1^2+M_2^2)+
  h^{(1,1)}(\tau) M_1M_2 + \dots \Big]
\end{eqnarray}
The coefficients $h^{(1)}$, $h^{(2)}$ and $h^{(1,1)}$ have been
evaluated numerically in~\cite{Marzani:2008az}\footnote{See Eq.~($36$) and
  Eq.~($38$) of that paper, but notice the differences in the notation,
  e.g. the definition of $\tau$.}. The only difference here is that, in
order to compute the \llx{} behaviour of all partonic subprocesses, we
must keep the contributions $h^{(2)}$ and $h^{(1,1)}$ separated.

It is then easy to substitute Eq.~(\ref{hexp}) into
Eqs.~(\ref{coeffuncsN}),~(\ref{coeffuncsN1}) and invert the $N$ Mellin transform to obtain the
result in $x$ space. Through \nnlo{}, the small $x$ limit of the
partonic coefficient functions can be written as follows:
\begin{equation}
\begin{split}
\Delta_{gg}&(x,\tau) = \delta(1-x) + \api \left[B^{(1)}_{gg}(\tau) - 2
  C_A l_F+ \mathcal{O}(x)\right] \\ &+ \left(\api\right)^2
\left[ \left(A^{(2)}_{gg}(\tau) - 2 C_A B^{(1)}_{gg}(\tau) l_F +2 C_A^2
  l_F^2 \right) \ln \frac{1}{x} +B^{(2)}_{gg}(\tau) + \mathcal{O}(x)
  \right]\,, \\ \Delta_{qg}&(x,\tau) = \api \left[
  B^{(1)}_{qg}(\tau) - C_F l_F+ \mathcal{O}(x)\right]\\ &+
\left(\api\right)^2 \Bigg[ \left(A^{(2)}_{qg}(\tau) -\frac{3}{2} C_F
  B^{(1)}_{gg}(\tau) l_F +\frac{3}{2} C_A C_F l_F^2 \right) \ln
  \frac{1}{x} + B^{(2)}_{qg}(\tau) + \mathcal{O}(x)
  \Bigg]\,, \\ 
\Delta_{q\bar q}&(x,\tau) =\Delta_{qq}(x,\tau) =\Delta_{qq'}(x,\tau) =\\
&\qquad=\left(\api\right)^2
\left[ \left(A^{(2)}_{qq}(\tau) - \frac{C_F^2}{C_A} B^{(1)}_{gg}(\tau)
  l_F + C_F^2 l_F^2 \right) \ln\frac{1}{x} +B^{(2)}_{qq}(\tau) +
  \mathcal{O}(x) \right]\,.
\end{split}
\end{equation}
Recall that we set $\muF=\muR$ in this section. The full $\muF$,
$\muR$-dependence is obtained by replacing
\begin{equation}
\begin{split}
\alpha_s\to \alpha_s(\muR^2)\left[ 1 
- \frac{\alpha_s(\muR^2)}{\pi}\beta_0\lFR 
+ \left(\frac{\alpha_s(\muR^2)}{\pi}\right)^2\left((\beta_0\lFR)^2 -
\beta_1\lFR\right)\right]
\end{split}
\end{equation}
in $\sigma_0\Delta_{\alpha\beta}$, where $\lFR = \ln(\muF^2/\muR^2)$ and
$\beta_0 = 23/12$, $\beta_1 = 29/12$.

The coefficients $A_{\alpha\beta}^{(2)}$ and
$B_{\alpha\beta}^{(1)}$ are provided in the form of numerical tables in
Table~\ref{tab::coefs}.  For all coefficients, the dependence on $\tau$
is very smooth and can safely be interpolated by straight lines, for
example. The \nnlo{} constants $B^{(2)}_{\alpha\beta}$ are currently
unknown; their influence on the final result will be studied at the end
of Section~\ref{sec::hadronic}.

\begin{table}
\begin{center}
\begin{tabular}{|c|c|c|}
\hline
$\tau$ & $B^{(1)}_{gg}$ &  $B^{(1)}_{qg}$ \\
\hline\hline
    1.0  & -0.8821&   -0.1960  \\
    1.5   & 2.9212   & 0.6492 \\
    2.0   & 5.0234   & 1.1163 \\
    2.5   & 6.5538   & 1.4564 \\
    3.0   & 7.7650    &1.7255 \\
    3.5   & 8.7693   & 1.9487 \\
    4.0    &9.6279    & 2.1395 \\
    4.5  & 10.3781   & 2.3062 \\
    5.0   &11.0444   & 2.4543 \\
    5.5  & 11.6437    & 2.5875 \\
    6.0   &12.1883    & 2.7085 \\
    6.5  & 12.6875   & 2.8194 \\
    7.0  & 13.1482  &  2.9218 \\
    7.5   &13.5760   & 3.0169 \\
    8.0   &13.9752    & 3.1056 \\
    8.5   &14.3495    & 3.1888 \\
    9.0   &14.7018    & 3.2671 \\
    9.5   &15.0345    & 3.3410 \\
   10.0   &15.3497    & 3.4110 \\
   10.5   &15.6491    & 3.4776 \\
   11.0   &15.9343    & 3.5410 \\
   11.5   &16.2065    & 3.6015 \\
   12.0  &16.4670   &  3.6593 \\
\hline
\end{tabular}
\qquad
\begin{tabular}{|c|c|c|c|}
\hline
$\tau$ & $A^{(2)}_{gg}$ &  $A^{(2)}_{qg}$&  $A^{(2)}_{qq}$ \\
\hline\hline
    1.0 &  33.0465  &  8.7703 &   1.2681   \\
    1.5 &   35.9907 & 9.5484   & 1.3782 \\
    2.0 &   44.2884  & 12.2677  &  2.1563 \\
    2.5 &   53.1336   &15.1924   & 3.0088 \\
    3.0 &   61.8029   & 18.0679   & 3.8524 \\
    3.5 &   70.1088   & 20.8272   & 4.6644 \\
    4.0 &   78.0127   & 23.4553   & 5.4393 \\
    4.5 &   85.5245   & 25.9547   & 6.1771 \\
    5.0 &   92.6698   & 28.3331   & 6.8798 \\
    5.5 &   99.4782   & 30.6002   & 7.5501 \\
    6.0 &  105.9788   & 32.7654   & 8.1907 \\
    6.5 &  112.1985   & 34.8374   & 8.8039 \\
    7.0 &  118.1616   & 36.8244   & 9.3922 \\
    7.5 &  123.8897   & 38.7333   & 9.9576 \\
    8.0 &  129.4021   & 40.5706  & 10.5019 \\
    8.5 &  134.7158   & 42.3420   & 11.0268 \\
    9.0 &  139.8461   & 44.0524   &11.5337 \\
    9.5 &  144.8064   & 45.7063   &12.0240 \\
   10.0 &  149.6090   & 47.3077   & 12.4989 \\
   10.5 &  154.2646   & 48.8603   & 12.9594 \\
   11.0 &  158.7829   & 50.3672   & 13.4064 \\
   11.5 &  163.1728   & 51.8315   & 13.8408 \\ 
   12.0 &  167.4422   & 53.2556   & 14.2633 \\
  \hline
\end{tabular}
\end{center}
\caption[]{\label{tab::coefs}Coefficients for the large-$\hat s$
  behaviour at \nlo{} (left table) and \nnlo{} (right).}
\end{table}

\subsection{Merging and partonic results}\label{sec::merging} 

\begin{figure}
  \begin{center}
    \begin{tabular}{cc}
      \includegraphics[bb=110 265 465
        560,width=.45\textwidth]{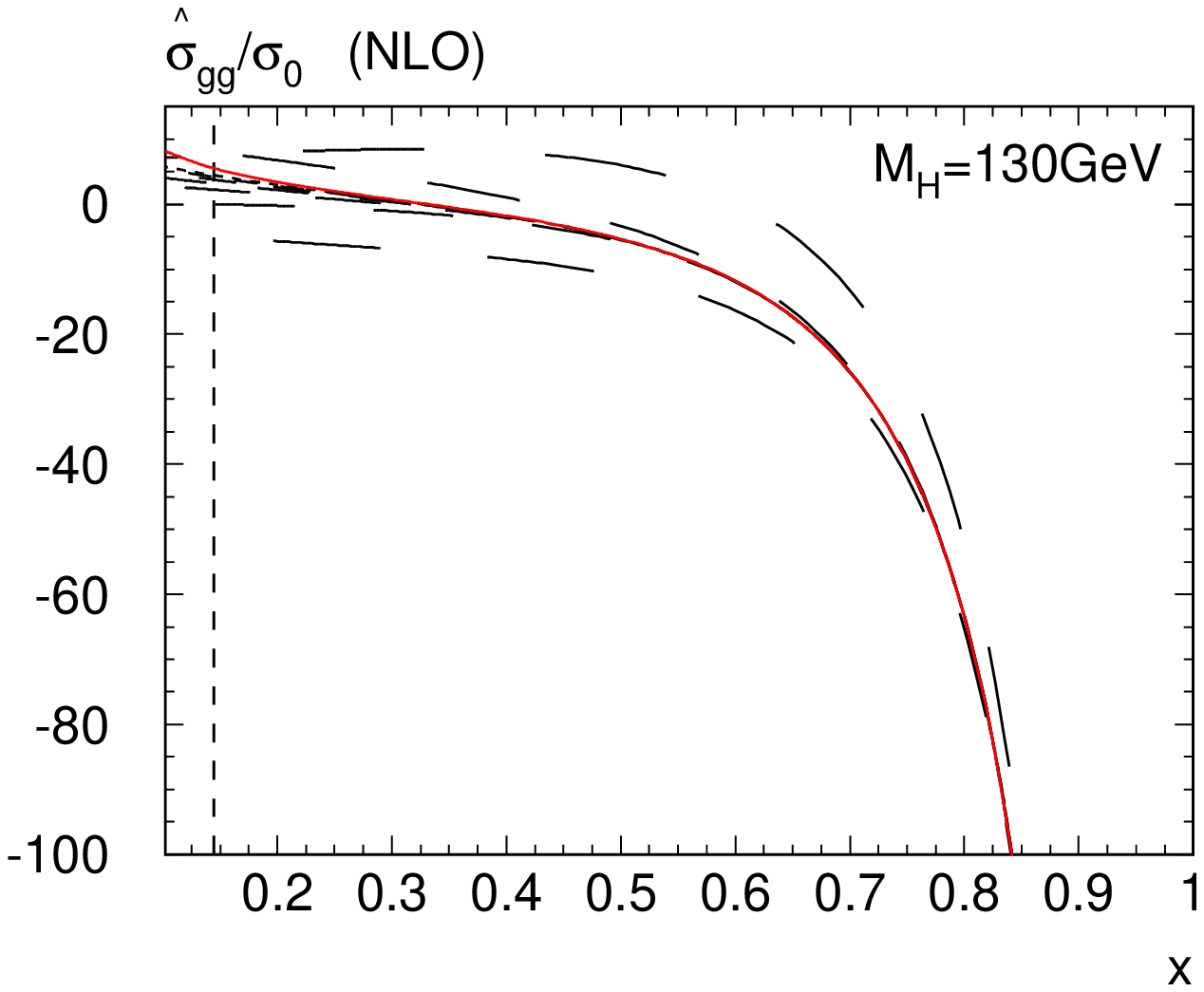} &
      \includegraphics[bb=110 265 465
        560,width=.45\textwidth]{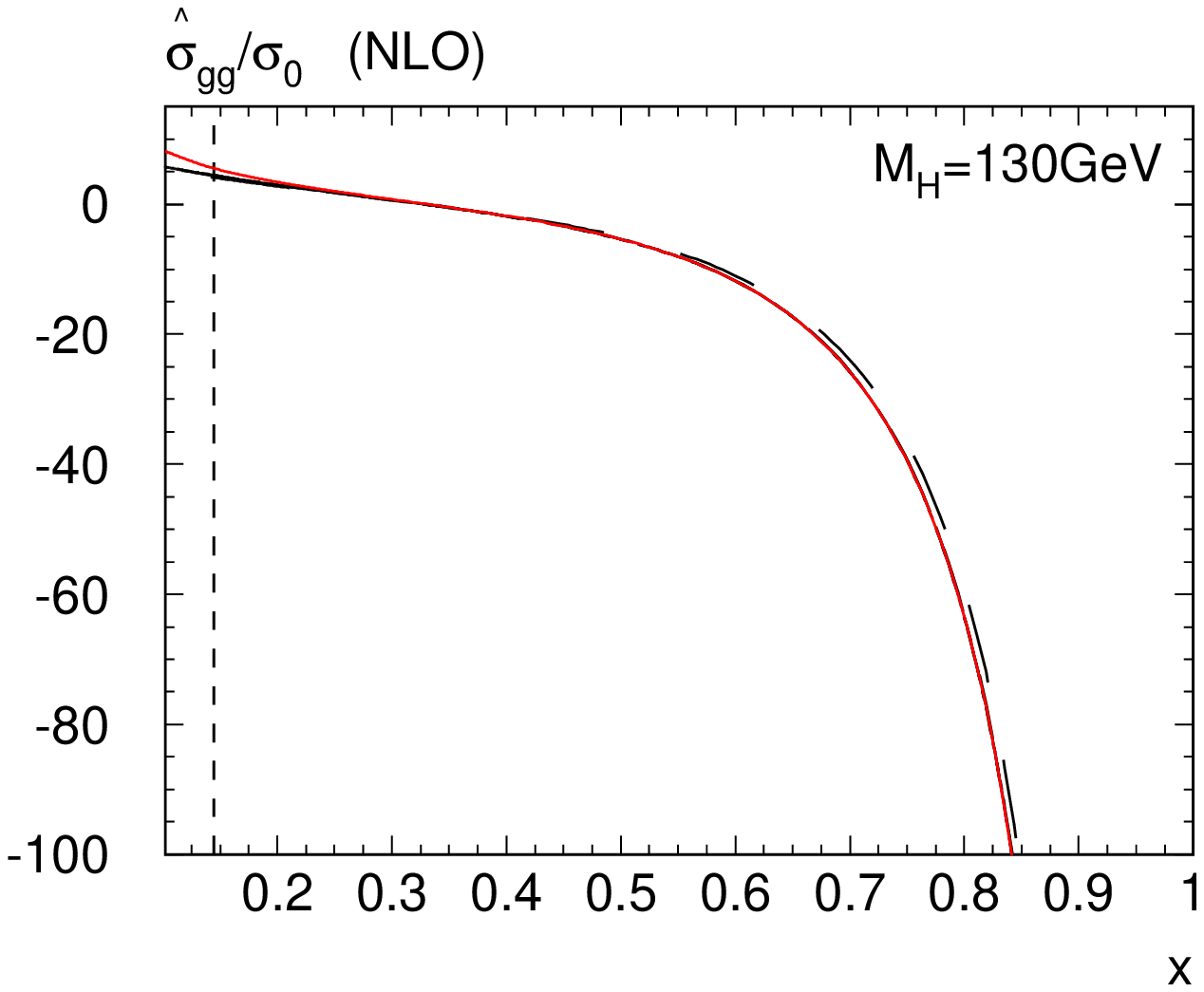} \\
      \includegraphics[bb=110 265 465
        560,width=.45\textwidth]{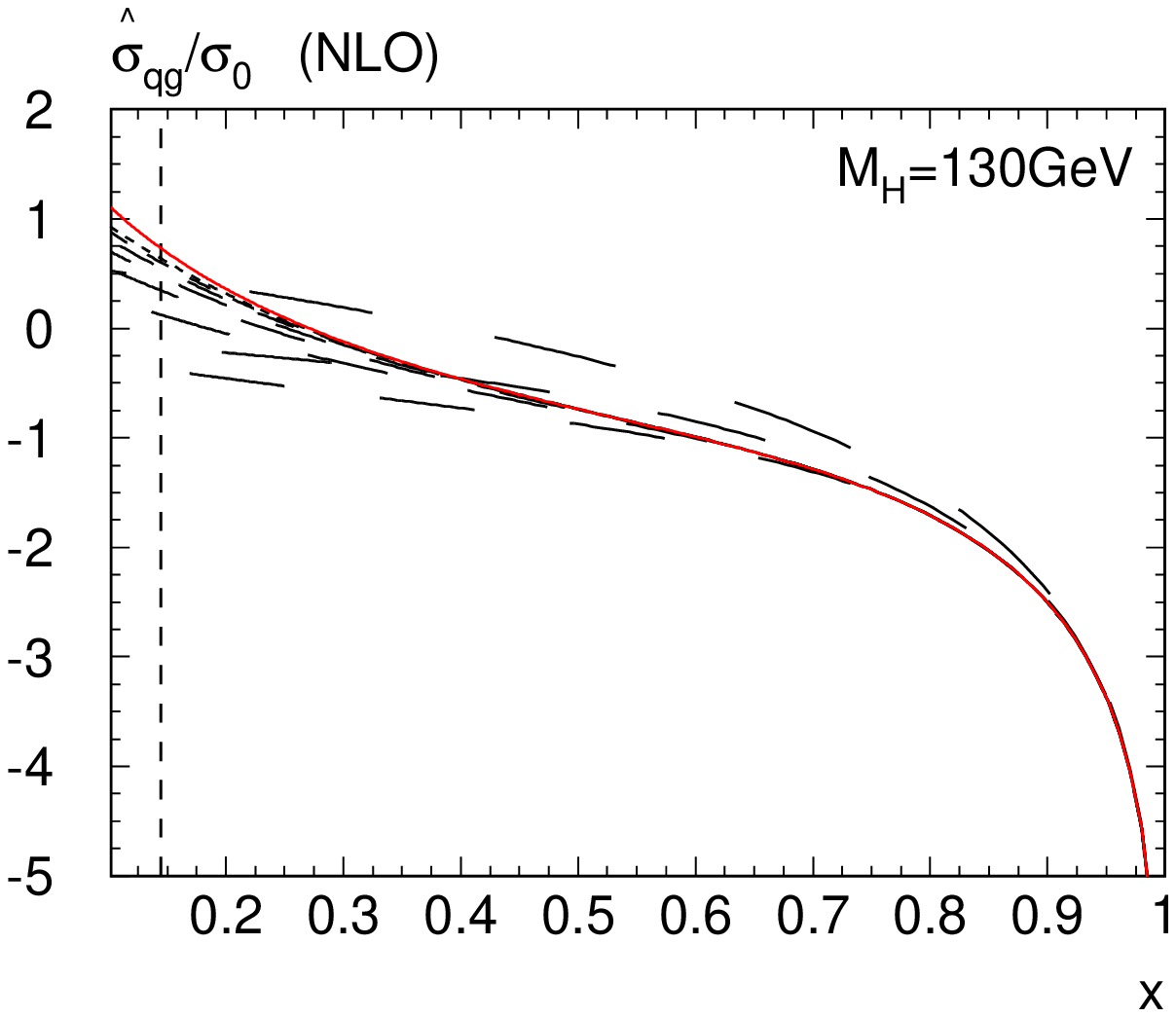} &
      \includegraphics[bb=110 265 465
        560,width=.45\textwidth]{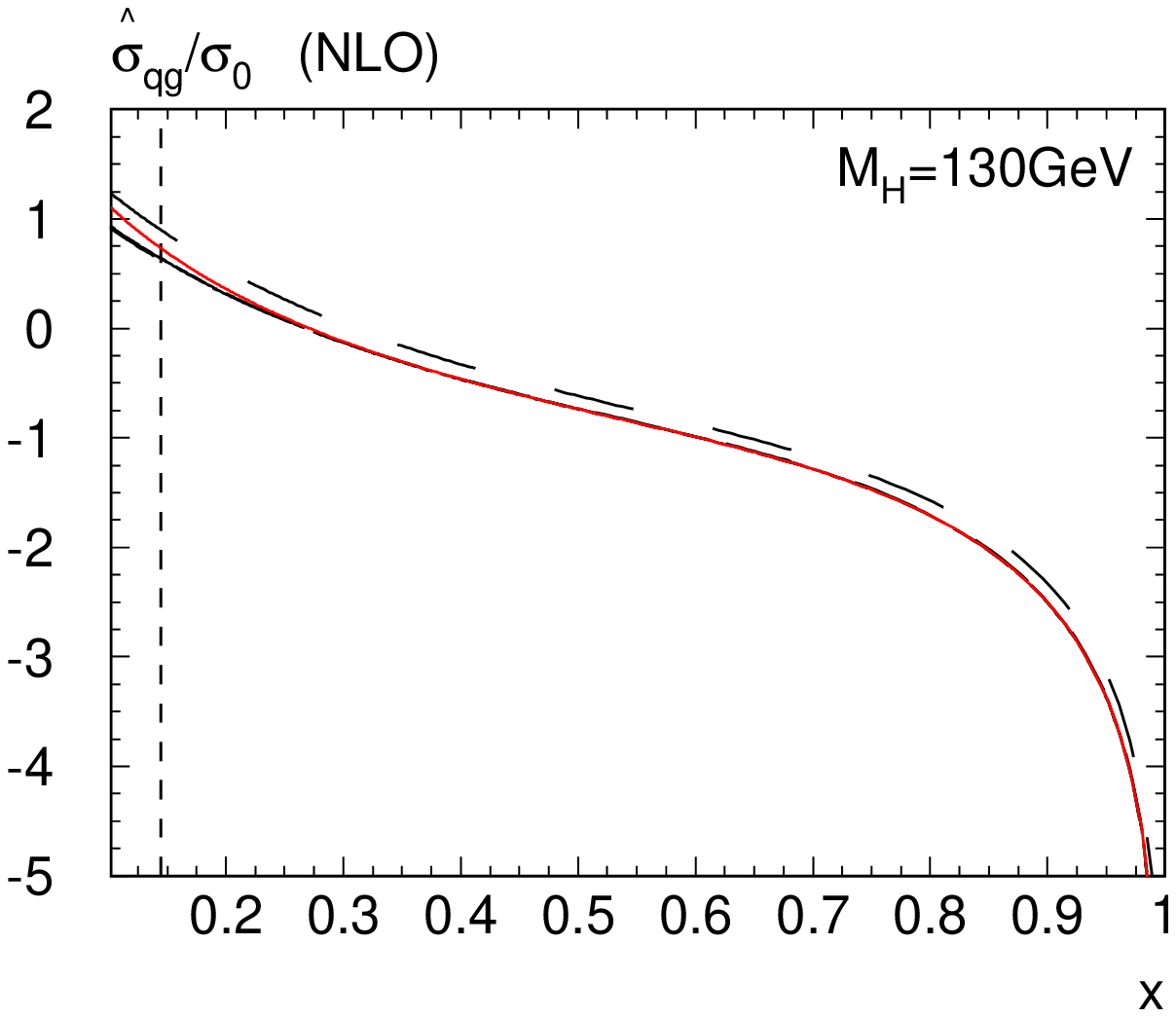} \\
      \includegraphics[bb=110 265 465
        560,width=.45\textwidth]{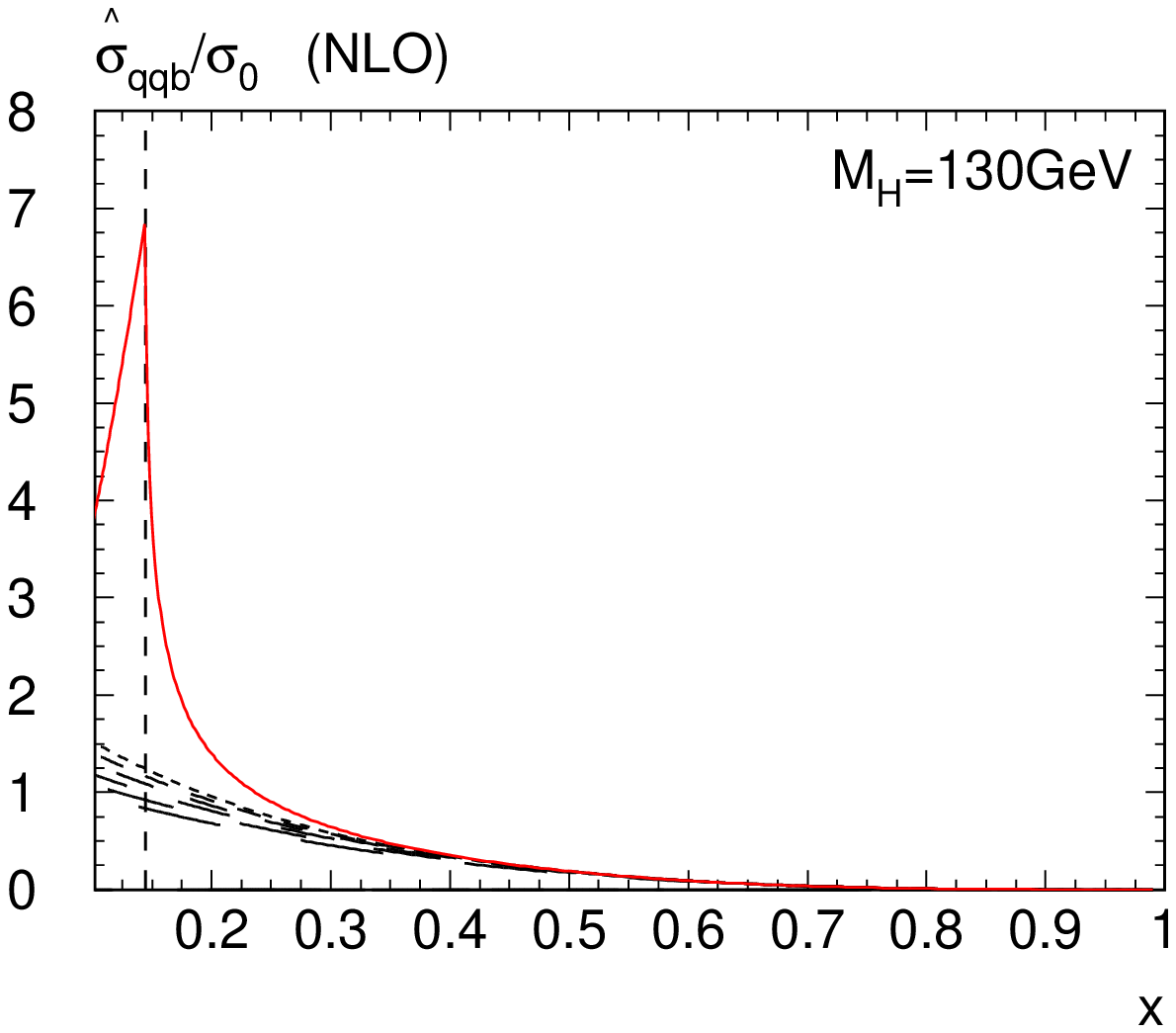} &
      \includegraphics[bb=110 265 465
        560,width=.45\textwidth]{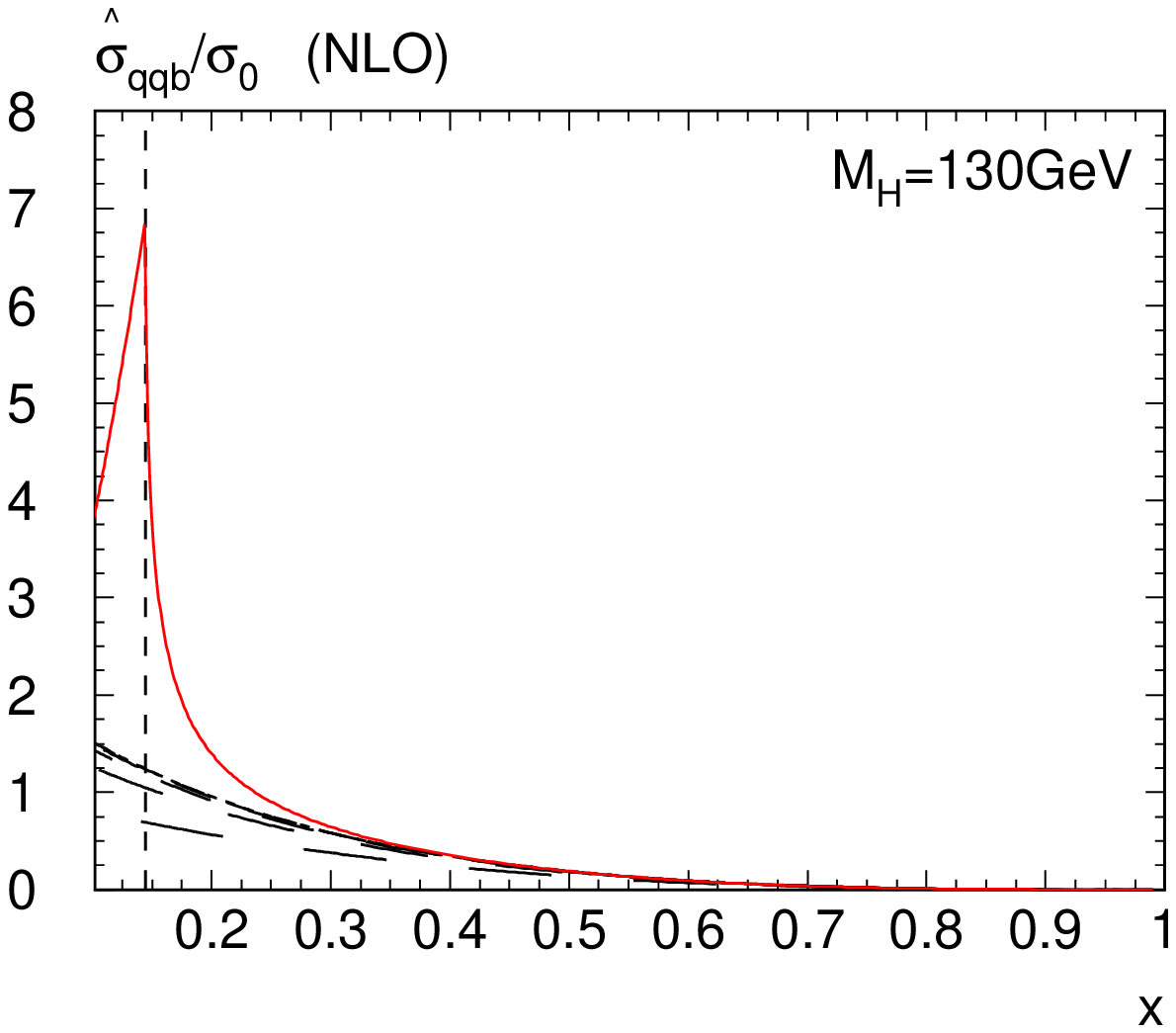}
    \end{tabular}
    \parbox{.9\textwidth}{
      \caption[]{\label{fig::sig1-130}\sloppy Partonic cross section at
        \nlo{} for $\mtop=170.9$ and $\mhiggs=130$\,GeV at various
        orders in the expansion parameters (increasing order corresponds
        to decreasing dash size of the lines). Left column:
        $\order{1/\mtop^{10}}$ and $\order{(1-x)^n}$, $n=0,\ldots,8$.
        Right column: $\order{(1-x)^8}$ and $\order{1/\mtop^{2n}}$,
        $n=0,\ldots,5$.  Solid: exact.  The dashed vertical line
        indicates the threshold.}}
  \end{center}
\end{figure}

\begin{figure}
  \begin{center}
    \begin{tabular}{cc}
      \includegraphics[bb=110 265 465
        560,width=.45\textwidth]{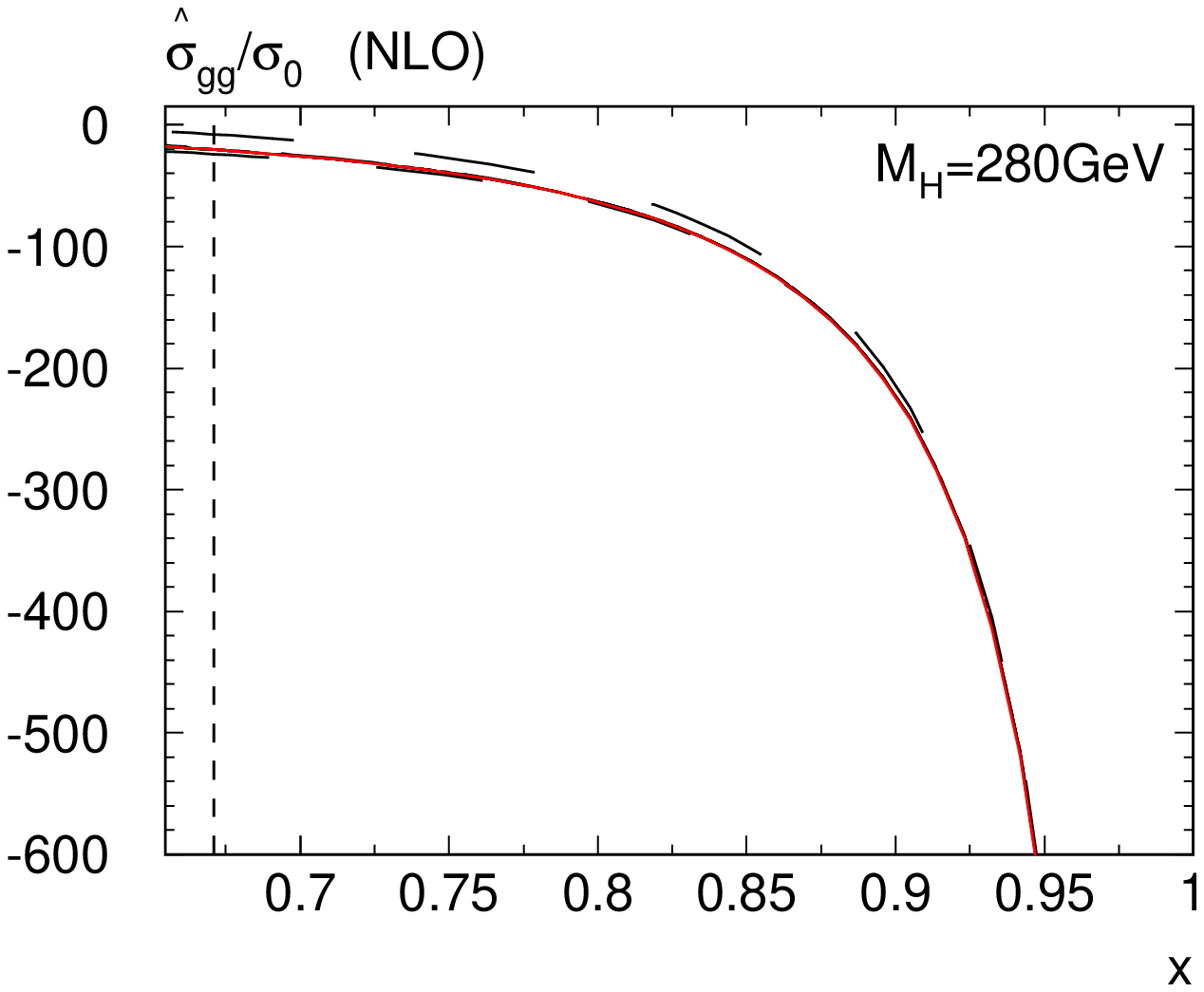} &
      \includegraphics[bb=110 265 465
        560,width=.45\textwidth]{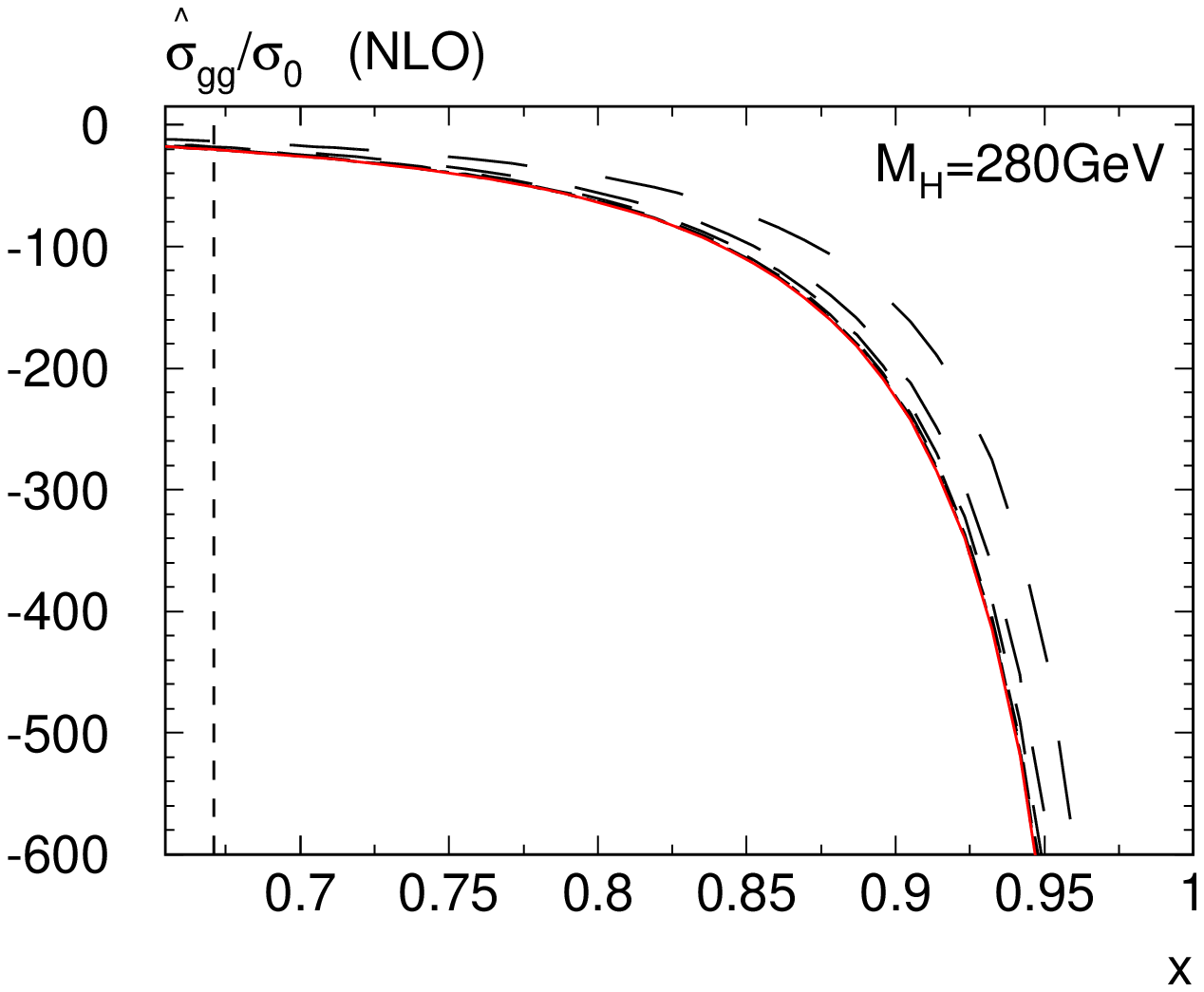} \\
      \includegraphics[bb=110 265 465
        560,width=.45\textwidth]{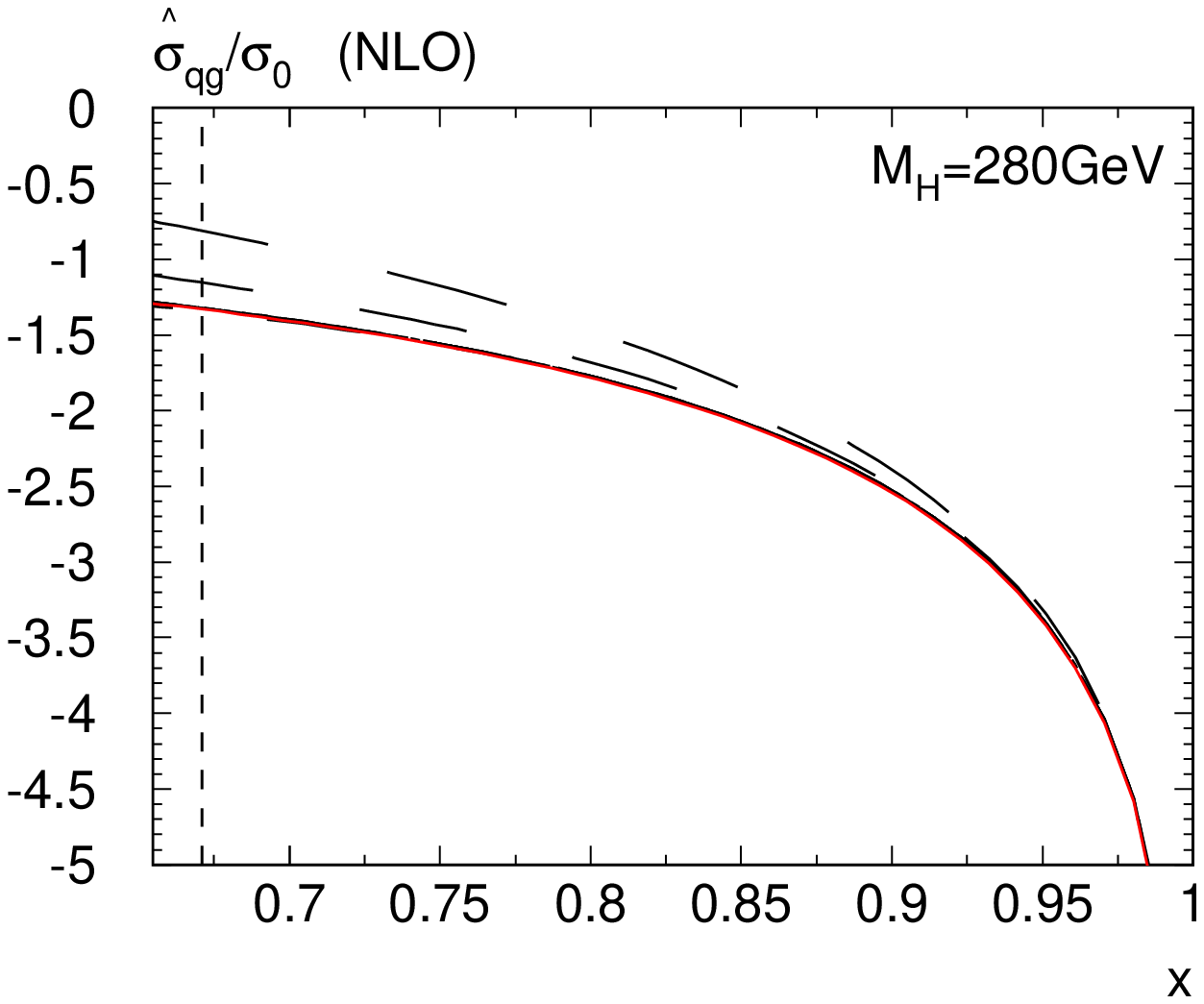} &
      \includegraphics[bb=110 265 465
        560,width=.45\textwidth]{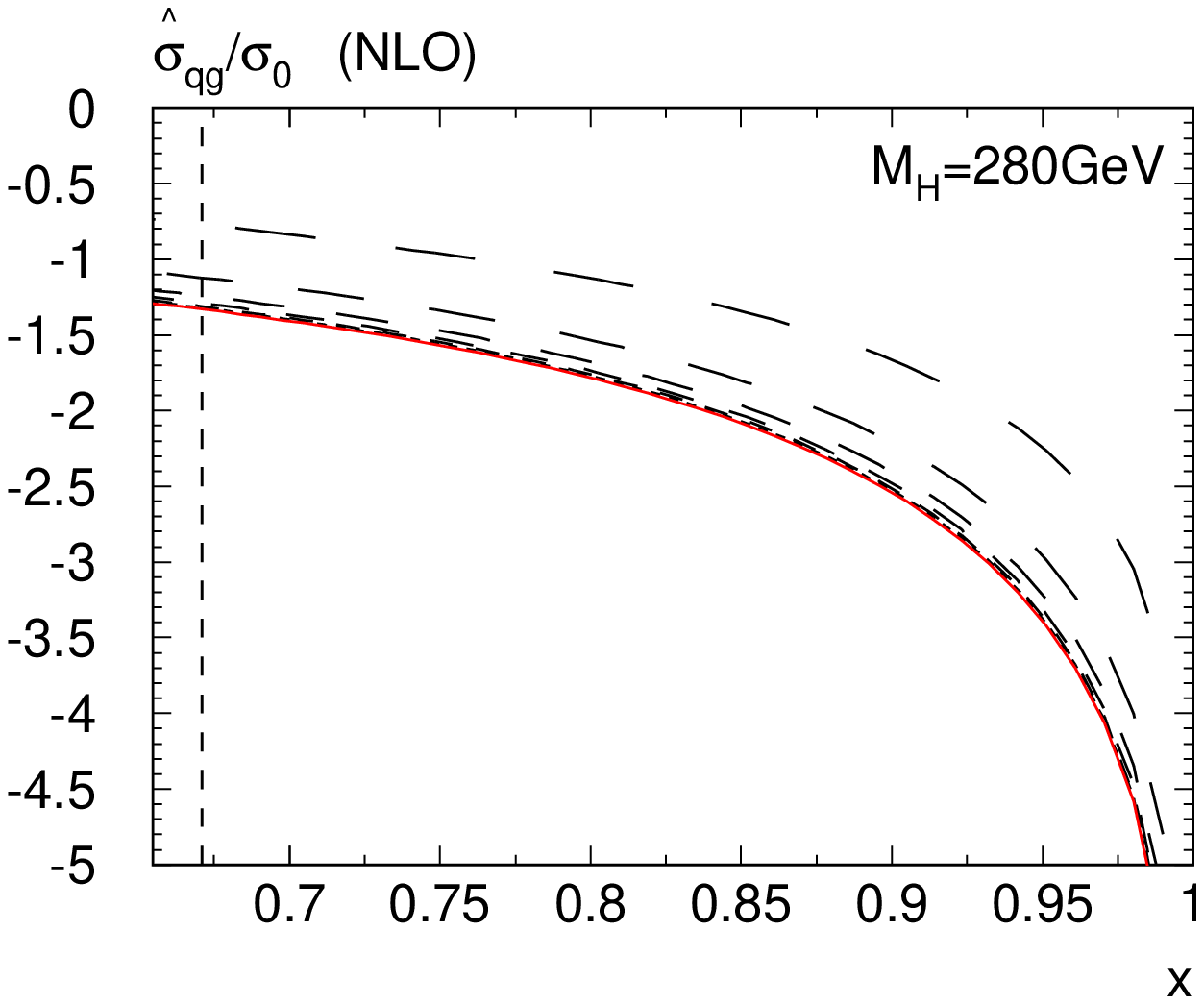} \\
      \includegraphics[bb=110 265 465
        560,width=.45\textwidth]{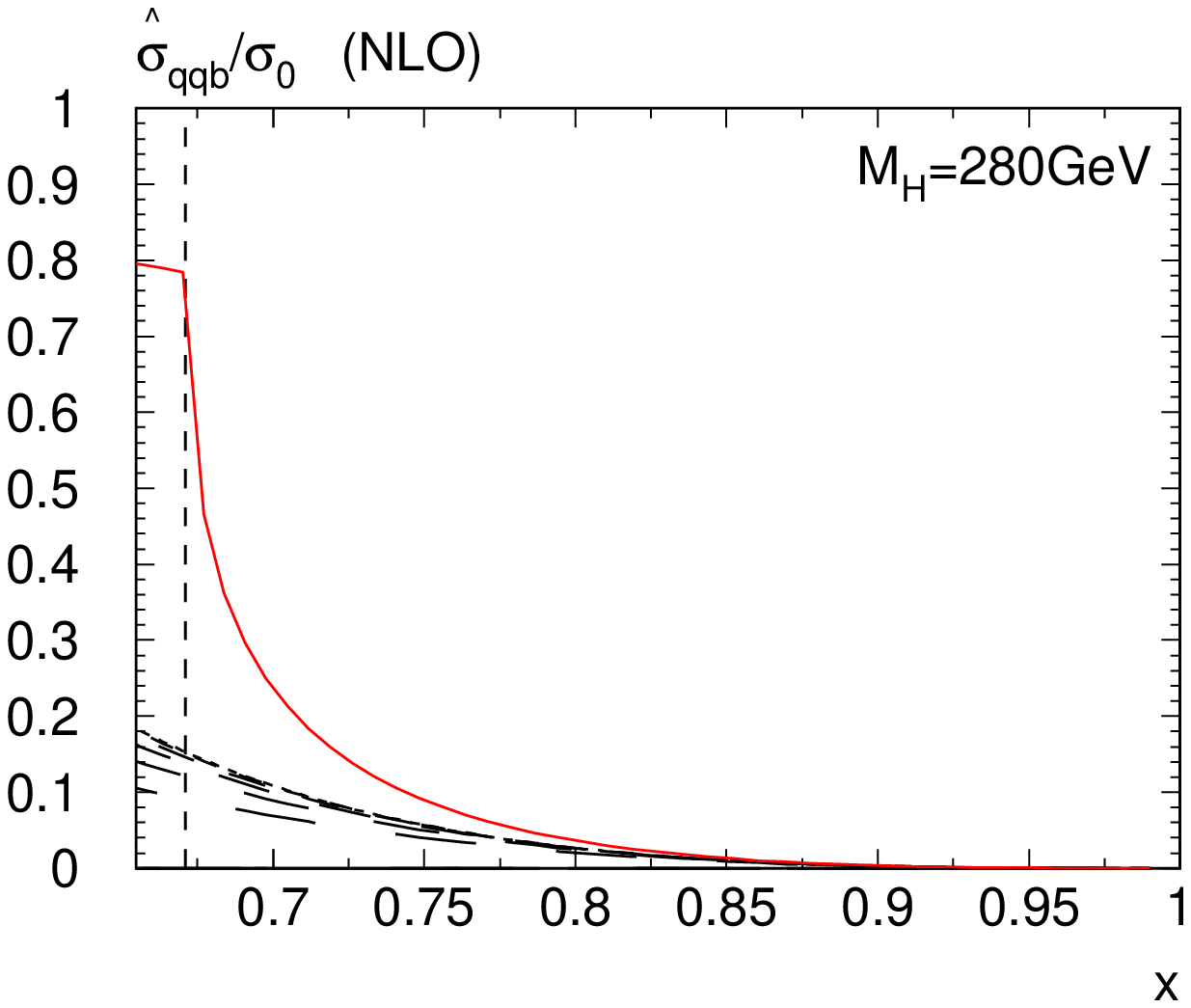} &
      \includegraphics[bb=110 265 465
        560,width=.45\textwidth]{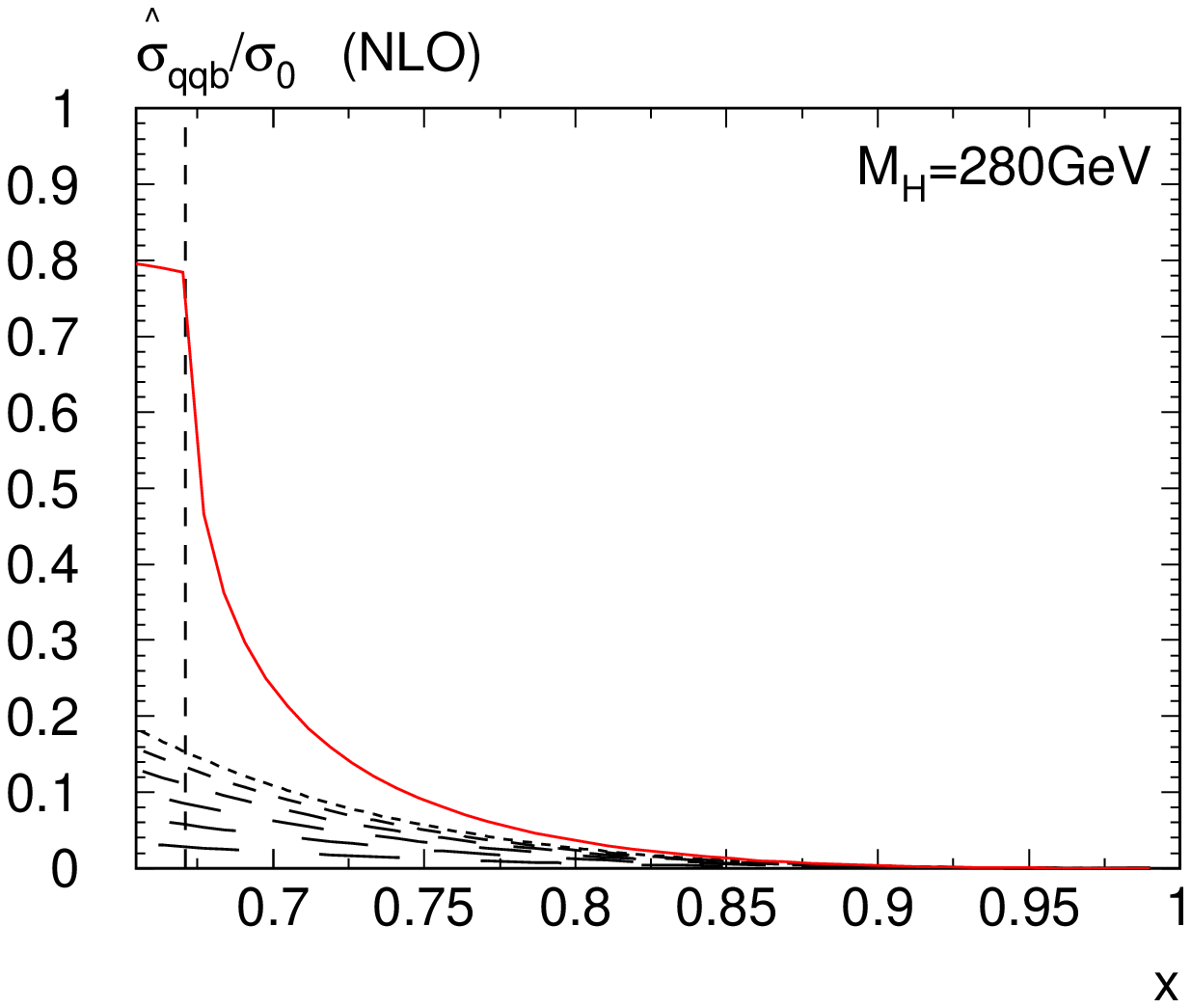}
    \end{tabular}
    \parbox{.9\textwidth}{
      \caption[]{\label{fig::sig1-280}\sloppy Same as
        \fig{fig::sig1-130}, but for $\mhiggs=280$\,GeV. }}
  \end{center}
\end{figure}

Let us recall the knowledge of the partonic cross section at \nnlo{}.
Below threshold ($\hat s<4\mtop^2$), the result is known in terms of an
expansion in $1/\mtop$ and
$(1-x)$~\cite{Harlander:2009mq,Pak:2009dg,Harlander:2009bw}\footnote{Recently,
  the full $x$ dependence was derived~\cite{Pak:2009bx}. However, as
  argued before, the $x$ dependence of the $1/\mtop$ expansion does not
  hold for $x<\mhiggs^2/(4\mtop^2)$.}. Both expansions are
expected to converge very well as long as $\mhiggs < 2\mtop\approx
340$\,GeV. This is indeed observed for the $gg$ and the $qg$ channels at
\nlo{} in Fig.\,\ref{fig::sig1-130} for $\mhiggs=130$\,GeV and in
Fig.\,\ref{fig::sig1-280} for $\mhiggs=280$\,GeV. They show the partonic
cross sections below threshold, keeping terms of order $(1-x)^a
(1/\mtop^2)^{b}$. In the left columns, $a=0,\ldots,8$ and $b=5$ (long to
short dashes), while in the right columns, $a=8$ and
$b=0,\ldots,5$. These figures compare the expansions to the exact result
which we derived using standard techniques (see, e.g.,
Ref.\,\cite{Ellis:2007qk}).  In fact, the behaviour of the expansions
suggests that, below threshold, the final result for the $gg$ and the
$qg$ channels is numerically almost equivalent to the full $\mtop$ and
$x$ dependence.

The \nlo{} $q\bar q$ channel, on the other hand, has a very peculiar
structure at threshold.  At this order only one diagram with $q\bar q$
annihilating into an $s$-channel gluon contributes. Such a diagram is
not enhanced in either the large- or small-$x$ region, leaving room for
a relatively pronounced structure at the threshold which cannot be
described properly in our approach.  However, the contribution of the
$q\bar q$ channel to the hadronic cross section is down by almost three
orders of magnitude relative to the $gg$ channel, and still a factor of
ten relative to the $qg$ channel.  We will nevertheless investigate its
influence on the final prediction in more detail below.  At higher
orders we expect this effect to be reduced, because other diagrams with
non-trivial high- or low-$x$ limits will contribute.

\begin{figure}
  \begin{center}
    \begin{tabular}{cc}
      \includegraphics[bb=110 265 465
        560,width=.45\textwidth]{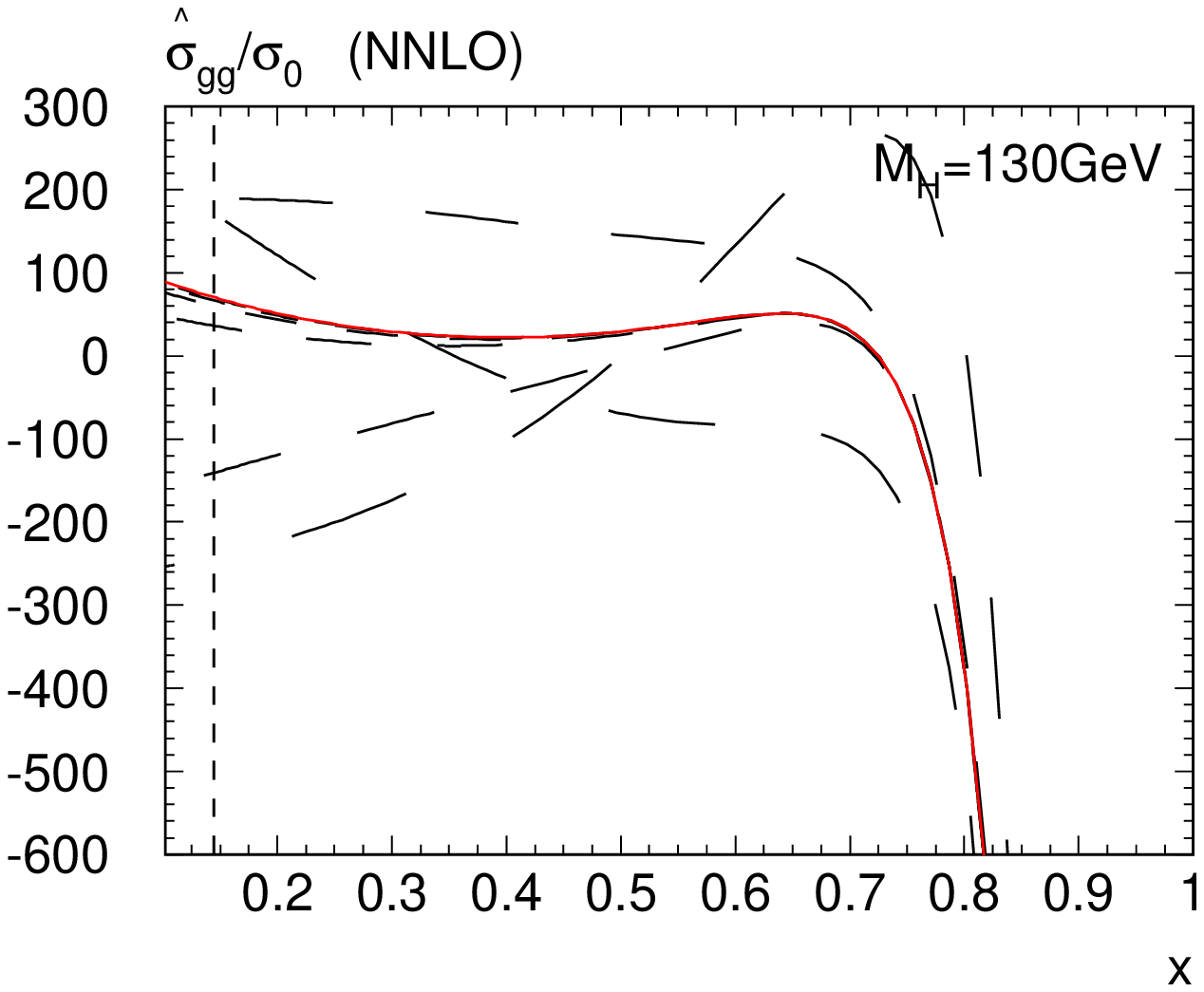} &
      \includegraphics[bb=110 265 465
        560,width=.45\textwidth]{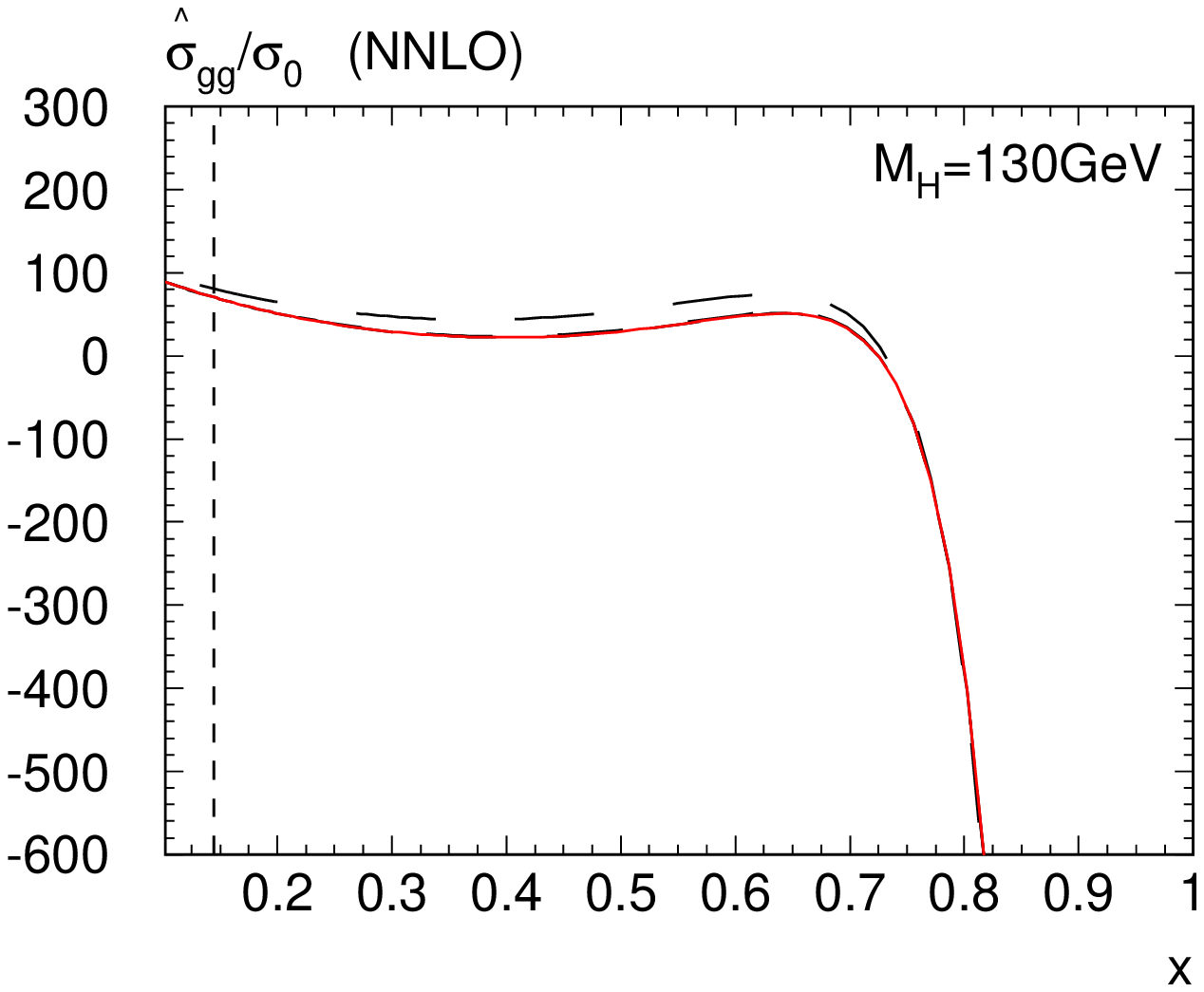} \\
      \includegraphics[bb=110 265 465
        560,width=.45\textwidth]{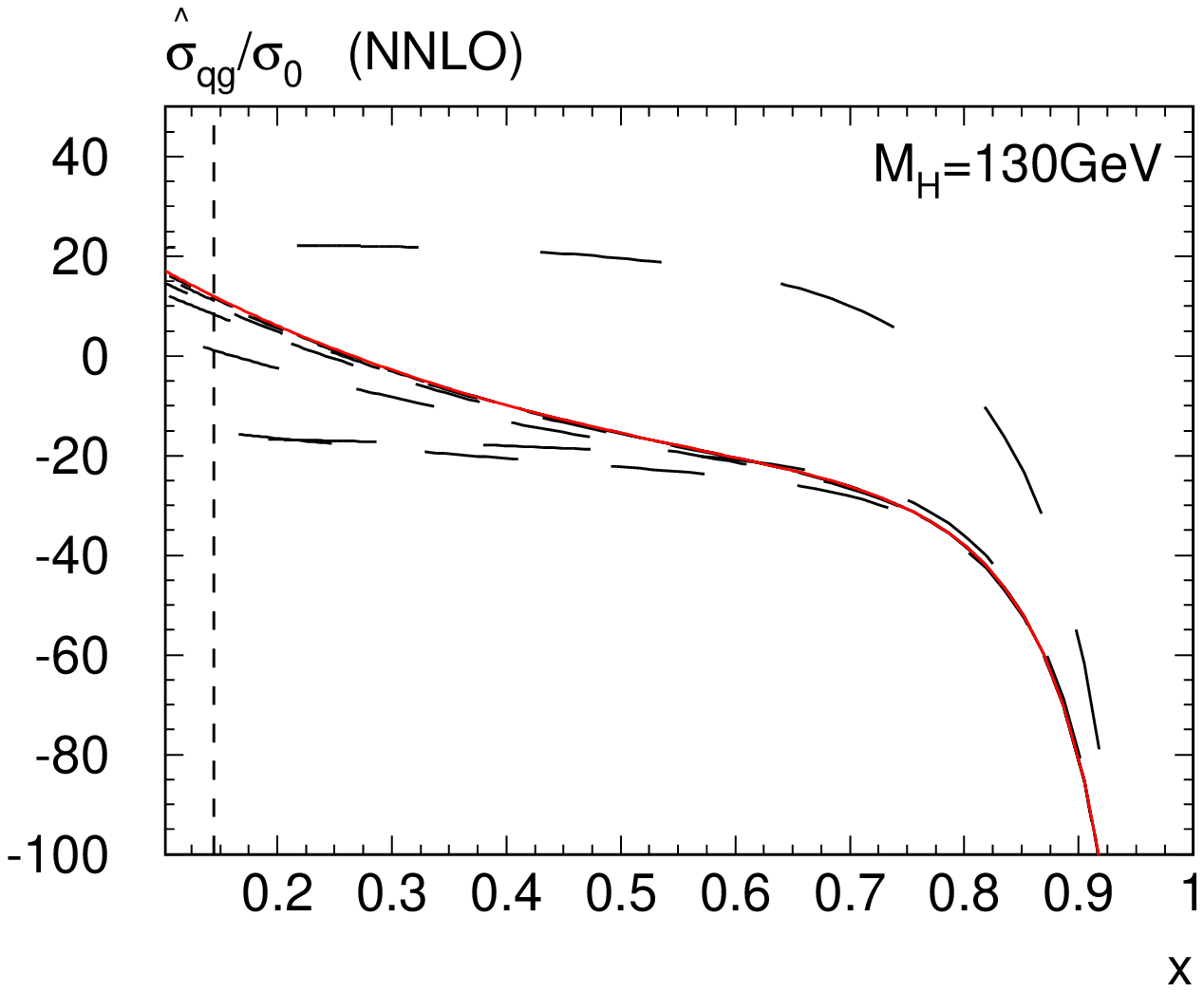} &
      \includegraphics[bb=110 265 465
        560,width=.45\textwidth]{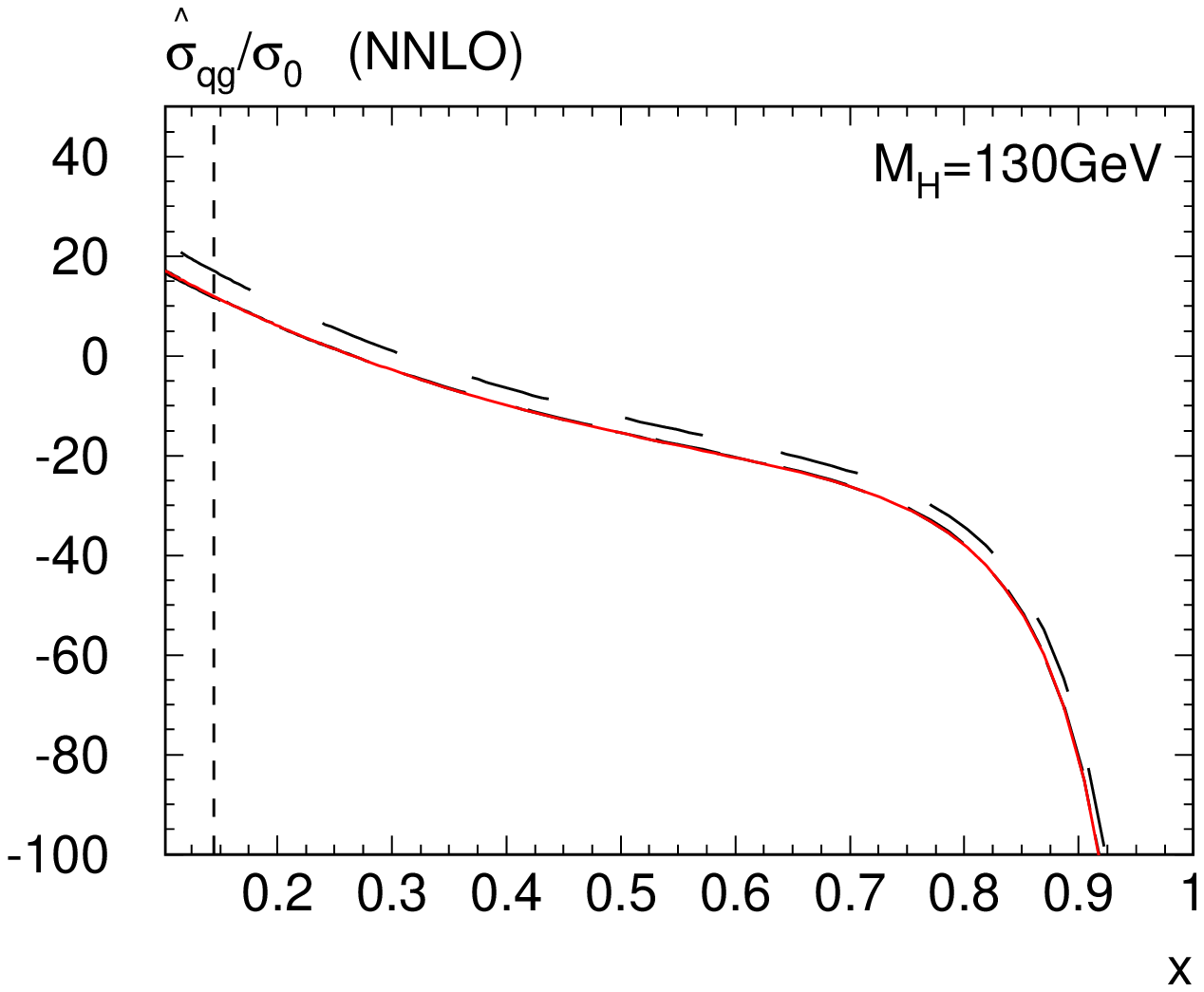}
    \end{tabular}
    \parbox{.9\textwidth}{
      \caption[]{\label{fig::sig2-130qg}\sloppy Partonic cross section
        ($gg$ and $qg$ channel) at \nnlo{} for $\mtop=170.9$ and
        $\mhiggs=130$\,GeV for various orders in the expansion
        parameters (increasing order corresponds to decreasing dash size
        of the lines). Left column: $\order{1/\mtop^{10}}$ and
        $\order{(1-x)^n}$, $n=0,\ldots,7$.  Right column:
        $\order{(1-x)^8}$ and $\order{1/\mtop^{2n}}$, $n=0,\ldots,2$.
        Solid: $\order{1/\mtop^{6}}$ and $\order{(1-x)^8}$.  The dashed
        vertical line indicates the threshold. For the $q\bar q$ and the
    $qq$ channel, see \fig{fig::sig2-130qq}.}}
  \end{center}
\end{figure}

\begin{figure}
  \begin{center}
    \begin{tabular}{cc}
      \includegraphics[bb=110 265 465
        560,width=.45\textwidth]{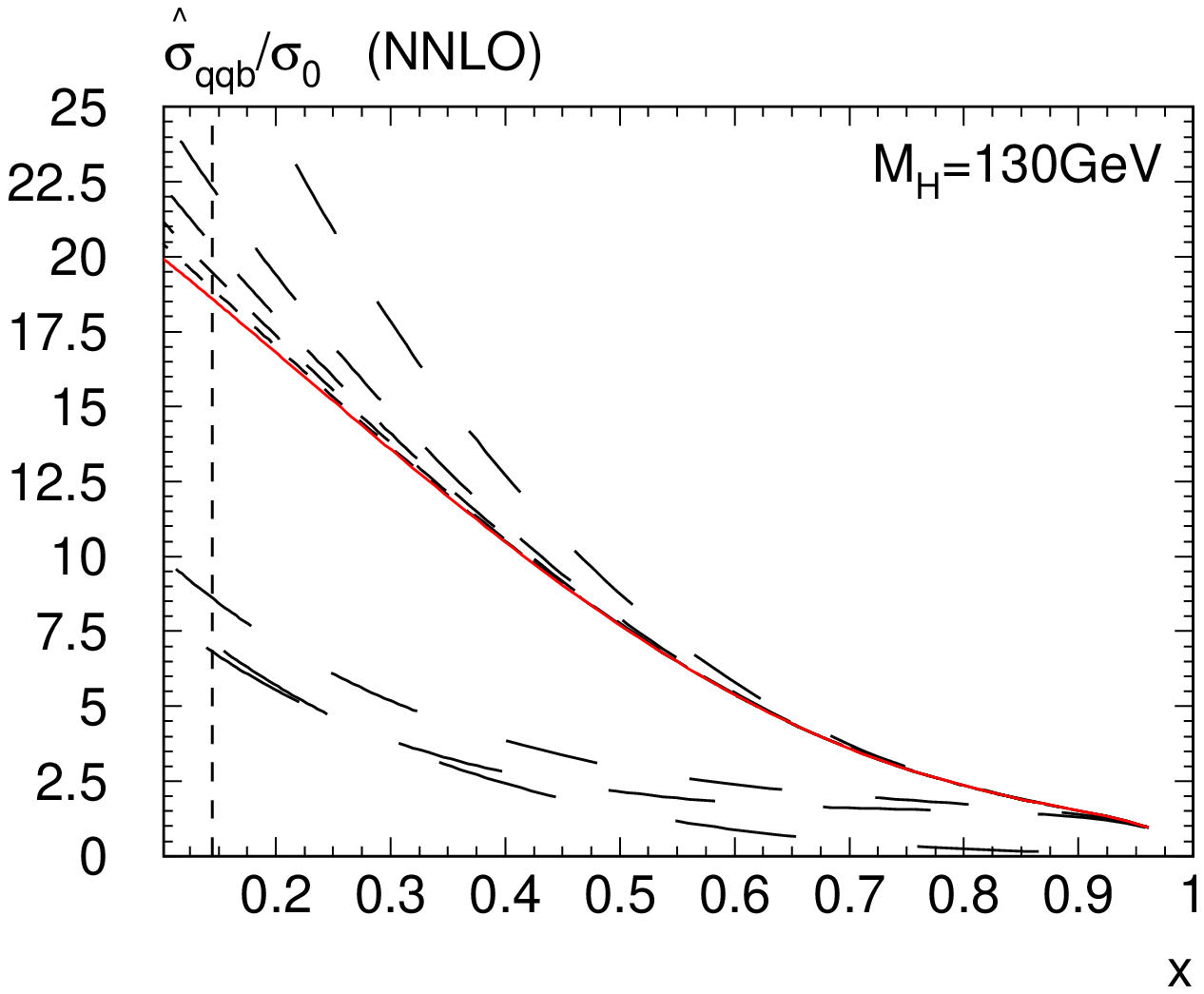} &
      \includegraphics[bb=110 265 465
        560,width=.45\textwidth]{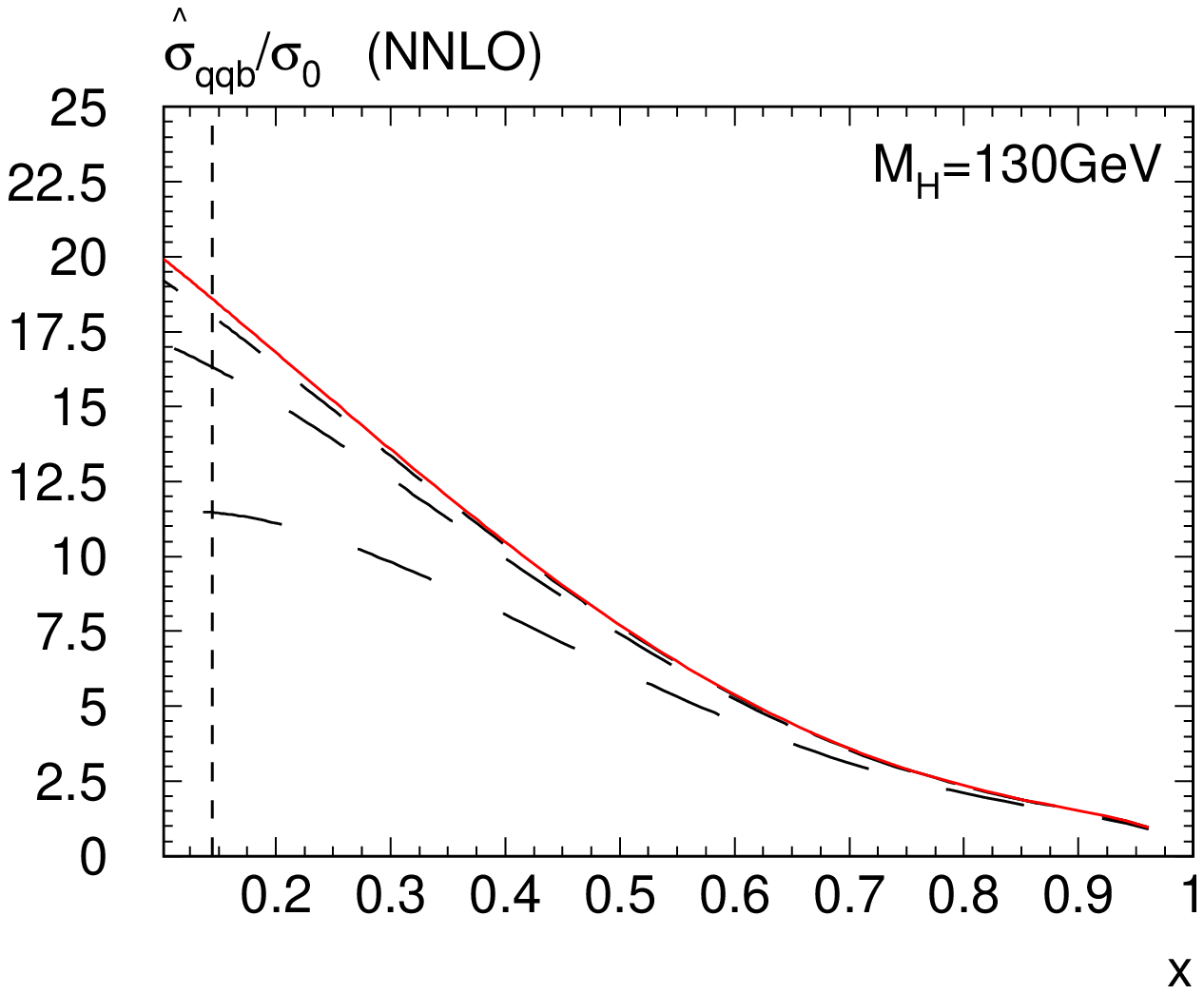} \\
      \includegraphics[bb=110 265 465
        560,width=.45\textwidth]{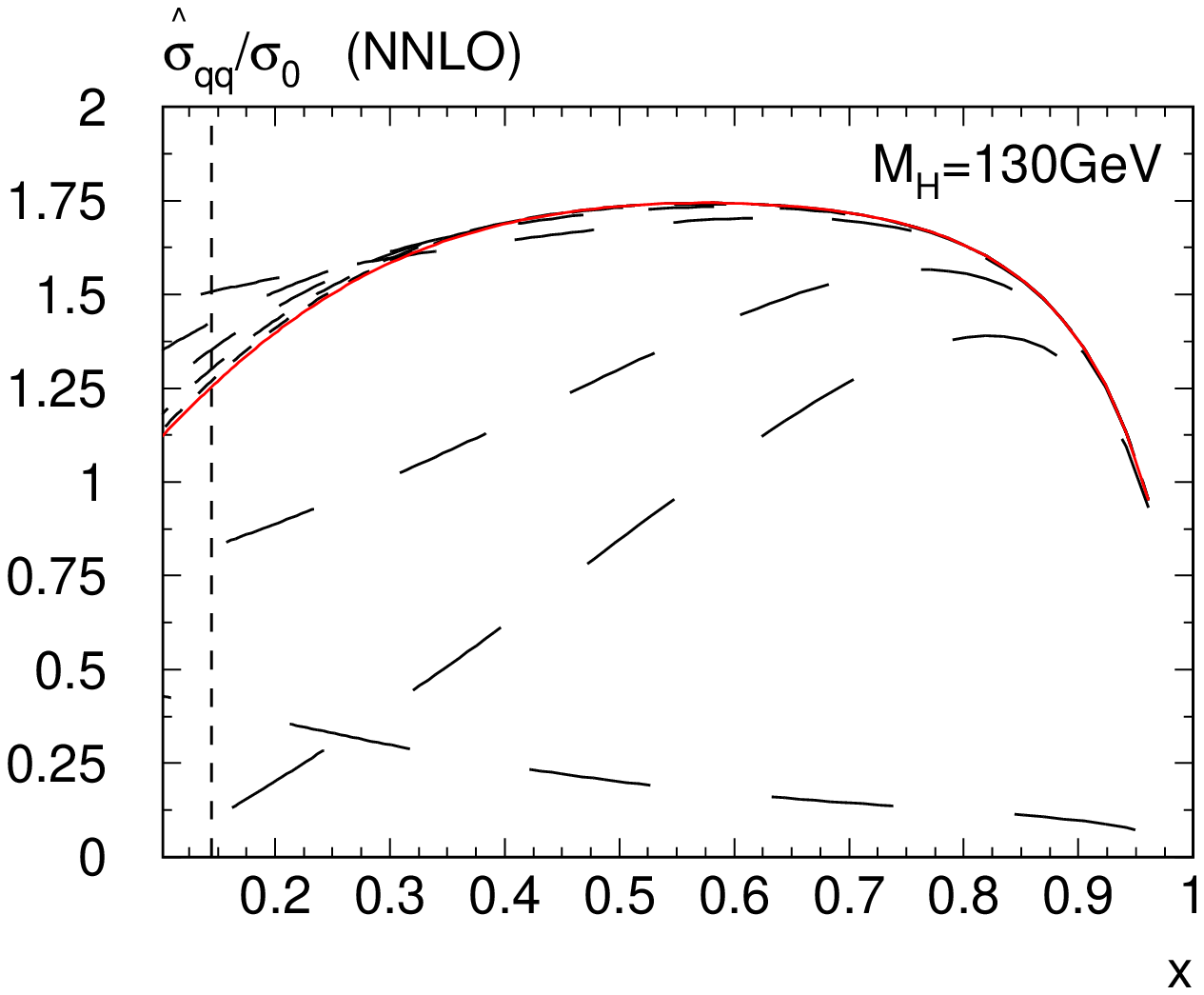} &
      \includegraphics[bb=110 265 465
        560,width=.45\textwidth]{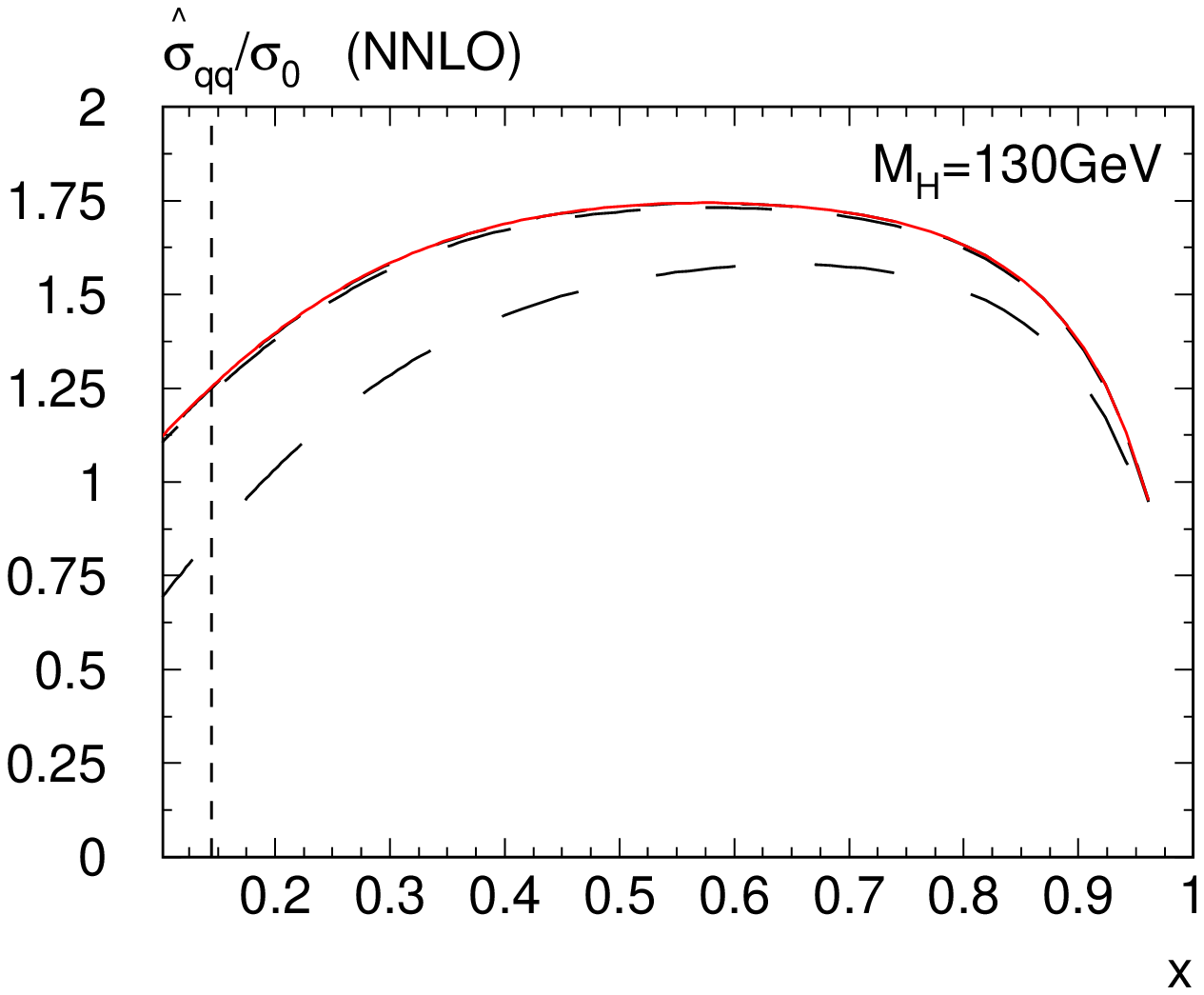}
    \end{tabular}
    \parbox{.9\textwidth}{
      \caption[]{\label{fig::sig2-130qq}\sloppy Same as
        \fig{fig::sig2-130qg}, but for the $q\bar q$ and the $qq$
        channels (identical quark flavors). The figure for the $qq'$
        channel (different quark flavours) is not shown since it is
        almost indistinguishable from the one for $qq$.}}
  \end{center}
\end{figure}

\begin{figure}
  \begin{center}
    \begin{tabular}{cc}
      \includegraphics[bb=110 265 465
        560,width=.45\textwidth]{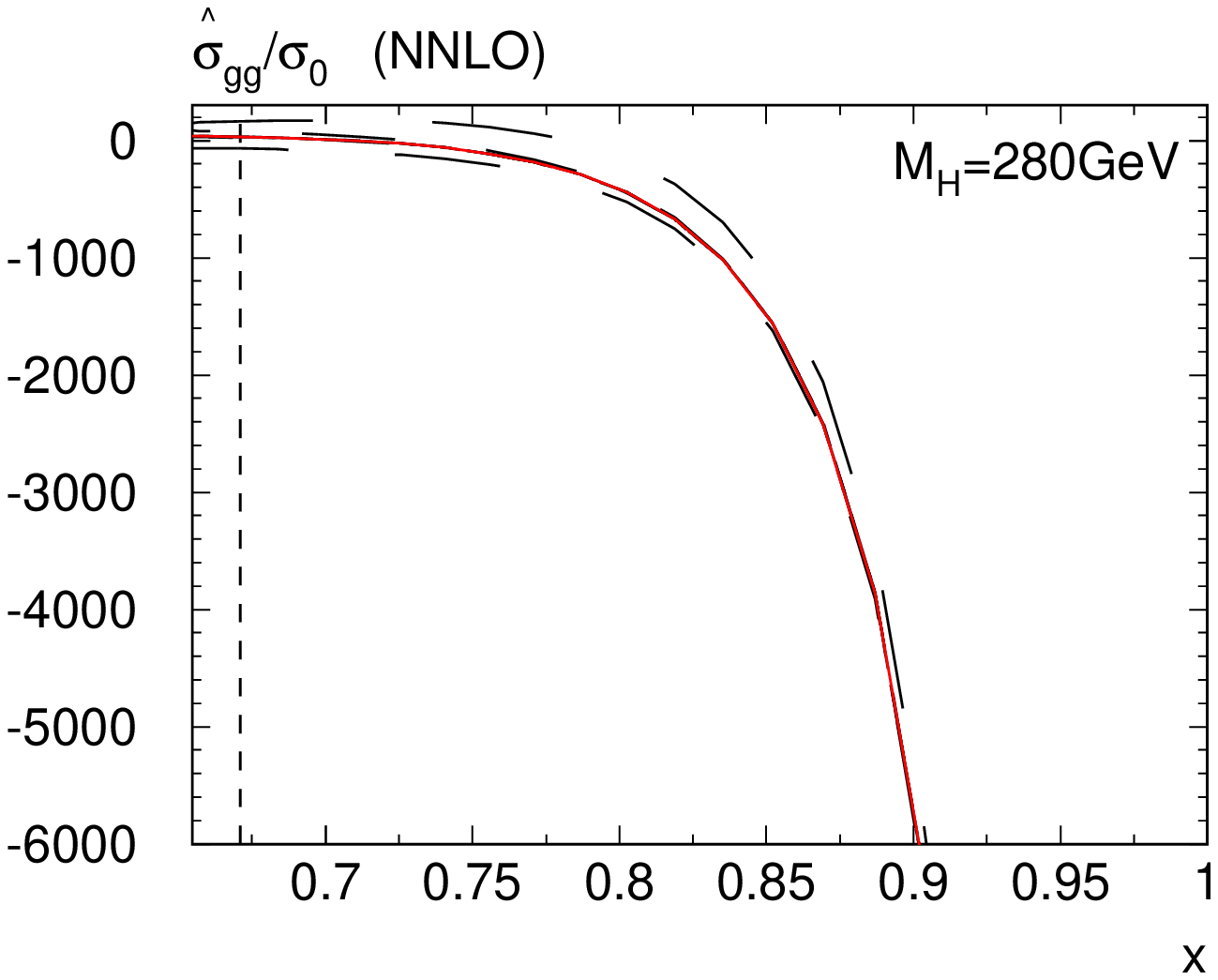} &
      \includegraphics[bb=110 265 465
        560,width=.45\textwidth]{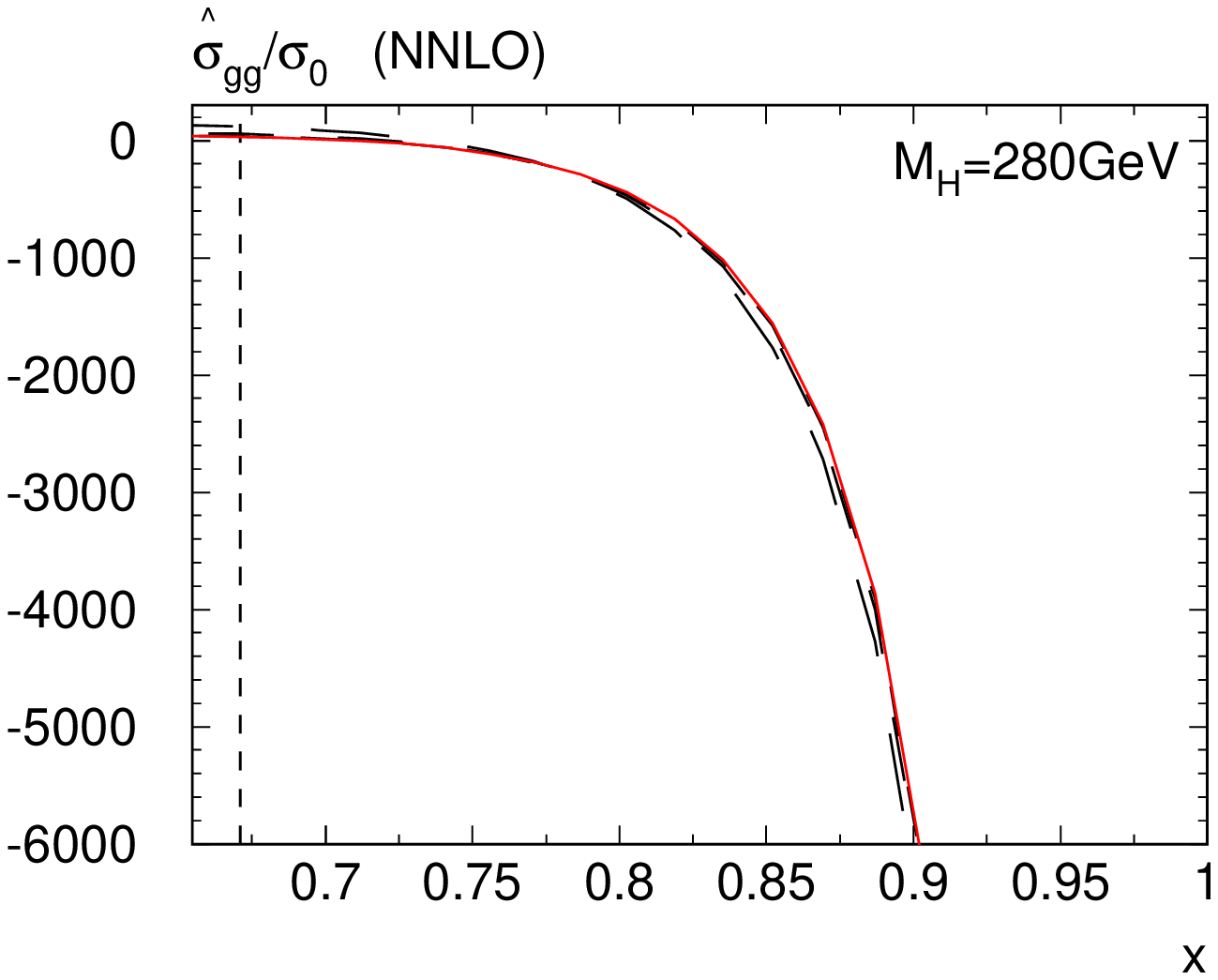} \\
      \includegraphics[bb=110 265 465
        560,width=.45\textwidth]{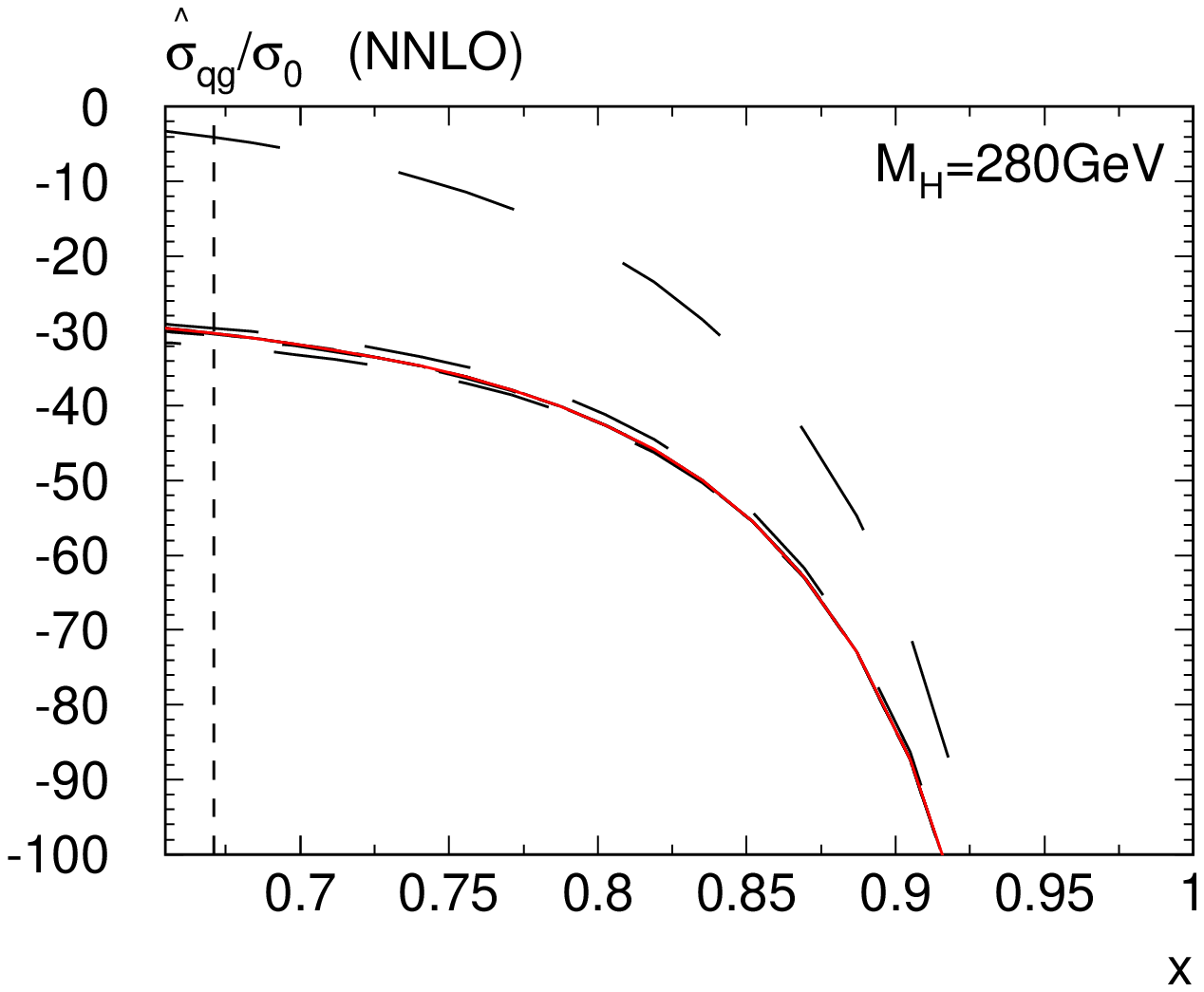} &
      \includegraphics[bb=110 265 465
        560,width=.45\textwidth]{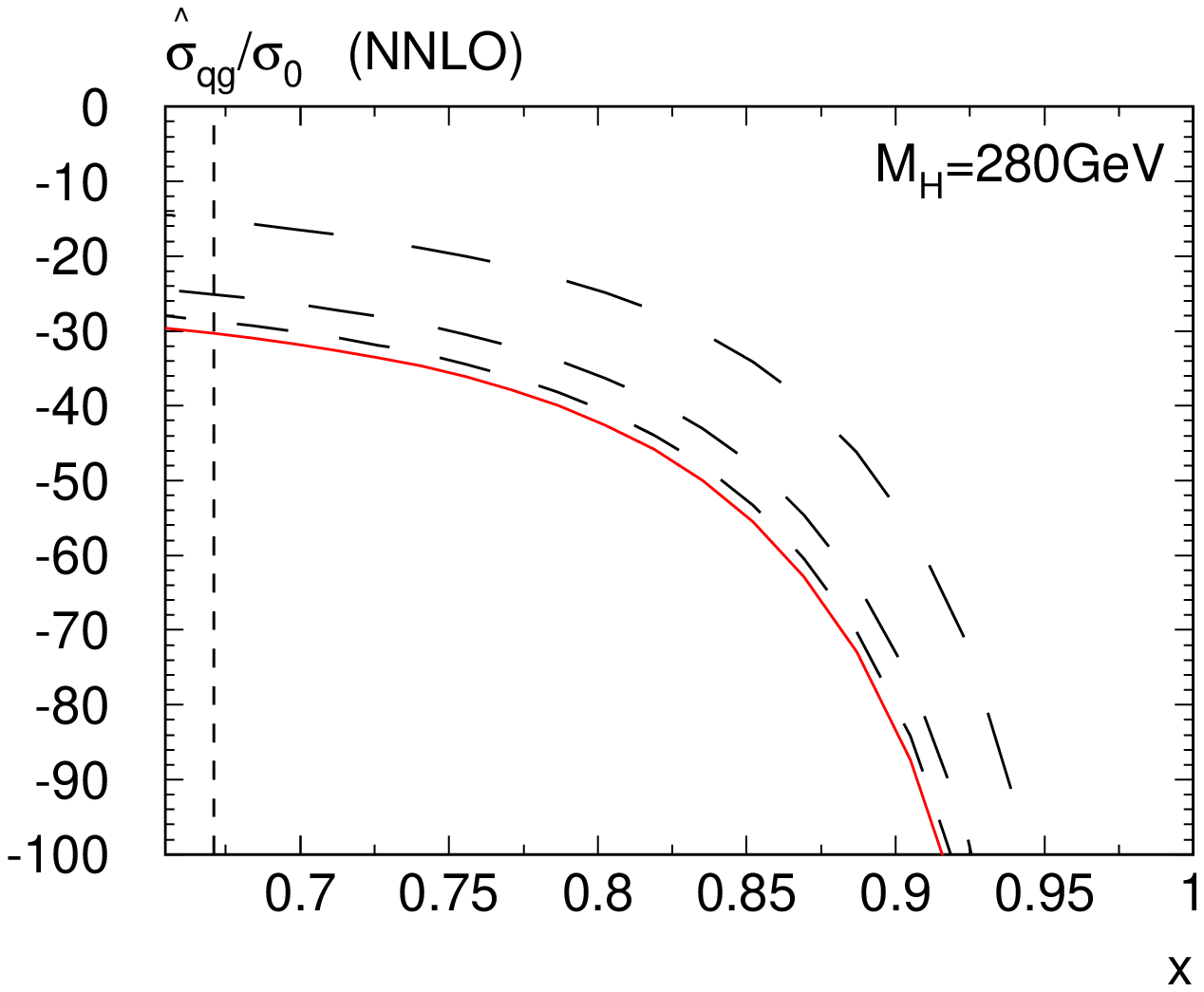}
    \end{tabular}
    \parbox{.9\textwidth}{
      \caption[]{\label{fig::sig2-280qg}\sloppy
        Same as \fig{fig::sig2-130qg}, but for $\mhiggs=280$\,GeV.
        }}
  \end{center}
\end{figure}

\begin{figure}
  \begin{center}
    \begin{tabular}{cc}
      \includegraphics[bb=110 265 465
        560,width=.45\textwidth]{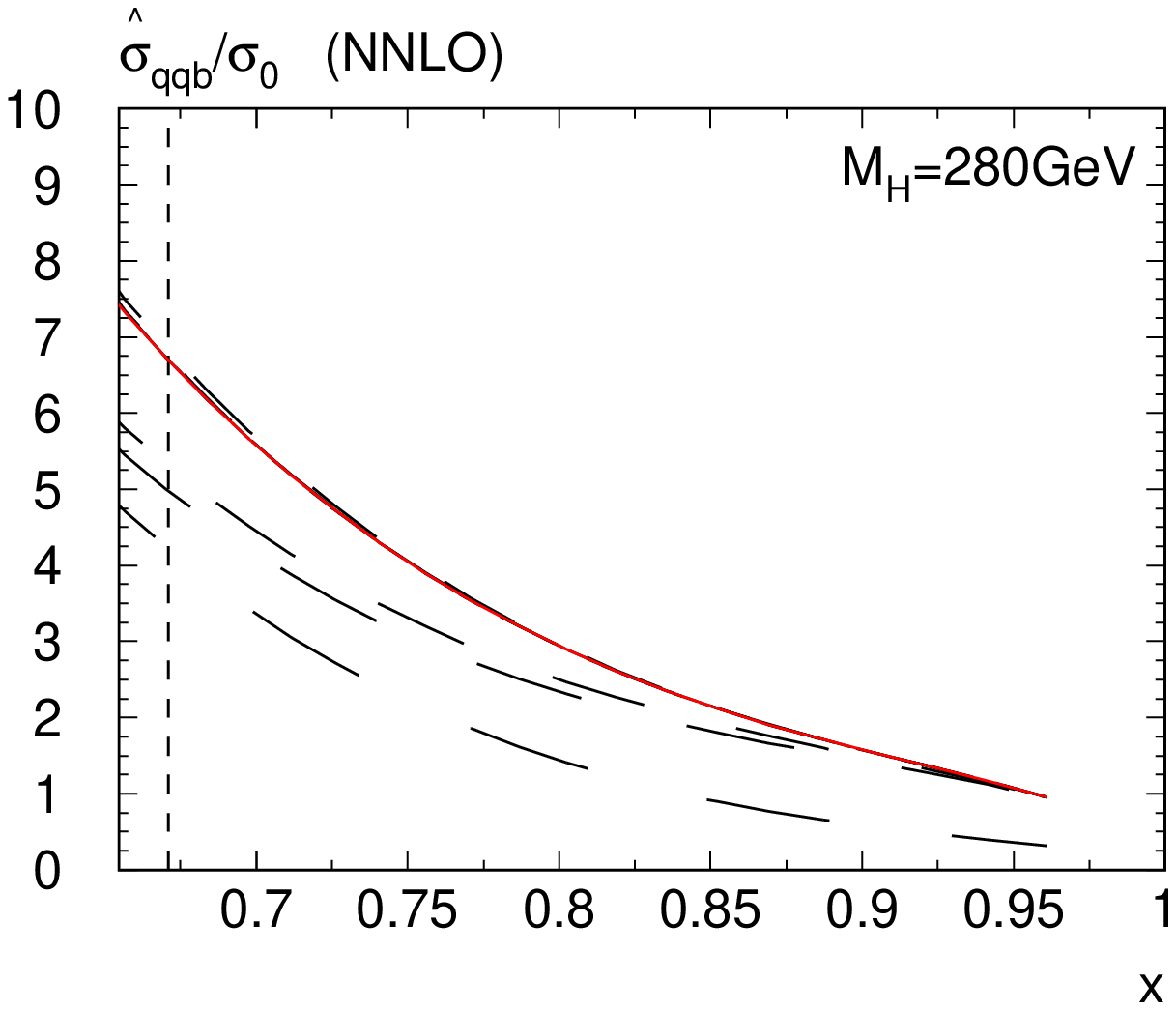} &
      \includegraphics[bb=110 265 465
        560,width=.45\textwidth]{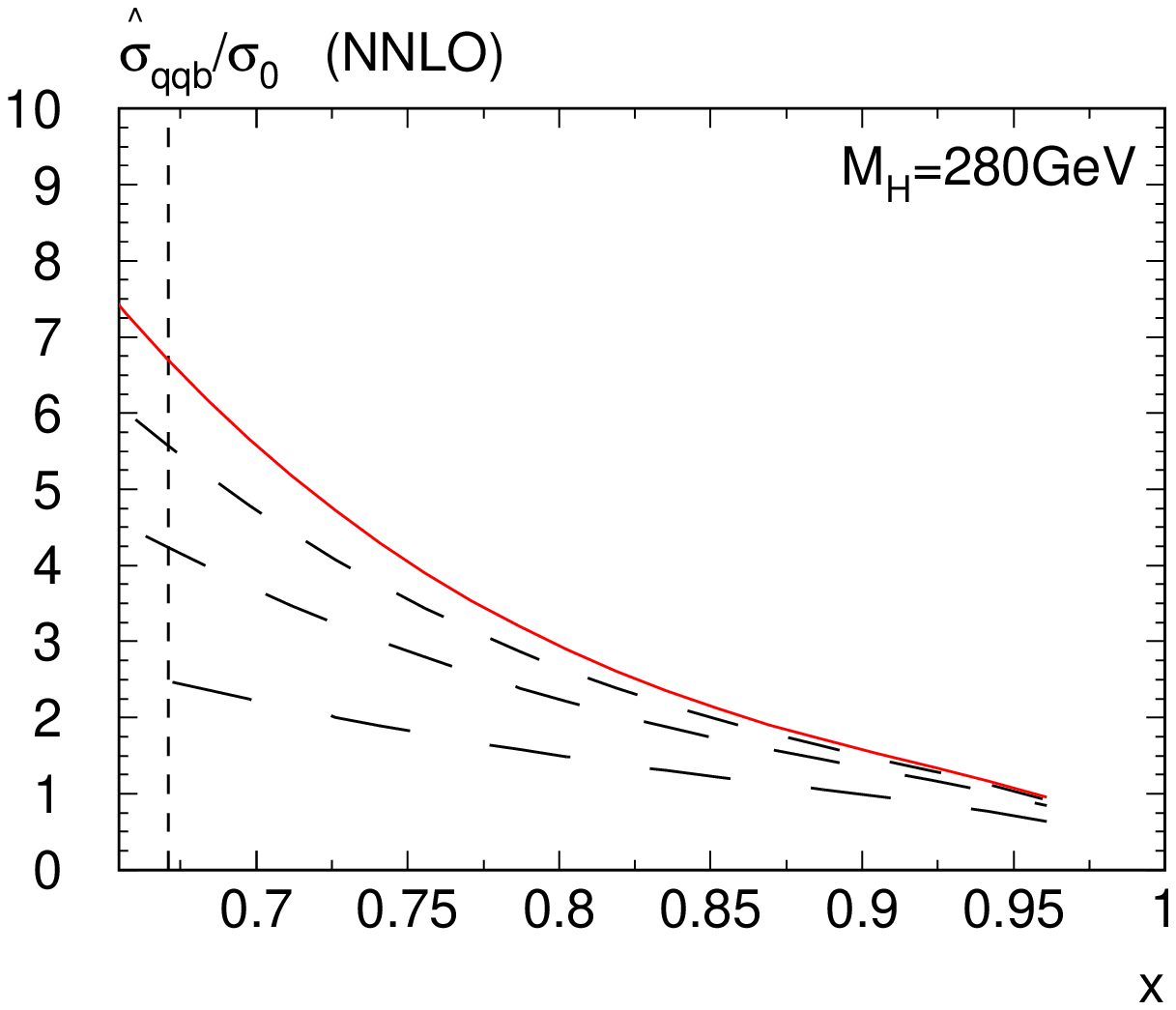} \\
      \includegraphics[bb=110 265 465
        560,width=.45\textwidth]{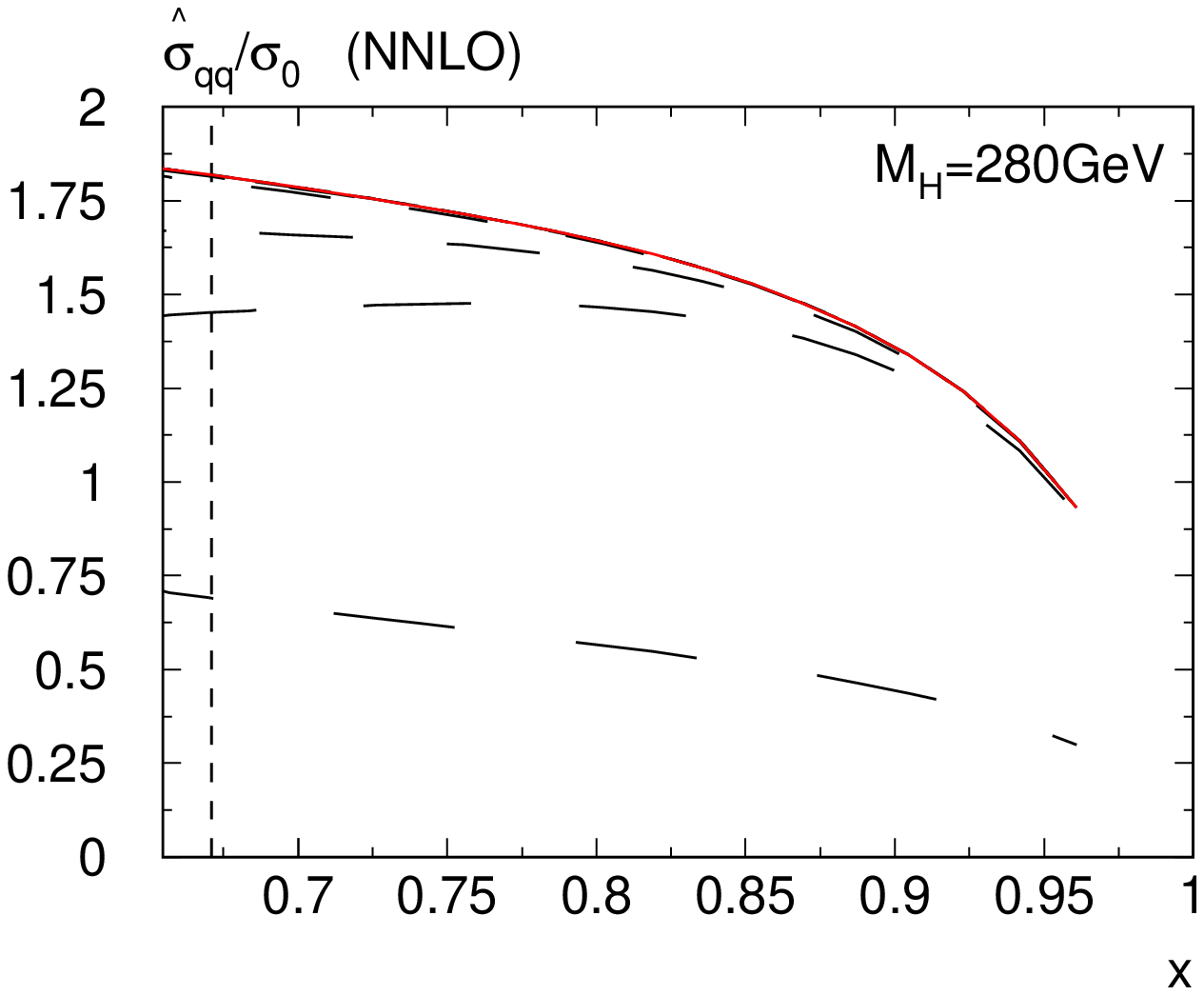} &
      \includegraphics[bb=110 265 465
        560,width=.45\textwidth]{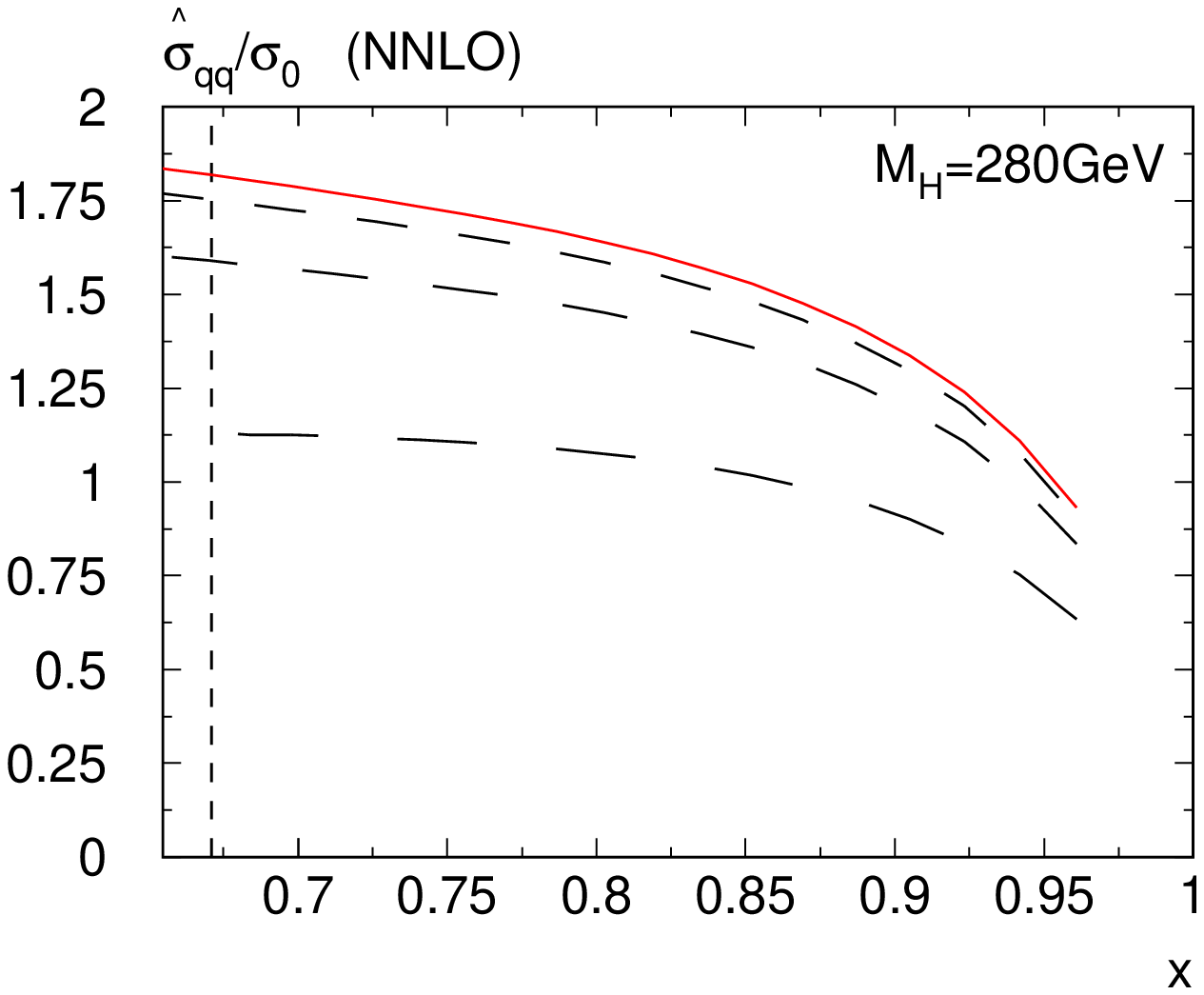}
    \end{tabular}
    \parbox{.9\textwidth}{
      \caption[]{\label{fig::sig2-280qq}\sloppy
        Same as \fig{fig::sig2-130qq}, but for $\mhiggs=280$\,GeV.
        }}
  \end{center}
\end{figure}

The corresponding curves at \nnlo{} are shown in
Figs.\,\ref{fig::sig2-130qg}--\ref{fig::sig2-280qq}. There is no exact
result that one could compare to, but the quality of the convergence
both of the $1/\mtop$ and the $(1-x)$ expansions below threshold
convincingly shows that they approximate the exact result to a very high
degree in this region.

\begin{figure}
  \begin{center}
    \begin{tabular}{cc}
      \includegraphics[bb=110 265 465
        560,width=.45\textwidth]{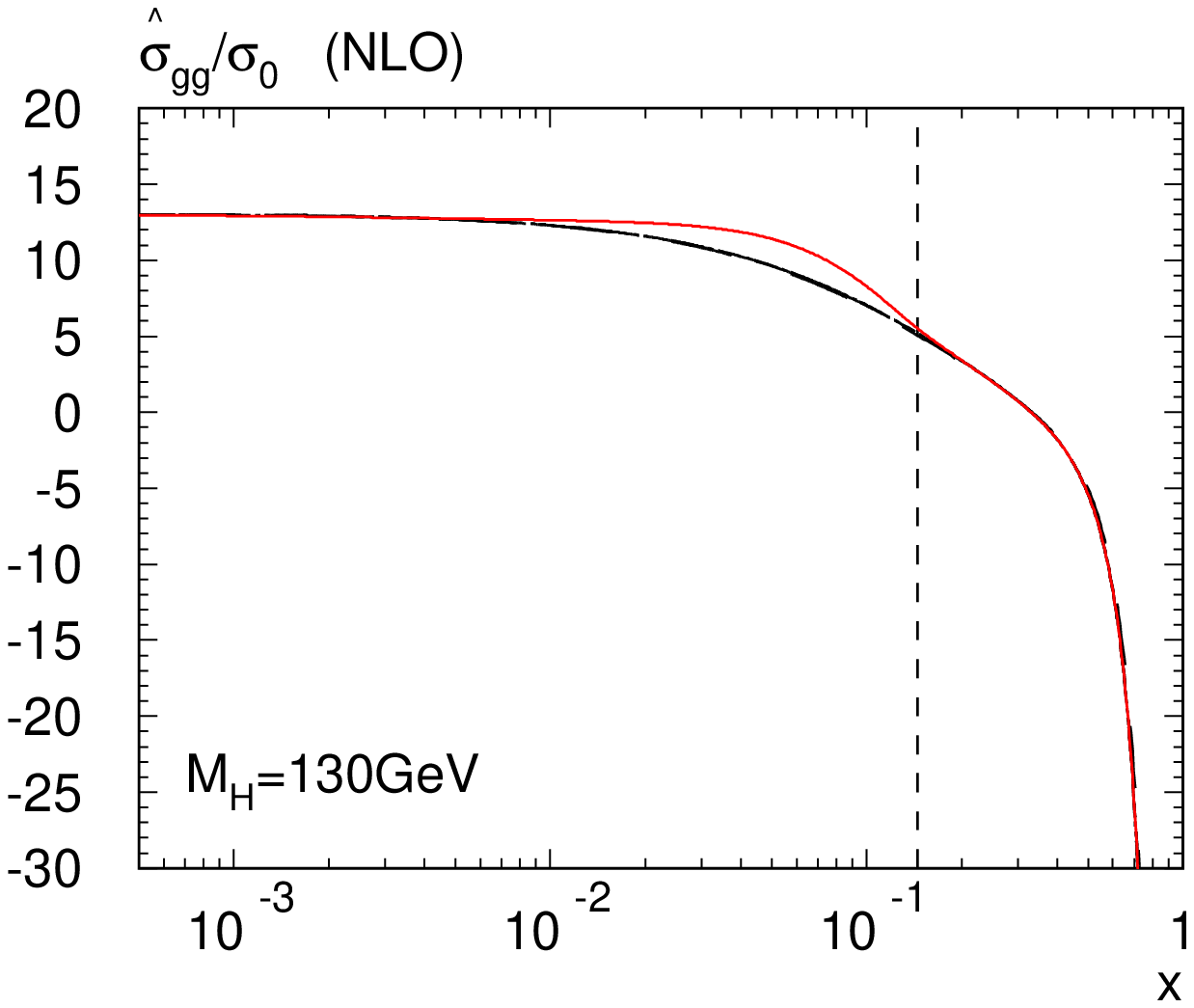} &
      \includegraphics[bb=110 265 465
        560,width=.45\textwidth]{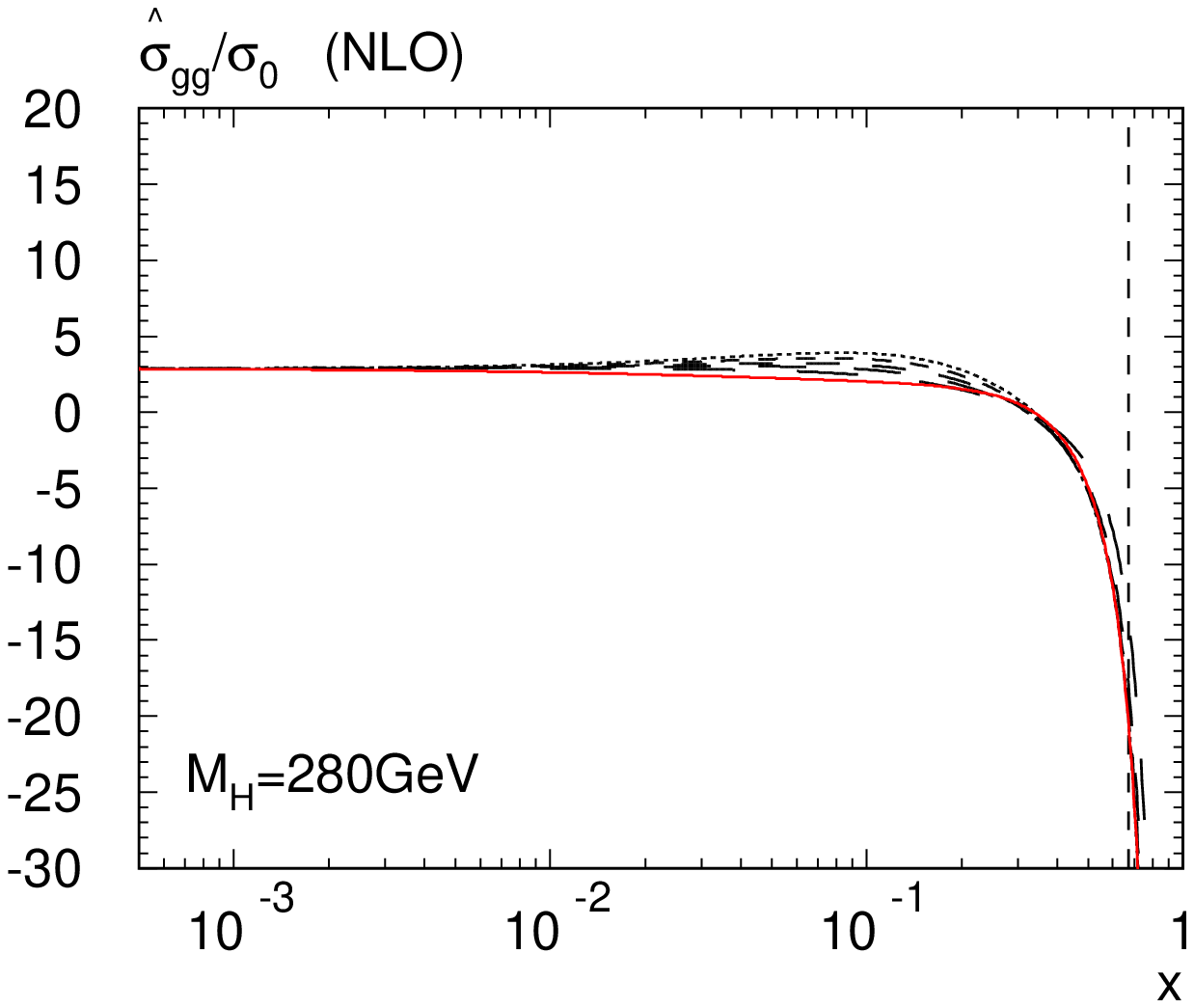} \\
      \includegraphics[bb=110 265 465
        560,width=.45\textwidth]{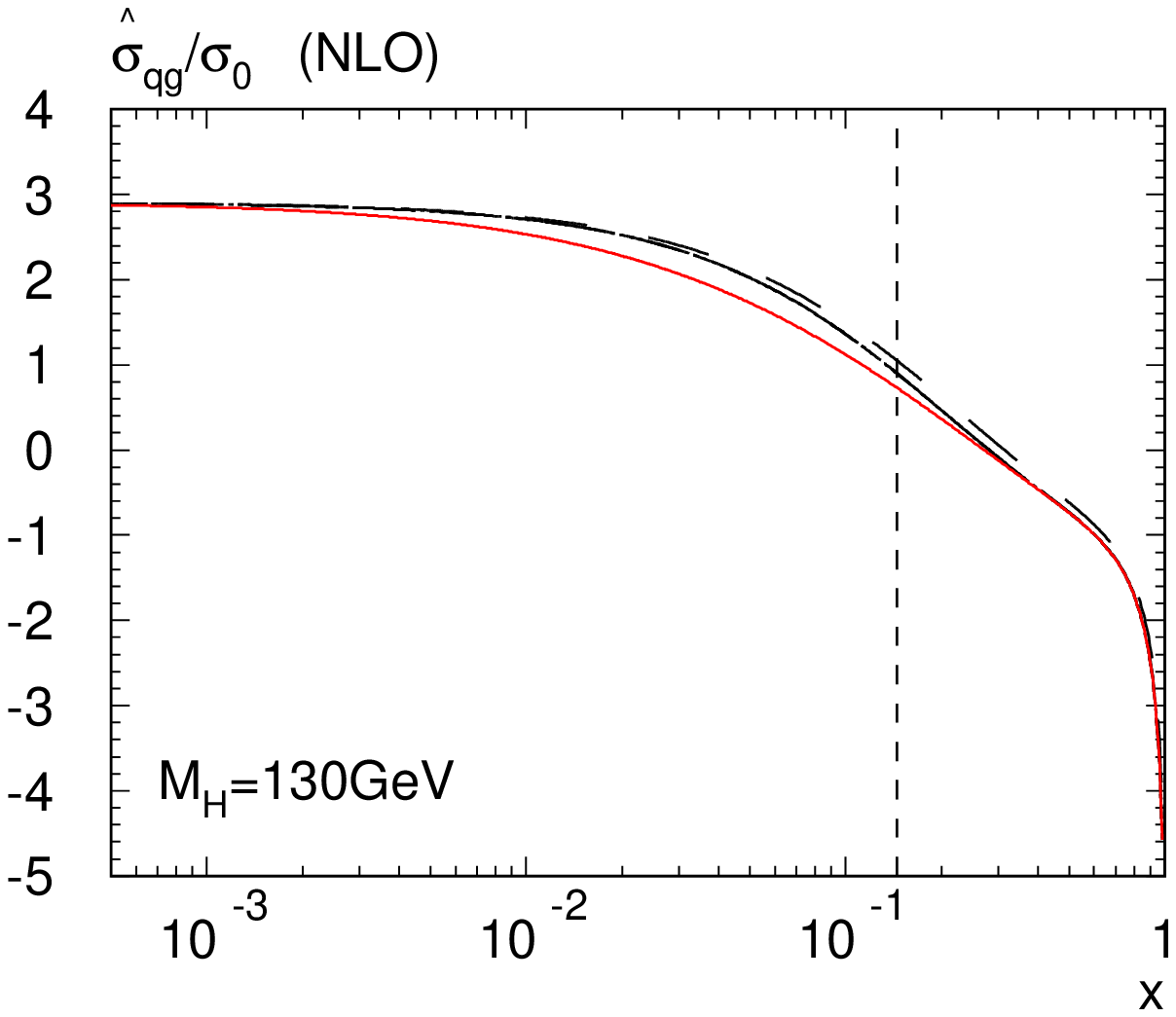} &
      \includegraphics[bb=110 265 465
        560,width=.45\textwidth]{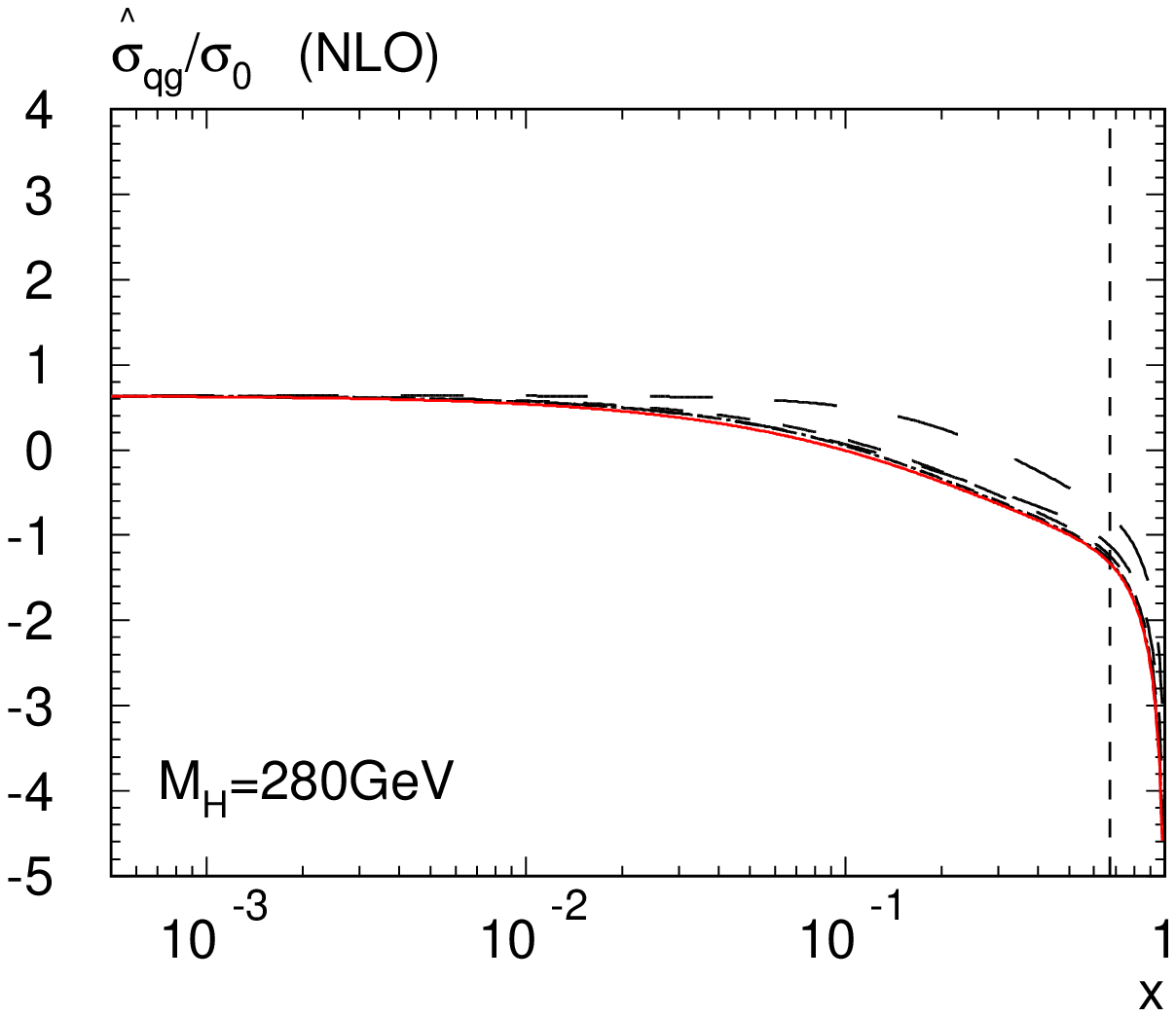} \\
      \includegraphics[bb=110 265 465
        560,width=.45\textwidth]{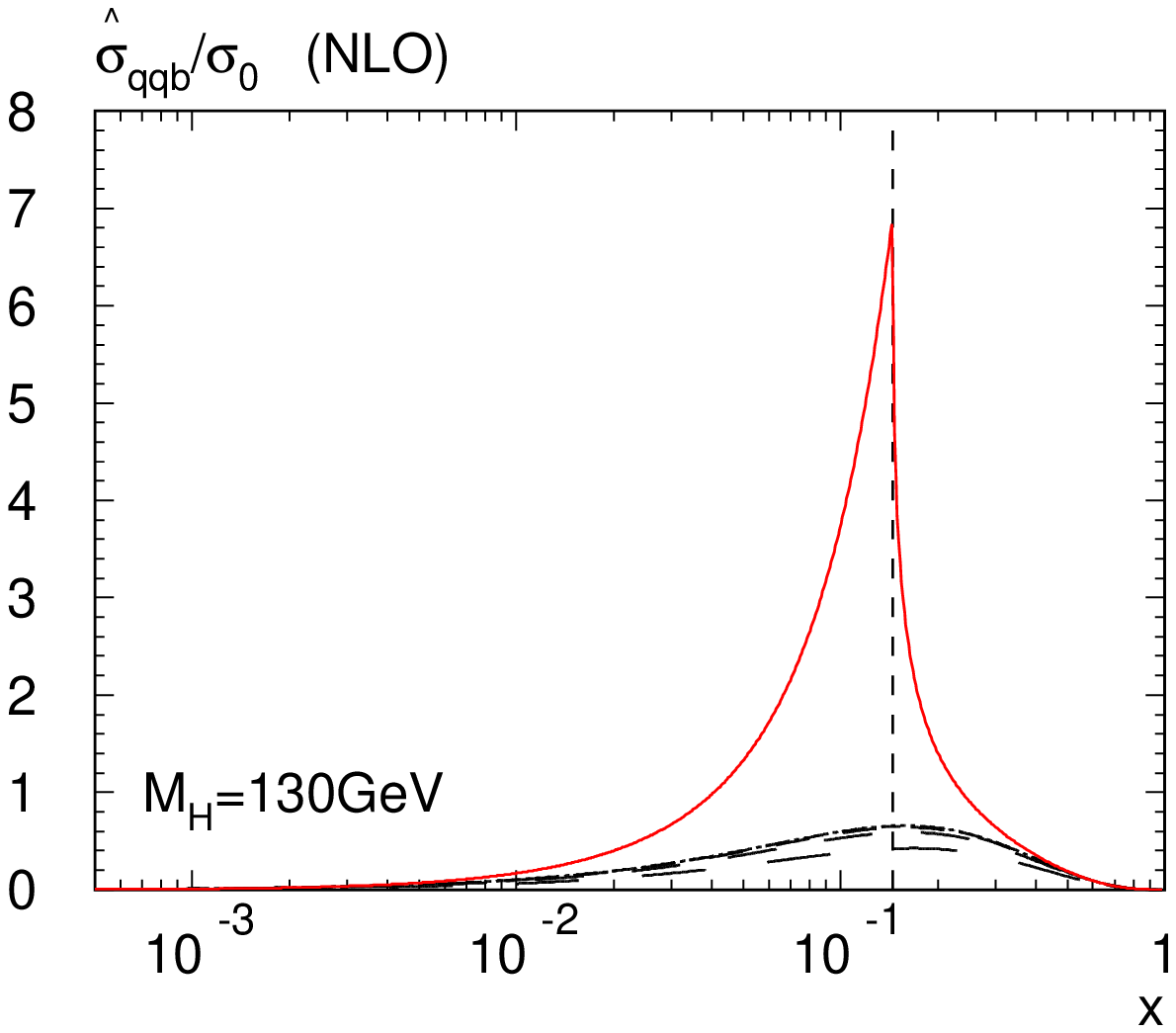} &
      \includegraphics[bb=110 265 465
        560,width=.45\textwidth]{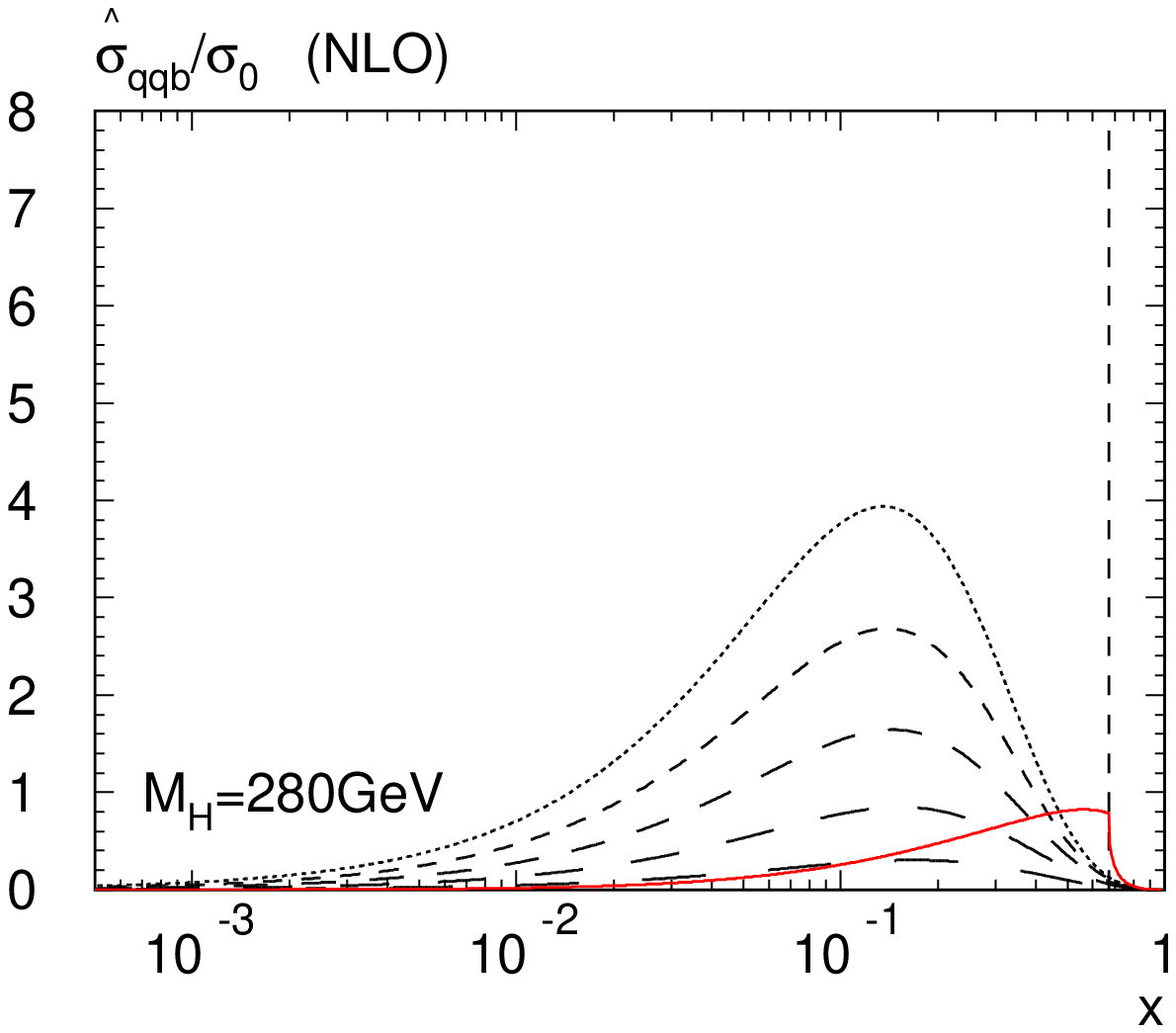}
    \end{tabular}
    \parbox{.9\textwidth}{
      \caption[]{\label{fig::sig1-match}\sloppy \nlo{} partonic cross
        sections as constructed from \eqn{eq::match} (with $N=8$) by
        including successively higher orders in $1/\mtop^2$. Dashed:
        $\order{\mtop^{2n}}$, $n=0,\ldots,5$.  Solid: exact. Left/right
        column: $\mhiggs=130$\,GeV/$\mhiggs=280$\,GeV.  The dashed
        vertical line indicates the threshold.}}
  \end{center}
\end{figure}

\begin{figure}
  \begin{center}
    \begin{tabular}{cc}
      \includegraphics[bb=110 265 465
        560,width=.45\textwidth]{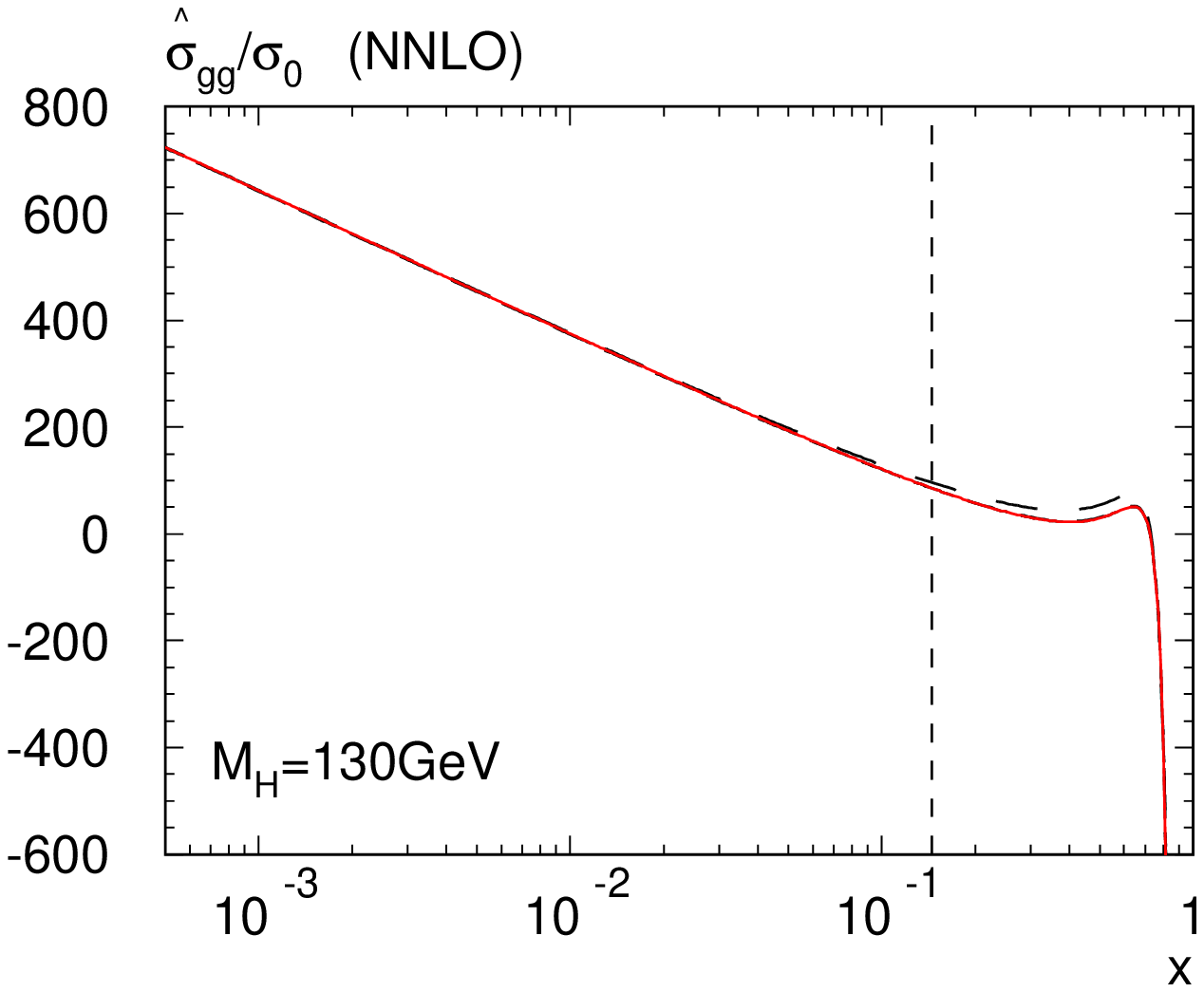} &
      \includegraphics[bb=110 265 465
        560,width=.45\textwidth]{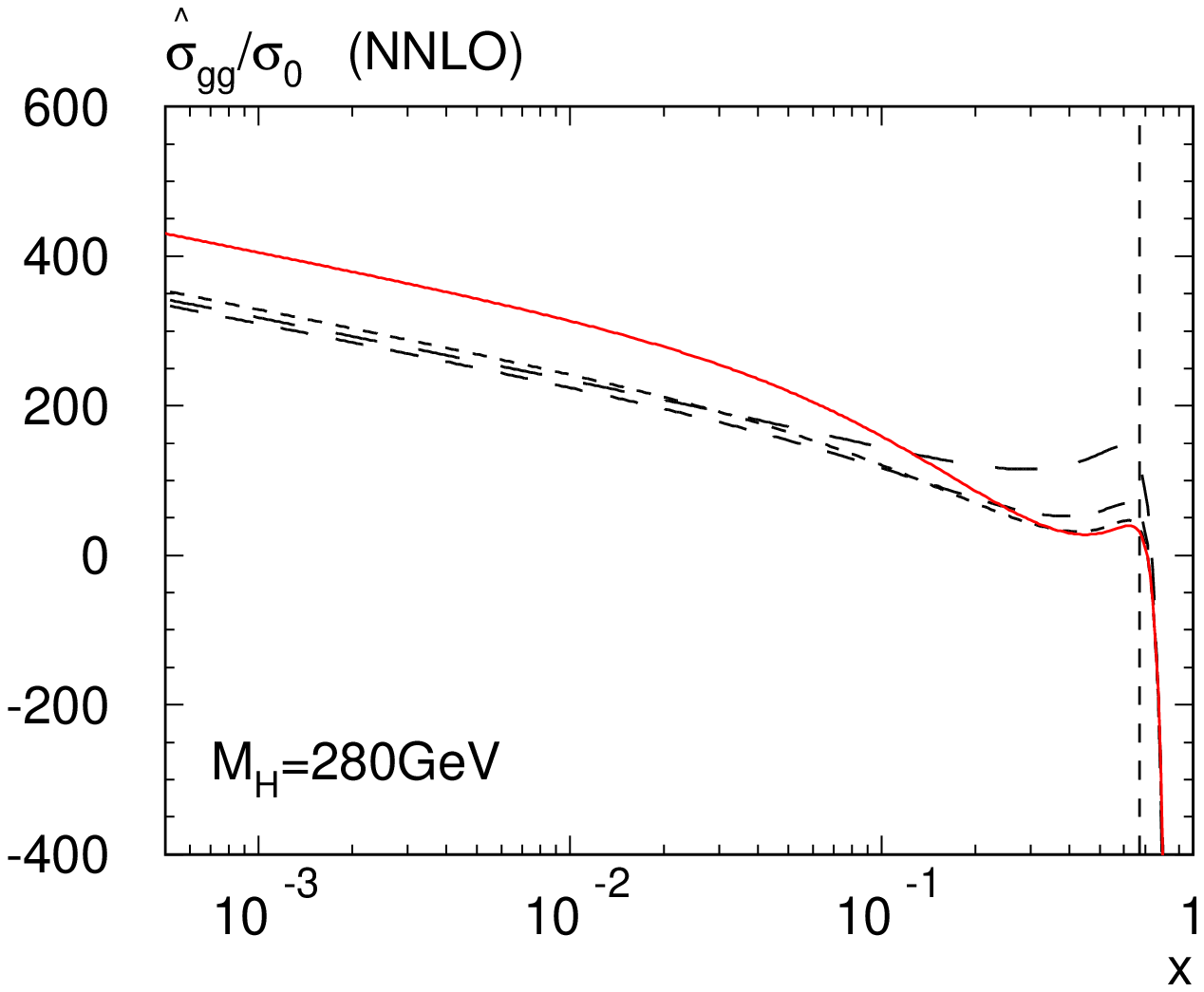} \\
      \includegraphics[bb=110 265 465
        560,width=.45\textwidth]{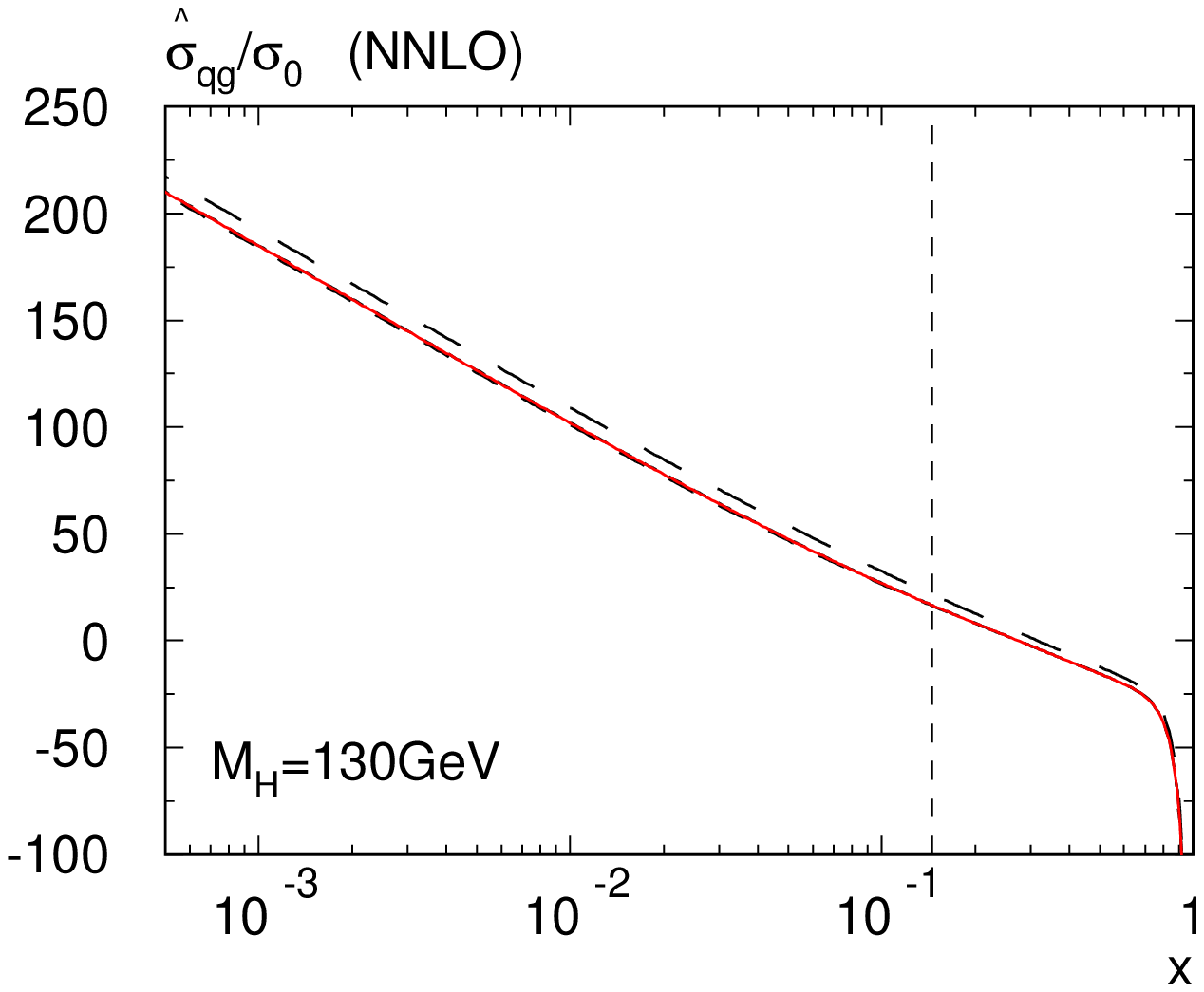} &
      \includegraphics[bb=110 265 465
        560,width=.45\textwidth]{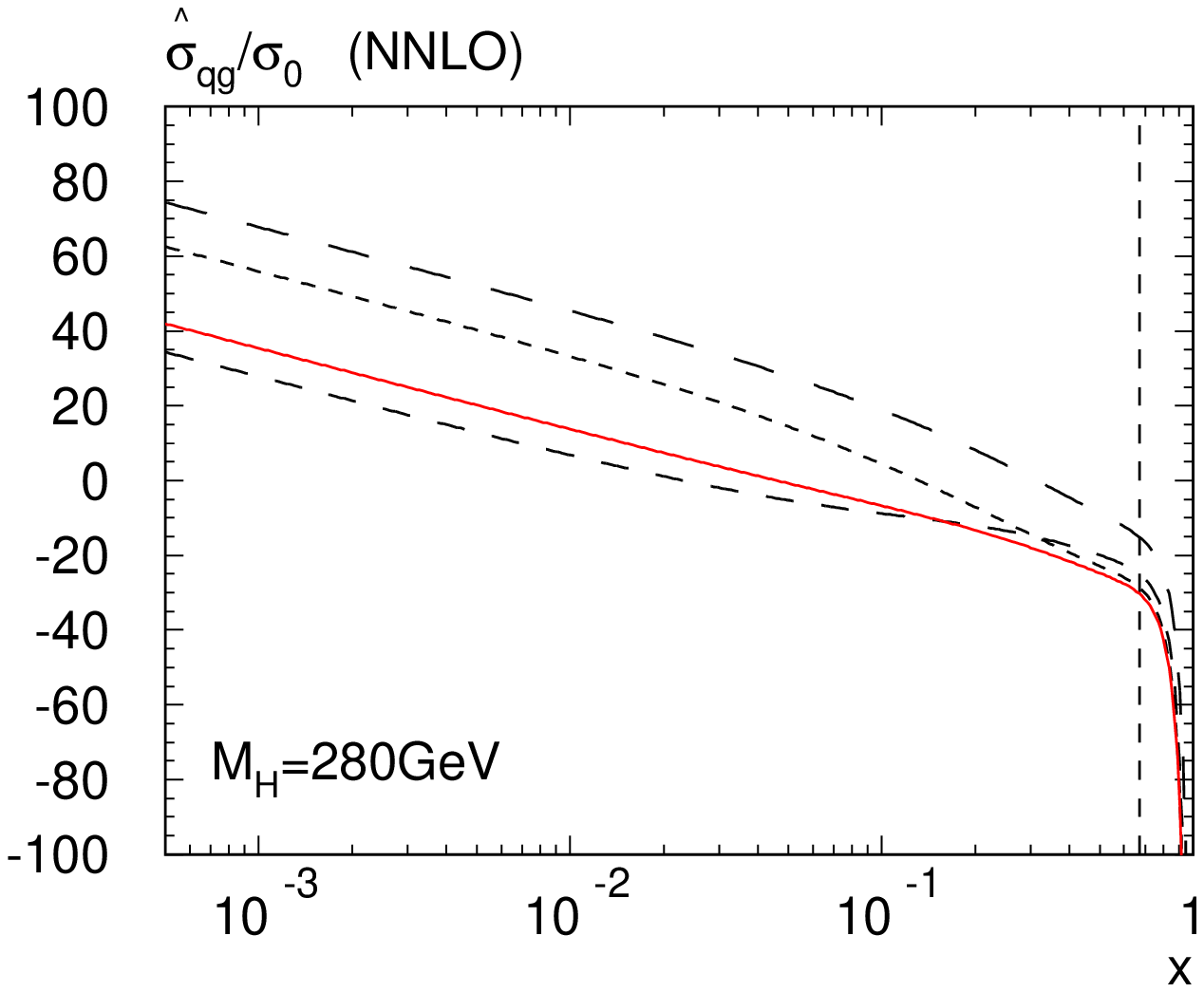}
    \end{tabular}
    \parbox{.9\textwidth}{
      \caption[]{\label{fig::sig2-matchqg}\sloppy \nnlo{} partonic cross
        sections ($gg$ and $qg$ channel) as constructed from
        \eqn{eq::match} (with $N=8$) by including successively higher
        orders in $1/\mtop^2$. Dashed: $\order{\mtop^{2n}}$, $n=0,1,2$.
        Solid: $n=3$. Left/right column:
        $\mhiggs=130$\,GeV/$\mhiggs=280$\,GeV.  The dashed vertical line
        indicates the threshold.  For the $q\bar q$ and the
        $qq$ channel, see \fig{fig::sig2-matchqq}.}}
  \end{center}
\end{figure}

\begin{figure}
  \begin{center}
    \begin{tabular}{cc}
      \includegraphics[bb=110 265 465
        560,width=.45\textwidth]{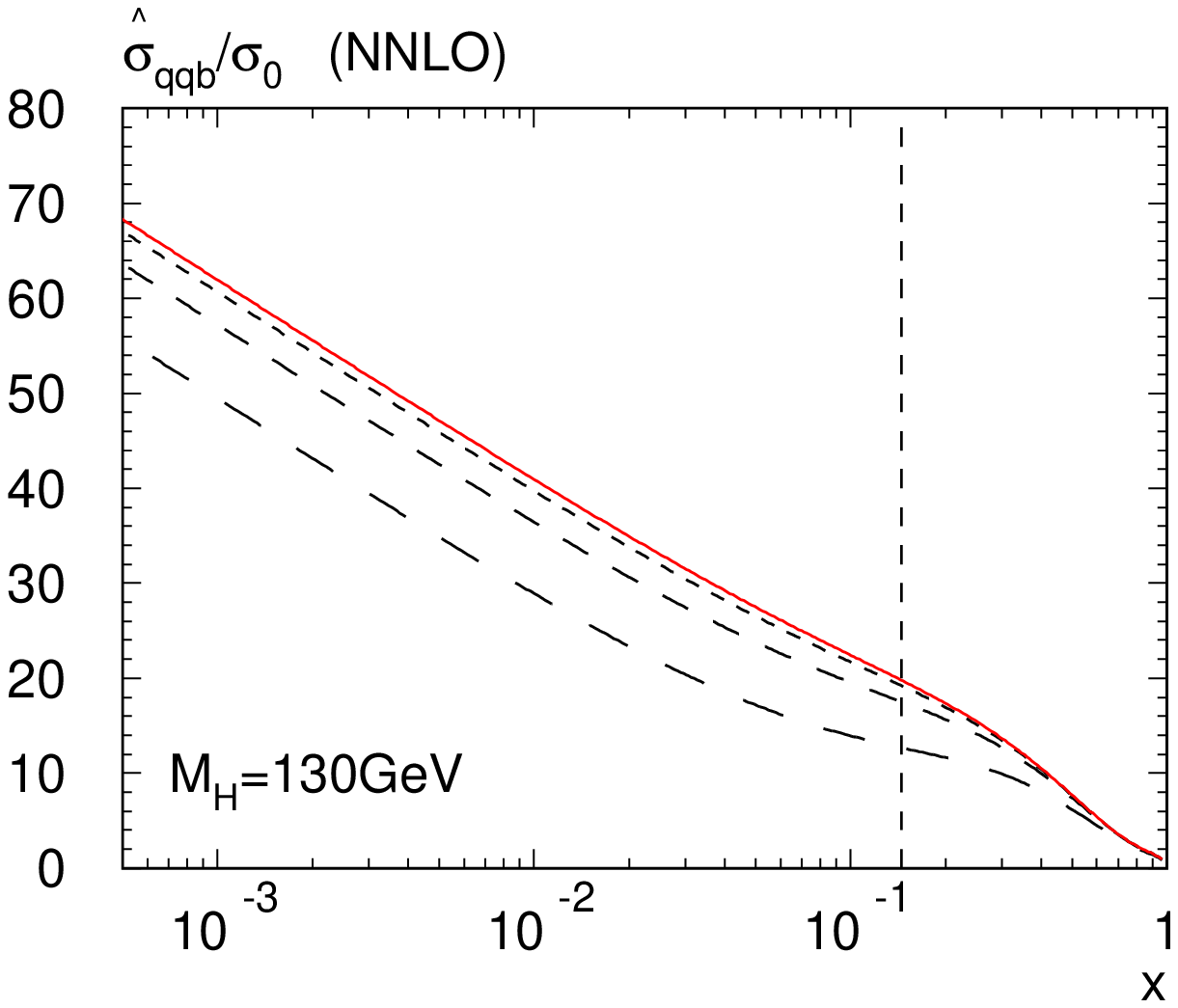} &
      \includegraphics[bb=110 265 465
        560,width=.45\textwidth]{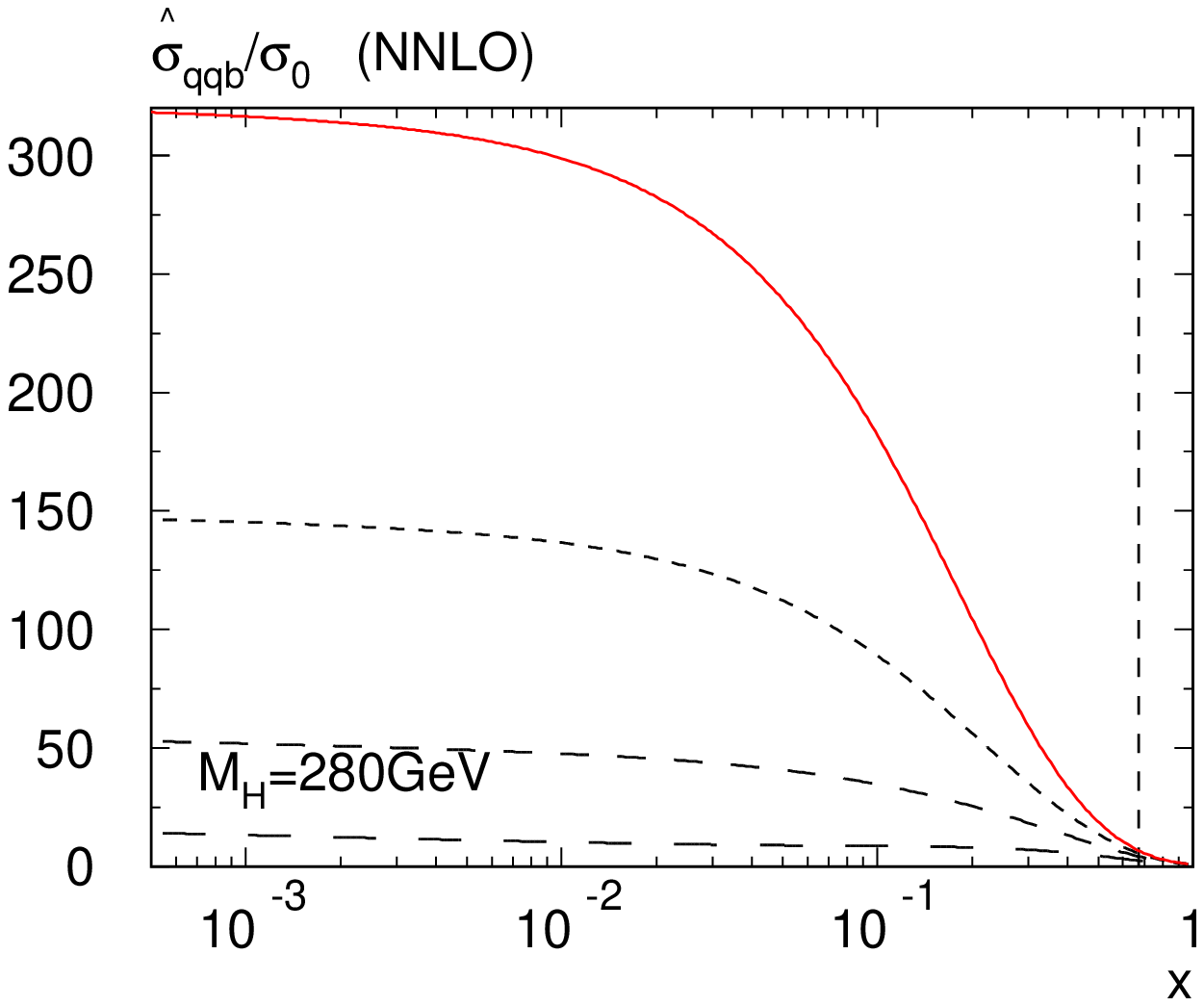} \\
      \includegraphics[bb=110 265 465
        560,width=.45\textwidth]{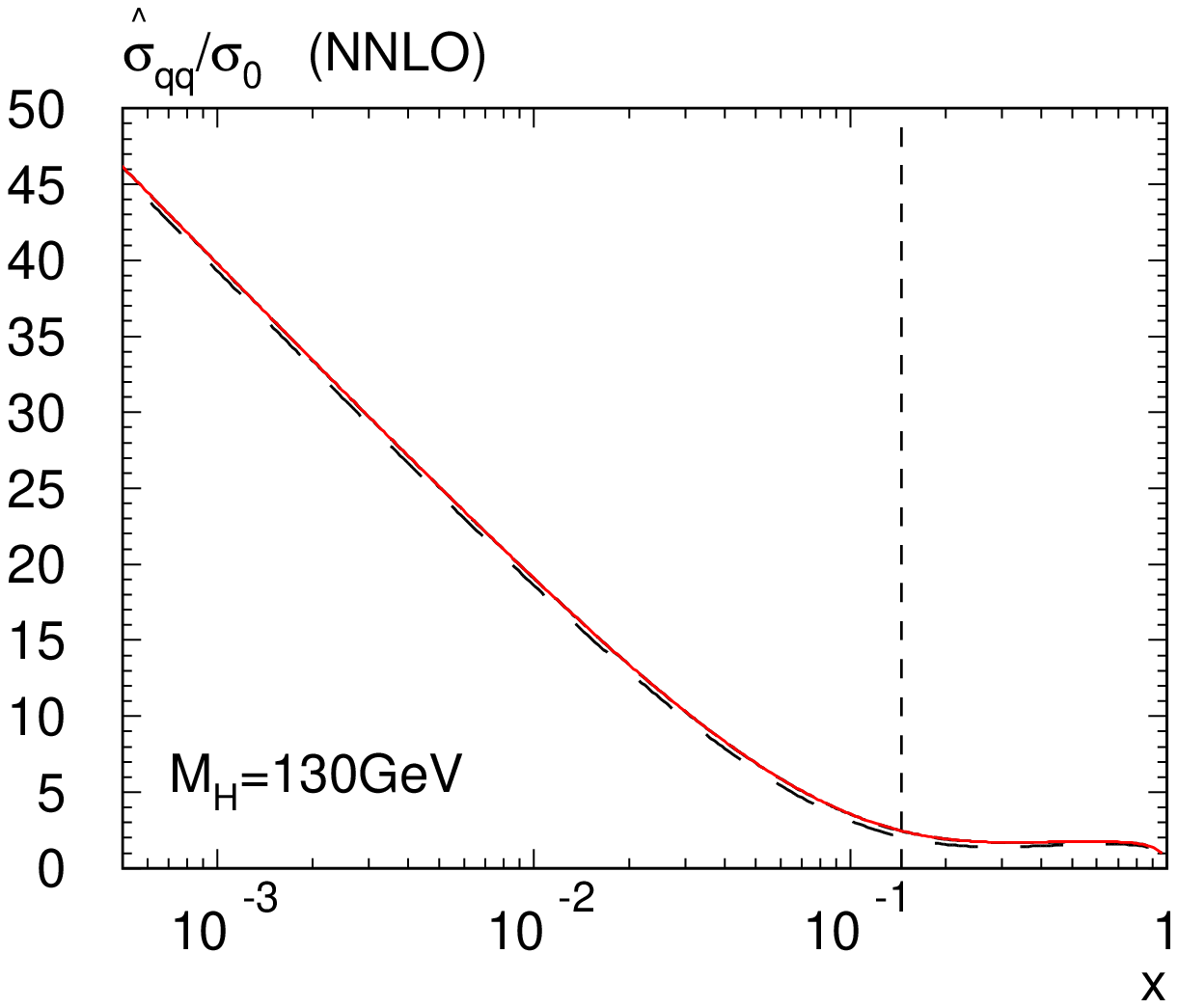} &
      \includegraphics[bb=110 265 465
        560,width=.45\textwidth]{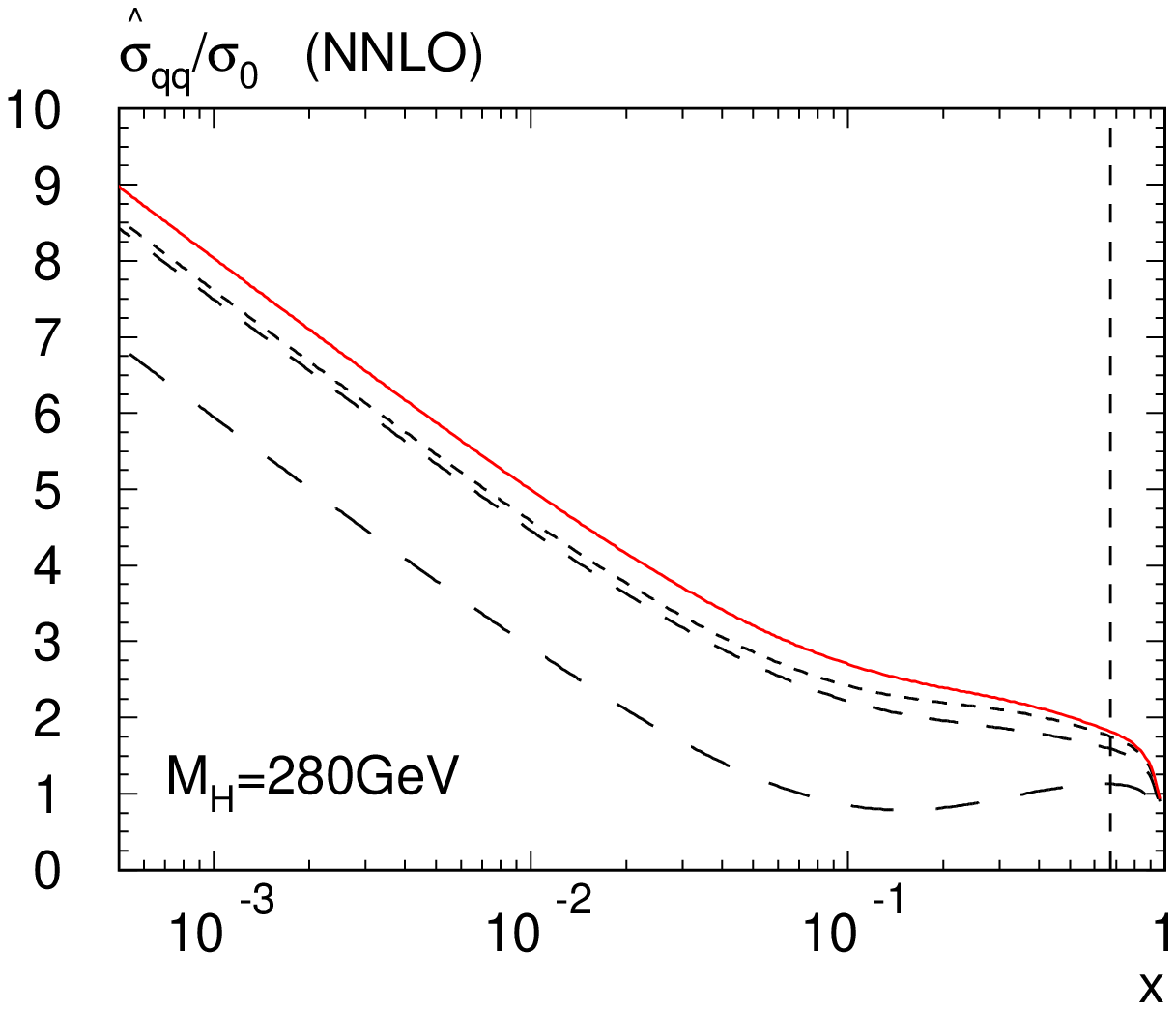}
    \end{tabular}
    \parbox{.9\textwidth}{
      \caption[]{\label{fig::sig2-matchqq}\sloppy Same as
        \fig{fig::sig2-matchqg}, but for the $q\bar q$ and the $qq$
        channels (identical quark flavors). The figure for the $qq'$
        channel (different quark flavours) is not shown since it is
        almost indistinguishable from the one for $qq$. }}
  \end{center}
\end{figure}

From Section~\ref{sec::smallxderiv} we know the leading high energy
behaviour for general values of $\mtop$ and $\mhiggs$. There are many
ways then to merge the available information into a smooth function with
the correct high- and low-energy behaviour (see, e.g.,
Refs.\,\cite{Marzani:2008az,Marzani:2008ih,Pak:2009dg}).  We decide to
use~\cite{Harlander:2009mq}
\begin{equation}
\begin{split}
\hat\sigma^{(n)}_{\alpha\beta}(x) = \hat\sigma^{(n)}_{\alpha\beta,N}(x)
&+ \,\sigma_0 A^{(n)}_{\alpha\beta}\left[ \ln \frac{1}{x} -
  \sum_{k=1}^N\frac{1}{k}(1-x)^k  \right]\\&
+ (1-x)^{N+1}\,\left[\sigma_0 B_{\alpha\beta}^{(n)} -
  \hat\sigma^{(n)}_{\alpha\beta,N}(0)
  \right]\,,
\label{eq::match}
\end{split}
\end{equation}
where $\hat\sigma^{(n)}_{\alpha\beta,N}(x)$ denotes the soft expansion
of the partonic cross section through order $(1-x)^{N}$.  For the
unknown constants at \nnlo{} we use the default values
$\sigma_0B^{(2)}_{\alpha\beta} = \hat\sigma^{(n)}_{\alpha\beta,N}(0)$,
but we will study their influence on the \nnlo{} hadronic results at the
end of Section~\ref{sec::hadronic}.

At \nlo{}, the resulting partonic cross sections are shown in
\fig{fig::sig1-match} for $\mhiggs=130$\,GeV and $\mhiggs=280$\,GeV. The
precision to which the $gg$ and $qg$ channels reproduce the exact result
is quite impressive. As expected, the $q\bar q$ channel is approximated
only very poorly though.

The corresponding plots at \nnlo{} are shown in \fig{fig::sig2-matchqg}
and \ref{fig::sig2-matchqq}, for the default values of the high energy
constant $B^{(2)}_{\alpha\beta}$. Note that due to this undetermined
constant, the curves do not all converge to the same point for $x\to 0$
as is the case at \nlo{}. Only the slope is determined by the
logarithmic coefficient $A^{(2)}_{\alpha\beta}$. Nevertheless,
convergence for the $gg$ and the $qg$ channel is very good, in
particular at low Higgs masses. At larger Higgs masses, the $x$
  dependence becomes more and more unreliable as more terms in $1/\mtop$
  are included. This leads to the observed variations of the final
  result at $\mhiggs=280$\,GeV.  As will be shown at the end of
  Section~\ref{sec::hadronic}, however, the variations affect the
  hadronic cross section by less than 1\%.

As expected, the $q\bar q$ channel does not seem to converge, but its
contribution to the hadronic cross section is negligible as are those of
the $qq$ and the $qq'$ channels (equal and different quark flavours,
respectively). The observed convergence of the latter is much better
though. In the following, we will include them in the total hadronic
cross section, but we will only discuss the $q\bar q$ channel as
representative of the pure quark channels (the one with the worst
convergence behaviour).

\section{Hadronic Results}\label{sec::hadronic}

In order to study the effect of the $1/\mtop$ terms on the hadronic
cross section, we define (see also Ref.\,\cite{Harlander:2009mq})
\begin{equation}
\begin{split}
\hat\sigma^\nlo_{\alpha\beta}(\mtop^n) &=
\sigma_0\,\delta_{\alpha g}\delta_{\beta g}\delta(1-x)
+ \hat\sigma_{\alpha\beta}^{(1)}(\mtop^n)\,,\\
\hat\sigma^\nnlo_{\alpha\beta}(\mtop^n) &=
\sigma_0\,\left[\delta_{\alpha g}\delta_{\beta g}\delta(1-x)
+ \Delta_{\alpha\beta,\infty}^{(1)}\right] 
+ \hat\sigma_{\alpha\beta}^{(2)}(\mtop^n)\,,
\label{eq::sigmtexp}
\end{split}
\end{equation}
where $\hat\sigma_{\alpha\beta}^{(k)}(\mtop^n)$ is the {\abbrev
  N}$^k$\lo{} contribution to the partonic cross section evaluated as an
expansion through $\order{1/\mtop^n}$, and matched to the low-$x$ limit
as described in
Section~\ref{sec::smallxderiv}. $\Delta_{\alpha\beta,\infty}^{(1)}$ is
the \eft{} result as defined in \eqn{eq::heavytop}.  Note that this
differs from an {\it extended \eft{} approach}, where
$\Delta_{\alpha\beta}^{(k)}$ would be expanded in terms of $1/\mtop$,
while the full $\tau$ dependence in $\sigma_0(\tau)$ is kept. We will
return to this latter approach at the end of this section.
The corresponding hadronic quantities derived from \eqn{eq::sigmtexp}
are denoted by $\sigma^\nlo_{\alpha\beta}(\mtop^n)$ and
$\sigma^\nnlo_{\alpha\beta}(\mtop^n)$.

\begin{figure}
  \begin{center}
    \begin{tabular}{cc}
      \subfigure[]{\includegraphics[bb=110 265 465
        560,width=.45\textwidth]{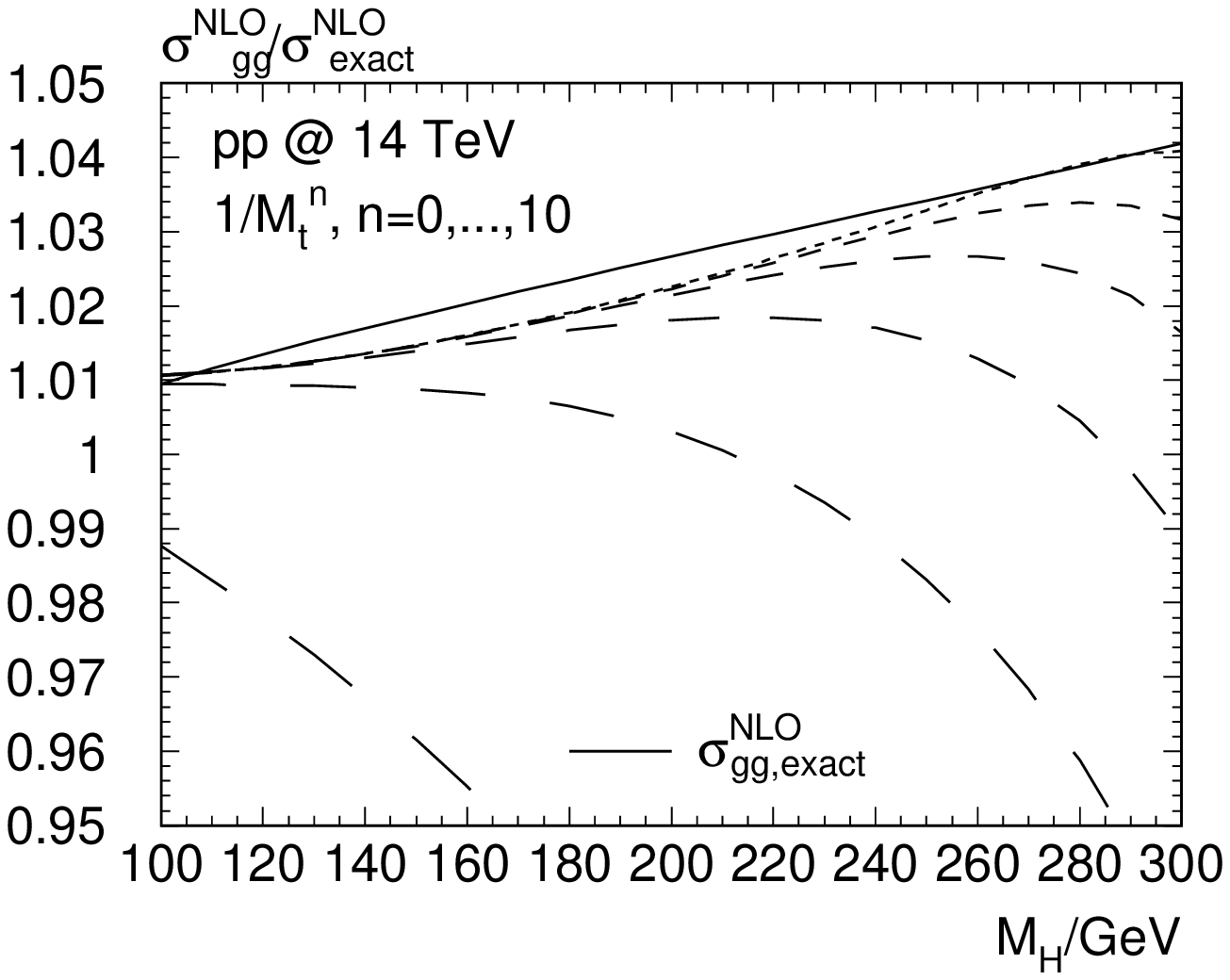}\label{subfig1}} &
      \subfigure[]{\includegraphics[bb=110 265 465
        560,width=.45\textwidth]{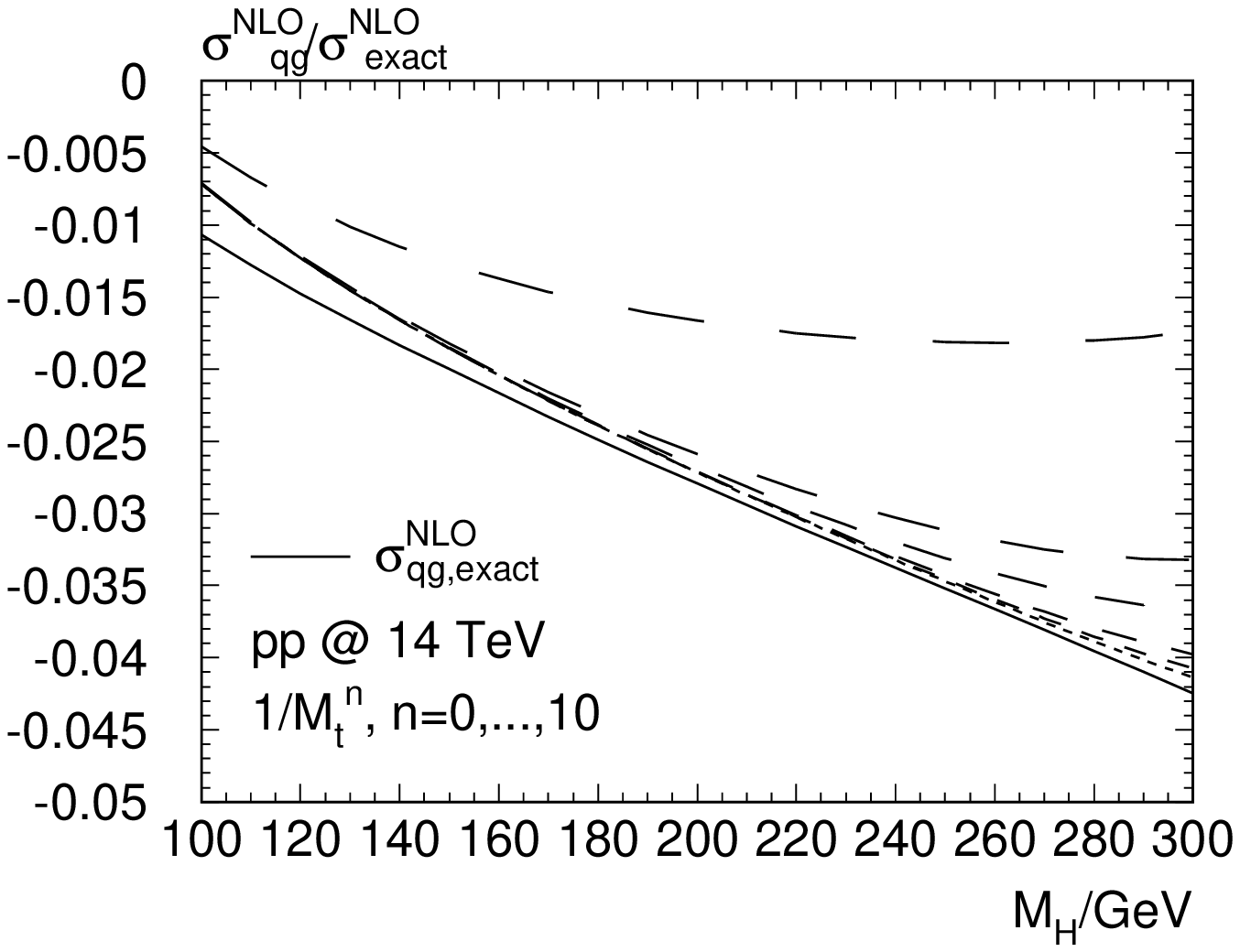}} \\
      \subfigure[]{\includegraphics[bb=110 265 465
        560,width=.45\textwidth]{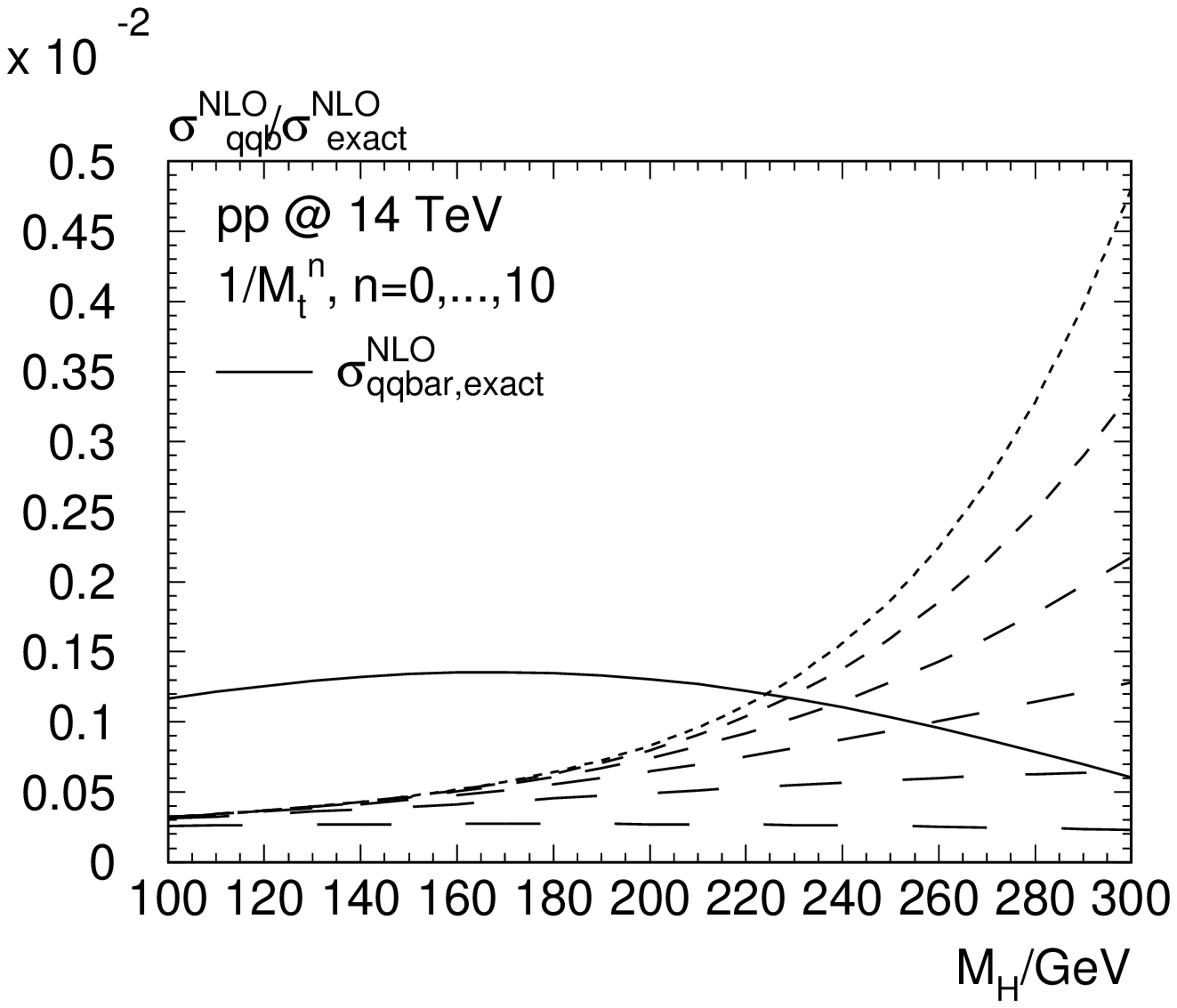}} &
      \subfigure[]{\includegraphics[bb=110 265 465
        560,width=.45\textwidth]{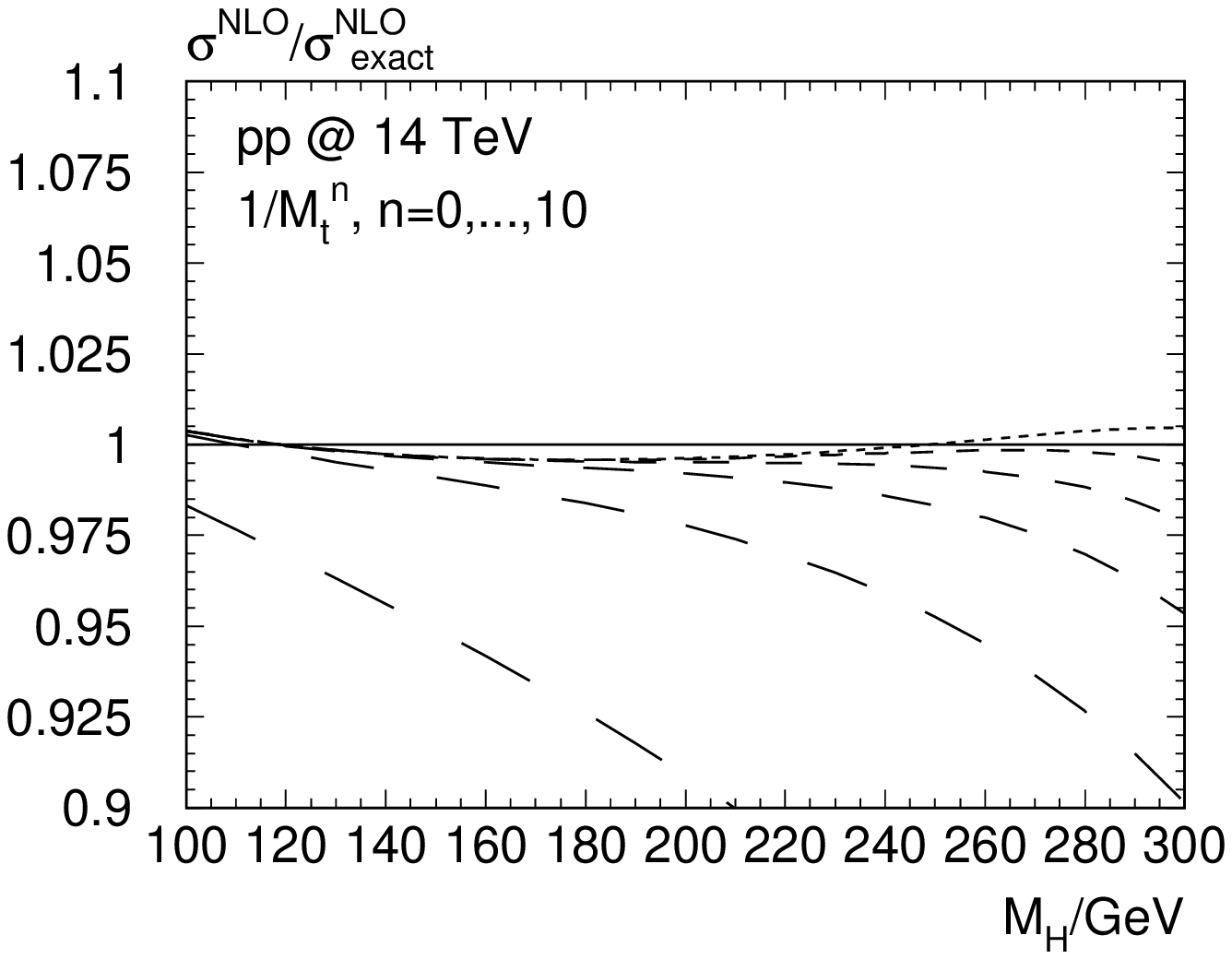}}
    \end{tabular}
    \parbox{.9\textwidth}{
      \caption[]{\label{fig::ratnlo}\sloppy (a)-(c) Sub-channel
        contributions to the hadronic cross section at \nlo{},
        normalized to the full result. Note that $gg$ includes the exact
        \lo{} contribution, cf.\,\eqn{eq::sigmtexp}. Dashed: including
        terms of order $1/\mtop{}^{2n}$ in the numerator ($n=0,\ldots,5$
        from long to short dashes). Solid: exact. (d) Sum over all
        sub-channels.  }}
  \end{center}
\end{figure}

\fig{fig::ratnlo} shows the relative $gg$, $qg$ and $q\bar q$
contribution $\sigma_{\alpha\beta}^\nlo(M_t^n)$ to the total hadronic
cross section. The dashed lines correspond to successively higher orders
in $1/\mtop$, while the solid line shows the exact result.  The curves
are all normalized to the exact \nlo{} cross section $\sigma^\nlo$. For
the $gg$ and the $qg$ channels, one observes excellent convergence
towards the exact result (solid line). The small deviations are
reflections of the deviations between the solid and the dashed lines in
\fig{fig::sig1-match}\,(a)-(d). As pointed out above, the mass effects
in the $qg$ channel are quite large, ranging from roughly a factor of
two to four in the relevant Higgs mass range. Of course, the overall
size of the $qg$ channel is below 5\%. As expected, the picture in the
$q\bar q$ channel is significantly worse. Only the order of magnitude is
captured, but there is no sign of convergence towards the exact result
whatsoever. Its contribution to the total cross section is only of order
$10^{-3}$ though and thus irrelevant.

\begin{figure}
  \begin{center}
    \subfigure[]{%
      \includegraphics[bb=110 265 465
        560,width=.45\textwidth]{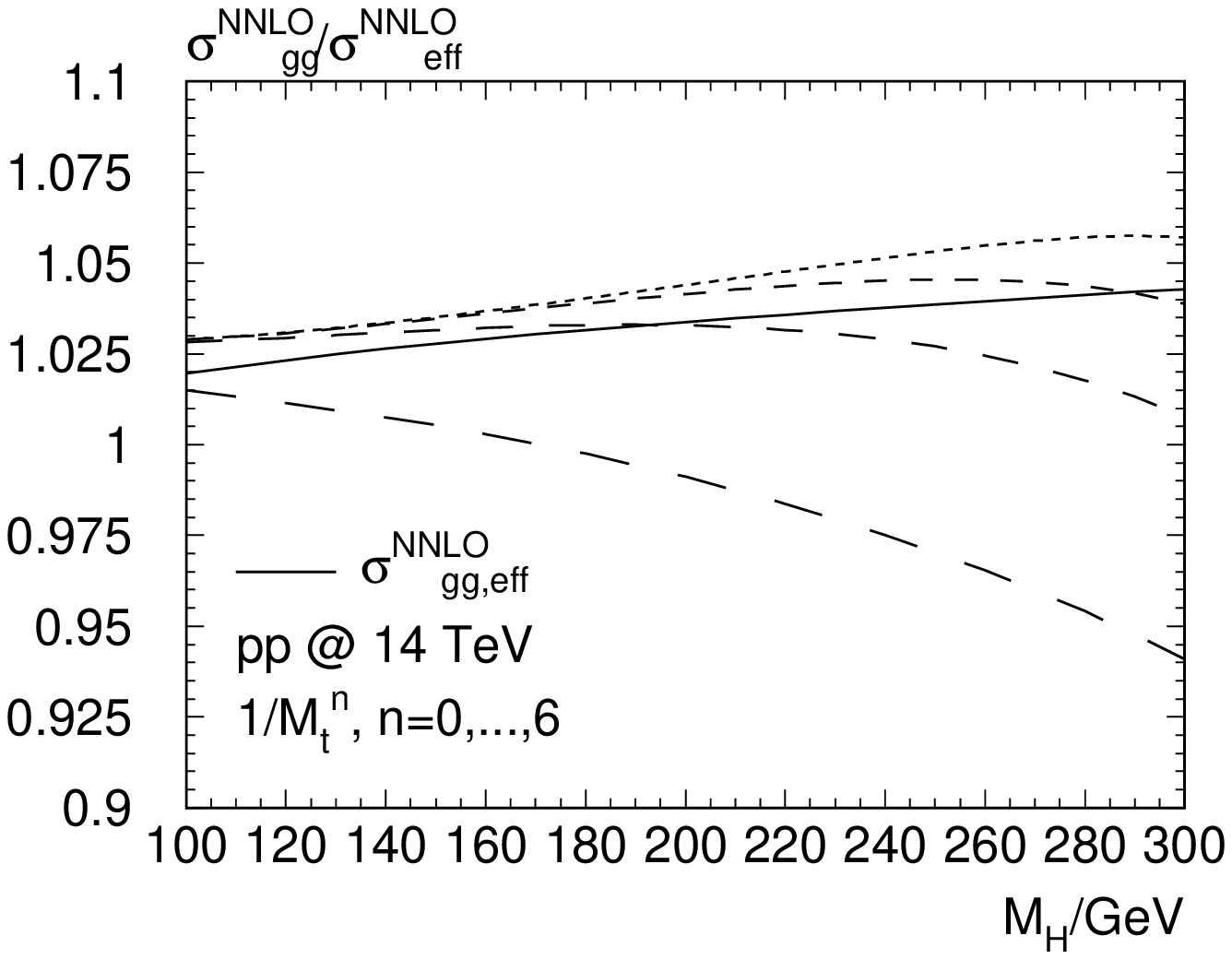}}\qquad
    \subfigure[]{%
      \includegraphics[bb=110 265 465
        560,width=.45\textwidth]{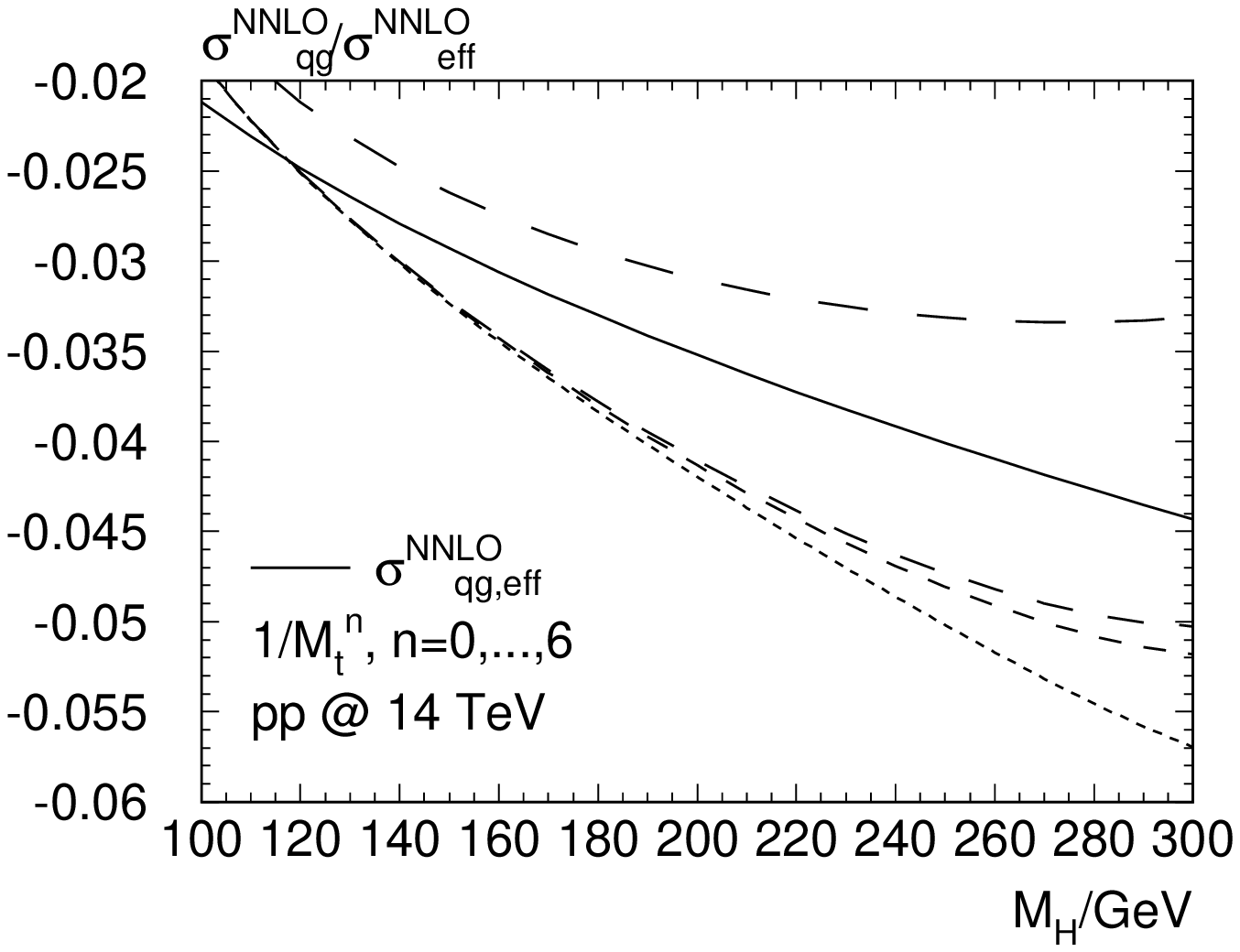}}
    \subfigure[]{%
      \includegraphics[bb=110 265 465
        560,width=.45\textwidth]{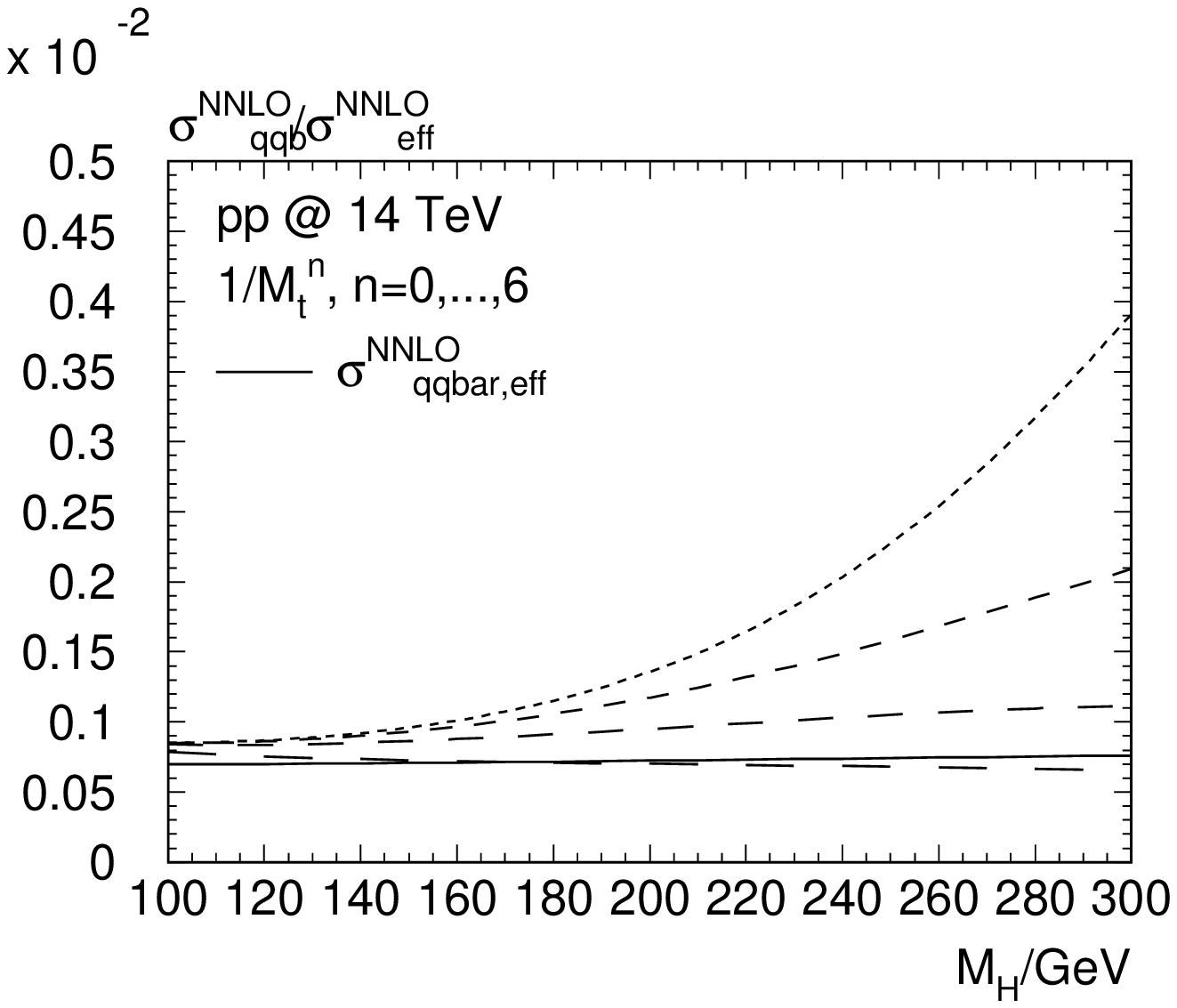}}\qquad
    \subfigure[]{%
      \includegraphics[bb=110 265 465
        560,width=.45\textwidth]{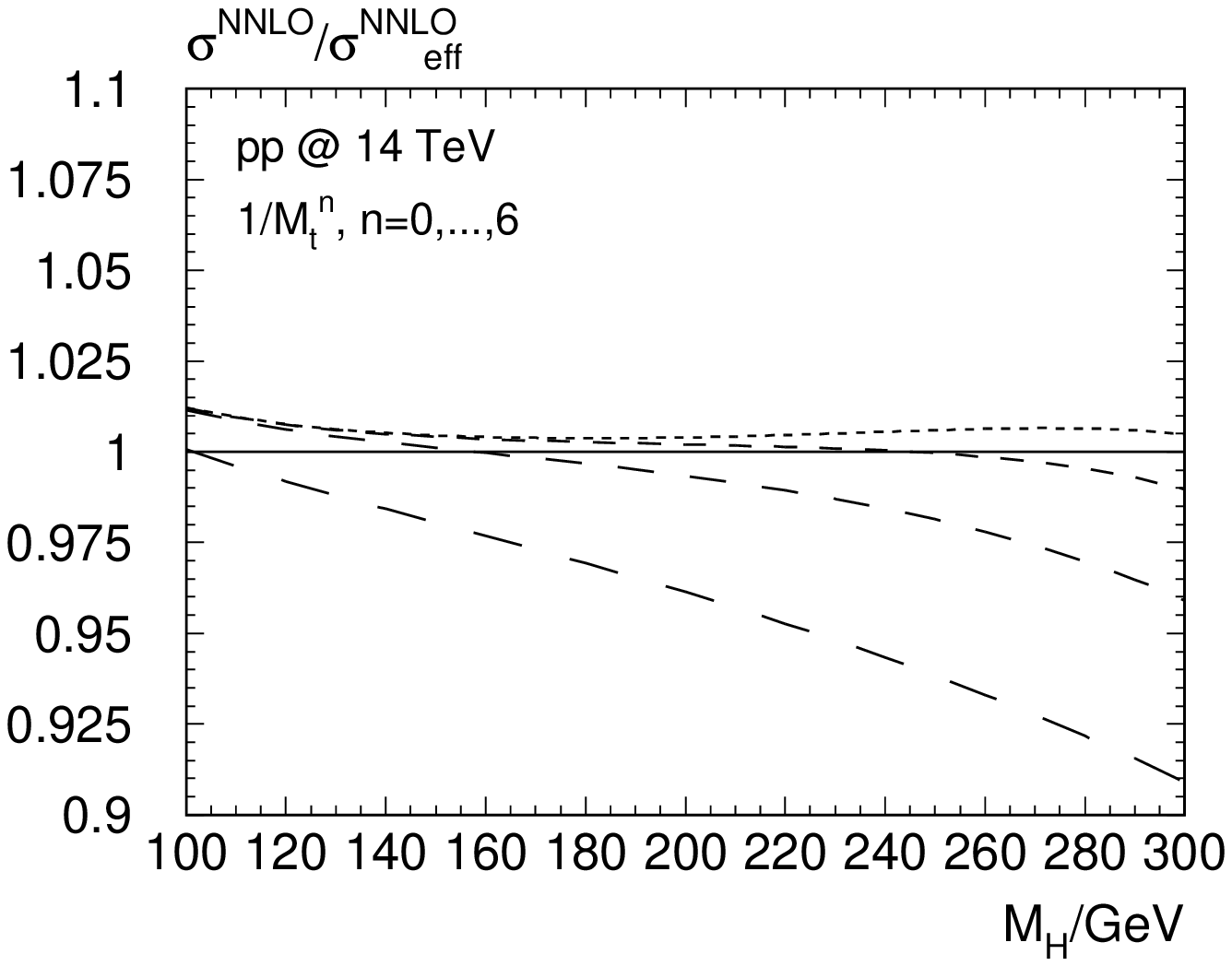}}
    \parbox{.9\textwidth}{
      \caption[]{\label{fig::ratnnlo}\sloppy (a)-(c) Sub-channel
        contributions to the hadronic cross section at \nnlo{},
        normalized to the full \nnlo{} \eft{} result (\lhc{}
        conditions). Note that all channels include their lower order
        contributions {\it in the \eft{} approach}
        (cf.~\eqn{eq::sigmtexp}). Dashed: including terms of order
        $1/\mtop{}^{2n}$ in the numerator ($n=0,1,2,3$ from long to
        short dashes). Solid: \eft{} result. (d) Sum over all
        sub-channels.  }}
  \end{center}
\end{figure}

The corresponding plots at \nnlo{} are shown in
\fig{fig::ratnnlo}. Since there is no exact result in this case, we
normalize the curves to the full \nnlo{} \eft{} result,
cf.\,\eqn{eq::heavytop}. Also, the solid lines always refer to the
subchannels evaluated in the \eft{} approach.  The observations are
quite similar as at \nlo{}: the difference between the \eft{} result and
the $1/\mtop$ expansion for the $gg$ channel is about 1\% which is of
the order of the accuracy to which we expect the capture the mass
effects. In the $qg$ channel, the relative difference between the \eft{}
result and the $1/\mtop$ expansion is significantly larger ($\sim
20\%$), but the influence of this effect on the total cross section is
again only of order $1\%$ due to the strong suppression of the $qg$
channel.  The $q\bar q$ channel does not seem to converge very well, but
is numerically negligible (the true mass effects are not expected to
change this).

\begin{figure}
  \begin{center}
    \subfigure[]{%
    \includegraphics[bb=110 265 465
      560,width=.45\textwidth]{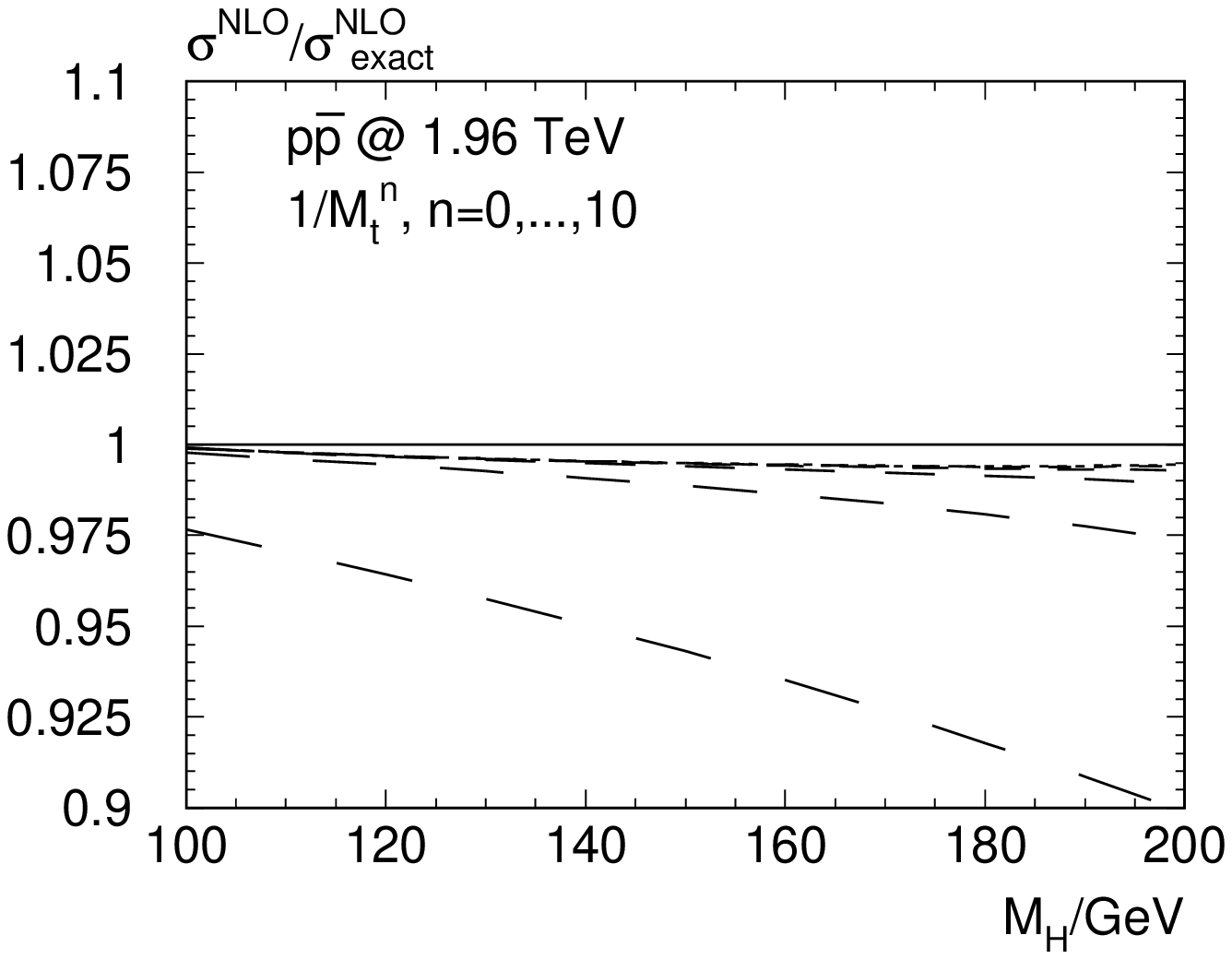}}\qquad
    \subfigure[]{%
    \includegraphics[bb=110 265 465
      560,width=.45\textwidth]{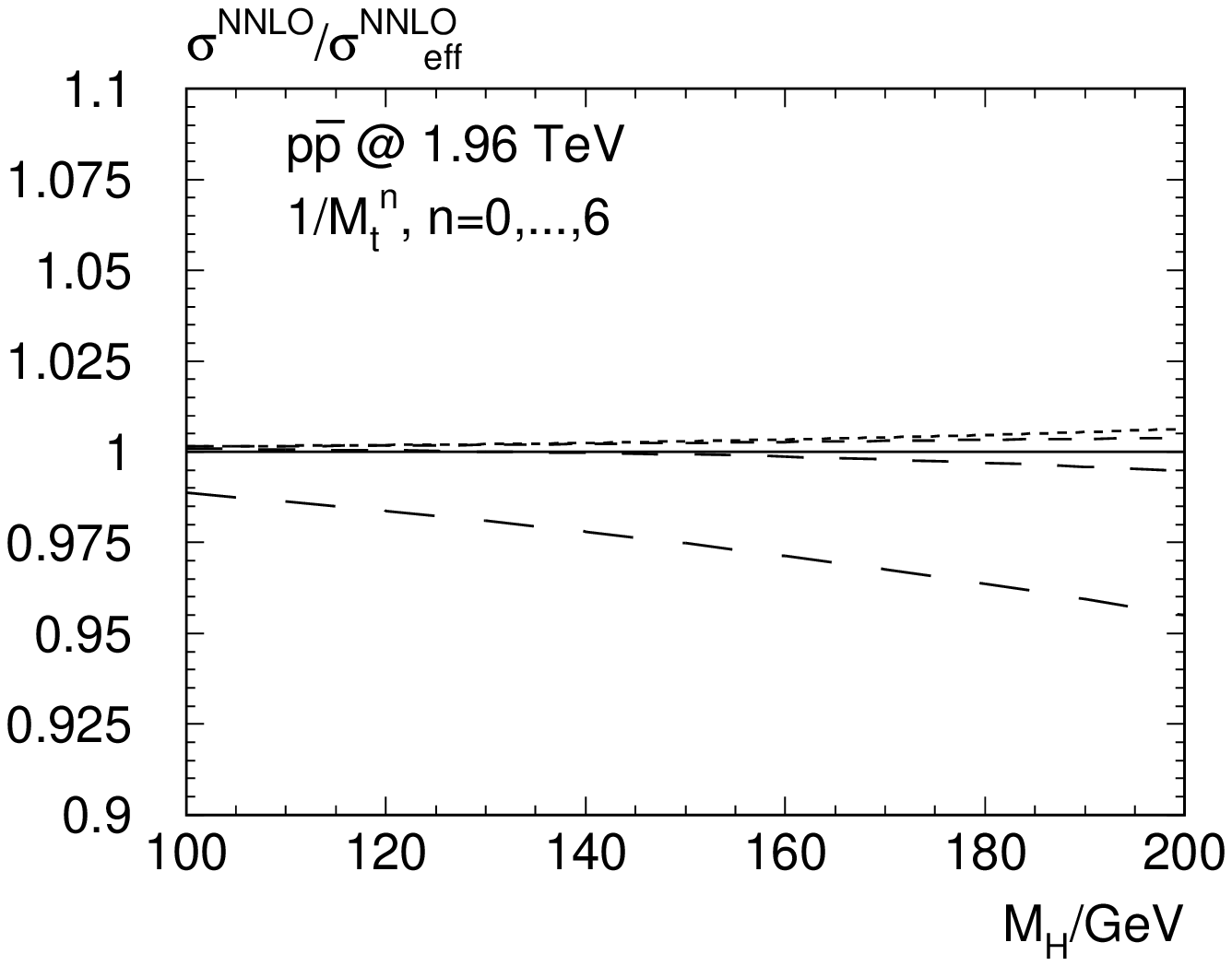}}
    \parbox{.9\textwidth}{
      \caption[]{\label{fig::ratmt-tev}\sloppy Hadronic cross section at
        (a)~\nlo{} and (b)~\nnlo{} for the Tevatron, normalized to the
        full \nnlo{} \eft{} result. Note that the lower order
        contributions are included {\it in the \eft{} approach}
        (cf.~\eqn{eq::sigmtexp}). Dashed: including terms of order
        $1/\mtop{}^{2n}$ in the numerator ($n=0,1,2,3$ from long to
        short dashes). Solid: \eft{} result.  }}
  \end{center}
\end{figure}

The hadronic results for the Tevatron are shown in \fig{fig::ratmt-tev}
at \nlo{} and \nnlo{}. The conclusions are very similar to those for the
\lhc{}, thus justifying the use of the \eft{} approximation for Higgs
searches also in this case~\cite{tevatronhiggs}.

Overall, we conclude that the final result for the \nnlo{} cross section
including top mass effects is within $1\%$ of the \eft{} result.

\paragraph{Dependence on \bld{B_{\alpha\beta}^{(2)}}.}
As pointed out above, the constants $B_{\alpha\beta}^{(2)}$ for the
large-$\hat s$ behaviour are currently unknown. From the curves in
\fig{fig::sig2-matchqg} and \ref{fig::sig2-matchqq}, our choice
$\sigma_0B_{\alpha\beta}^{(2)} = \hat\sigma_{\alpha\beta}^{(2)}(0)$
seems to be reasonable, leading to rather smooth curves over the full
$x$-range.  Nevertheless, in order to estimate the uncertainty induced
by this unknown constant, we set $\sigma_0B_{gg}^{(2)} =
t\times\hat\sigma_{gg}^{(2)}(0)$ and find that the dependence of the
hadronic cross section on $t$ is very well described by a linear
function:
\begin{equation}
\begin{split}
\sigma^\nnlo\bigg|_t &\approx
\left(1 - 0.01\,t\right)
\sigma^\nnlo
\end{split}
\end{equation}
Again recalling the smoothness of the curves in \fig{fig::sig2-matchqg}
and \ref{fig::sig2-matchqq}, we do not expect the parameter $t$ to be
significantly larger than one. The resulting uncertainty is therefore at
most at the percent level and therefore much smaller than the scale
uncertainty of the \nnlo{} result.

\paragraph{Is the heavy-top limit a coincidence?}
Let us conclude this section with a remark on the {\it extended \eft{}
  approach} as mentioned in the discussion after \eqn{eq::sigmtexp}. It
would be possible that the high quality of the \eft{} approach is a
coincidence, in the sense that there is an accidental cancellation among
the higher order terms in the $1/\mtop{}$ expansion of the
$\Delta_{\alpha\beta}$. This would have a significant effect on the
applicability of the \eft{} approach to other quantities, of course.

However, we have checked that this is not the
case. All the curves of the extended \eft{} approach lie within 1\% of
our final result.

\section{Conclusions}
The hadronic Higgs production cross section due to gluon fusion was
presented including effects from a finite top quark mass. We have
extended previous analyses by deriving the high-energy limits of all
partonic sub-channels and combining them with the known $1/\mtop$
expansions.  Although the mass effects on the absolute size of the $qg$
channel are large, they have no significant effect on the total hadronic
cross section. Therefore, the main conclusions of previous
analyses~\cite{Harlander:2009bw,Pak:2009bx} remain valid, and the \eft{}
approach is still justified.

\paragraph{Acknowledgments.}
We would like to thank Stefano Forte for useful comments on the
manuscript.  This work was supported by {\abbrev DFG} contract
HA~2990/3-1, and by the {\it Helmholtz Alliance ``Physics at the
  Terascale''}.  SM would like to thank Bergische Universit\"at
Wuppertal for the kind hospitality.  The work of SM is supported by UK's
STFC.

\def\app#1#2#3{{\it Act.~Phys.~Pol.~}\jref{\bf B #1}{#2}{#3}}
\def\apa#1#2#3{{\it Act.~Phys.~Austr.~}\jref{\bf#1}{#2}{#3}}
\def\annphys#1#2#3{{\it Ann.~Phys.~}\jref{\bf #1}{#2}{#3}}
\def\cmp#1#2#3{{\it Comm.~Math.~Phys.~}\jref{\bf #1}{#2}{#3}}
\def\cpc#1#2#3{{\it Comp.~Phys.~Commun.~}\jref{\bf #1}{#2}{#3}}
\def\epjc#1#2#3{{\it Eur.\ Phys.\ J.\ }\jref{\bf C #1}{#2}{#3}}
\def\fortp#1#2#3{{\it Fortschr.~Phys.~}\jref{\bf#1}{#2}{#3}}
\def\ijmpc#1#2#3{{\it Int.~J.~Mod.~Phys.~}\jref{\bf C #1}{#2}{#3}}
\def\ijmpa#1#2#3{{\it Int.~J.~Mod.~Phys.~}\jref{\bf A #1}{#2}{#3}}
\def\jcp#1#2#3{{\it J.~Comp.~Phys.~}\jref{\bf #1}{#2}{#3}}
\def\jetp#1#2#3{{\it JETP~Lett.~}\jref{\bf #1}{#2}{#3}}
\def\jphysg#1#2#3{{\small\it J.~Phys.~G~}\jref{\bf #1}{#2}{#3}}
\def\jhep#1#2#3{{\small\it JHEP~}\jref{\bf #1}{#2}{#3}}
\def\mpl#1#2#3{{\it Mod.~Phys.~Lett.~}\jref{\bf A #1}{#2}{#3}}
\def\nima#1#2#3{{\it Nucl.~Inst.~Meth.~}\jref{\bf A #1}{#2}{#3}}
\def\npb#1#2#3{{\it Nucl.~Phys.~}\jref{\bf B #1}{#2}{#3}}
\def\nca#1#2#3{{\it Nuovo~Cim.~}\jref{\bf #1A}{#2}{#3}}
\def\plb#1#2#3{{\it Phys.~Lett.~}\jref{\bf B #1}{#2}{#3}}
\def\prc#1#2#3{{\it Phys.~Reports }\jref{\bf #1}{#2}{#3}}
\def\prd#1#2#3{{\it Phys.~Rev.~}\jref{\bf D #1}{#2}{#3}}
\def\pR#1#2#3{{\it Phys.~Rev.~}\jref{\bf #1}{#2}{#3}}
\def\prl#1#2#3{{\it Phys.~Rev.~Lett.~}\jref{\bf #1}{#2}{#3}}
\def\pr#1#2#3{{\it Phys.~Reports }\jref{\bf #1}{#2}{#3}}
\def\ptp#1#2#3{{\it Prog.~Theor.~Phys.~}\jref{\bf #1}{#2}{#3}}
\def\ppnp#1#2#3{{\it Prog.~Part.~Nucl.~Phys.~}\jref{\bf #1}{#2}{#3}}
\def\rmp#1#2#3{{\it Rev.~Mod.~Phys.~}\jref{\bf #1}{#2}{#3}}
\def\sovnp#1#2#3{{\it Sov.~J.~Nucl.~Phys.~}\jref{\bf #1}{#2}{#3}}
\def\sovus#1#2#3{{\it Sov.~Phys.~Usp.~}\jref{\bf #1}{#2}{#3}}
\def\tmf#1#2#3{{\it Teor.~Mat.~Fiz.~}\jref{\bf #1}{#2}{#3}}
\def\tmp#1#2#3{{\it Theor.~Math.~Phys.~}\jref{\bf #1}{#2}{#3}}
\def\yadfiz#1#2#3{{\it Yad.~Fiz.~}\jref{\bf #1}{#2}{#3}}
\def\zpc#1#2#3{{\it Z.~Phys.~}\jref{\bf C #1}{#2}{#3}}
\def\ibid#1#2#3{{ibid.~}\jref{\bf #1}{#2}{#3}}
\def\otherjournal#1#2#3#4{{\it #1}\jref{\bf #2}{#3}{#4}}
\newcommand{\jref}[3]{{\bf #1} (#2) #3}
\newcommand{\bibentry}[4]{#1, {\it #2}, #3\ifthenelse{\equal{#4}{}}{}{, }#4.}
\newcommand{\hepph}[1]{{\tt hep-ph/#1}}
\newcommand{\mathph}[1]{[math-ph/#1]}
\newcommand{\arxiv}[2]{{\tt arXiv:#1}}

\end{document}